\documentclass[journal]{IEEEtran}
\usepackage{lineno}
\usepackage{cite}
\usepackage{hyperref}
\usepackage{amsmath,amssymb,amsfonts}
\usepackage{amsmath}
\usepackage{amsthm}

\usepackage{graphicx}
\usepackage{textcomp}
\usepackage{xcolor}
\usepackage{graphicx}
\usepackage{float}
\usepackage{subfigure}
\usepackage{amsmath}
\usepackage{amsfonts,amssymb}
\usepackage{mathrsfs}
\usepackage{mathtools}
\usepackage{algorithm}
\usepackage{algorithmicx}
\usepackage{algpseudocode}
\usepackage{bm}
\usepackage{multirow}
\usepackage{array}
\usepackage{amssymb}
\usepackage{amsmath}
\usepackage{cite}
\usepackage{url}
\usepackage{xcolor}
\usepackage{cite,graphicx,amsmath,amssymb}
\usepackage{subfigure}
\usepackage{fancyhdr}
\usepackage{mdwmath}
\usepackage{mdwtab}
\usepackage{caption}
\usepackage{amsthm}
\usepackage{setspace}
\usepackage{bm}
\usepackage{algorithm}
\usepackage{algpseudocode}
\usepackage{mathtools}
\usepackage{dsfont}
\usepackage{bbm}
\usepackage{cases}
\usepackage{stfloats}
\newtheorem{remark}{Remark}
\theoremstyle{definition}
\newtheorem{theorem}{Theorem}

\newtheorem{lemma}{Lemma}

\newtheorem{corollary}{Corollary}

\makeatletter
\newcommand{\biggg}{\bBigg@{3}}
\newcommand{\Biggg}{\bBigg@{3.5}}
\makeatother
\begin{document}
	
	\title{Channel Capacity of Near-Field Line-of-Sight Multiuser Communications}
	\author{Boqun~Zhao, Chongjun~Ouyang, Xingqi~Zhang, and Yuanwei~Liu\vspace{-10pt}
		\thanks{An earlier version of this paper is to be presented in part at the IEEE Global Communications Conference, Cape Town, South Africa, December 2024 \cite{globecom_2024}.}
		\thanks{B. Zhao and X. Zhang are with Department of Electrical and Computer Engineering, University of Alberta, Edmonton AB, T6G 2R3, Canada (email: \{boqun1, xingqi.zhang\}@ualberta.ca).}
		\thanks{C. Ouyang is with the School of Electrical and Electronic Engineering, University College Dublin, Dublin, D04 V1W8, Ireland, and also with the School of Electronic Engineering and Computer Science, Queen Mary University of London, London, E1 4NS, U.K. (e-mail: chongjun.ouyang@ucd.ie).}
		\thanks{Y. Liu is with the Department of Electrical and Electronic Engineering, The University of Hong Kong, Hong Kong (e-mail: yuanwei@hku.hk).}
	}

	\maketitle
	
	\begin{abstract}
		The channel capacity of near-field (NF) communications is characterized by considering three types of line-of-sight multiuser channels: \romannumeral1) multiple access channel (MAC), \romannumeral2) broadcast channel (BC), and \romannumeral3) multicast channel (MC). For NF MAC and BC, closed-form expressions are derived for the sum-rate capacity as well as the capacity region under a two-user scenario. These results are further extended to scenarios with an arbitrary number of users. For NF MC, closed-form expressions are derived for the two-user channel capacity and the capacity upper bound with more users. Further insights are gleaned by exploring special cases, including scenarios with infinitely large array apertures, co-directional users, and linear arrays. For comparison, the MAC and BC sum-rates achieved by typical linear combiners and precoders are also analyzed. Theoretical and numerical results are presented and compared with far-field communications to demonstrate that: \romannumeral1) the NF capacity of these three channels converges to finite values rather than growing unboundedly as the number of array elements increases; \romannumeral2) the capacity of the MAC and BC with co-directional users can be improved by using the additional range dimensions in NF channels to reduce inter-user interference (IUI); and \romannumeral3) the MC capacity benefits less from the NF effect compared to the MAC and BC, as multicasting is less sensitive to IUI.
	\end{abstract}
	
	
	\vspace{-5pt}
	\section{Introduction}
	In light of recent developments in wireless networks, emerging technical trends, such as the application of extremely large-scale antenna arrays and tremendously high frequencies, significantly expand the near-field (NF) region, even to hundreds of meters \cite{liu2023near,wangzhe}. It is important to emphasize that electromagnetic (EM) waves exhibit distinct propagation characteristics in the NF region compared to the far field where EM waves can be adequately approximated as planar waves. In the near field, a more precise spherical wave-based model becomes necessary \cite{NF_overview}. Therefore, it is imperative to reevaluate the performance of multiuser systems from an NF perspective, which ensures that modeling and analysis accurately reflect these distinct propagation characteristics.
	
	By leveraging the additional range dimensions introduced by spherical wave propagation \cite{liu2023near}, near-field communications (NFC) can manage inter-user interference (IUI) more flexibly. This capability has inspired a considerable amount of research focused on multiuser NF beamforming design; see \cite{MU_beamforming_1,MU_beamforming_2,MU_beamforming_3,MU_beamforming_4,liu2024near} for more details. On the other hand, the fundamental performance limits of NF multiuser communications have not been adequately studied. Only the achievable rates for an uplink NF multiuser channel were studied in \cite{zeng_MU}, where the maximum-ratio transmission, zero-forcing, and minimum mean-squared error beamforming strategies are considered. By now, one of the most fundamental problems in NF multiuser communications remains unsolved: \emph{channel capacity characterization}. This issue has long been of significant value and interest within the realm of multiuser multiple-antenna communications \cite{goldsmith}.
	
	Research on multiuser communications has focused on several primary models that play fundamental roles in shaping the theoretical and practical landscapes of communication networks \cite{el1980multiple,el2011network}. Among these models, three classical ones are outlined here, which are also the main focus of this article: the \emph{multiple access channel (MAC)} \cite{yu2004iterative,MAC_asy}, where multiple transmitters communicate with a single receiver; the \emph{broadcast channel (BC)} \cite{weingarten2006capacity,BC_twc}, where a single transmitter broadcasts different messages to multiple receivers; and the \emph{multicast channel (MC)} \cite{jindal2006capacity,multicast_tsp}, where a single transmitter sends a common message to multiple receivers. The MAC addresses the challenges of uplink communications, while the BC and MC capture the essence of downlink communications. Prior studies have extensively characterized the \emph{(sum-rate) capacity and capacity regions} of these channels under various conditions, which unveil the nature of multiuser communications and its capacity limits; see \cite{goldsmith,el1980multiple,el2011network} and the references therein for further details.
		
	Some existing literature has analyzed the channel capacity of NFC. For example, the capacity of a point-to-point multiple-input multiple-output (MIMO) channel is analyzed from a circuit perspective \cite{NF_capacity_2}. However, this work sheds few insights on the NF effect on channel capacity. As an advancement, the authors of \cite{liu2023near} analyzed the channel capacity of an NF multiple-input single-output (MISO) system and revealed the impact of the NF effect on the capacity scaling law. Further work by the authors of \cite{NF_capacity_1} approximated the capacity of a linear arrays-based NF MIMO channel from a degrees-of-freedom (DoFs) perspective. In \cite{NF_capacity_3}, the asymptotic NF capacity for extremely large-scale MIMO (XL-MIMO) achieved by a beamspace modulation strategy is studied. Leveraging the NF property for capacity improvement, the authors of \cite{NF_capacity_4} proposed a distance-aware precoding architecture and the corresponding precoding algorithm for XL-MIMO. Additionally, in \cite{NF_capacity_5}, the authors proposed a generalized NF channel modeling for point-to-point holographic MIMO systems and studied the capacity limit. It is important to note that all existing works regarding NF capacity focus only on single-user scenarios, while the more general and complex scenarios involving multiple users remain unexplored.

	Motivated by existing research gaps, this article analyzes the channel capacity of NF multiuser communications in terms of the aforementioned three fundamental channels: \emph{MAC}, \emph{BC}, and \emph{MC}\footnote{It is worth noting that the capacity regions of the MAC and BC are formally equivalent to those classically studied in information theory. The main contributions of this paper lie in revisiting these results within the NF setup, which allows us to explore the impact of NF effects on channel capacity. More specifically, while the basic expressions for the capacity of these channels are well known and can be found in many textbooks, their specific forms under NF channel models have not been presented in the current literature. Additionally, a detailed discussion of the effects of NF and spherical-wave propagation on channel capacity is still lacking. This gap in the literature has motivated our work.}. To facilitate theoretical investigations into fundamental performance limits and asymptotic behaviors, this discussion is focused on line-of-sight (LoS) channels. Extensions to non-LoS NF channels will be addressed in future work. The main contributions are summarized as follows.
	\begin{itemize}
		\item We propose a transmission framework for planar array-based multiuser NFC. This framework models NF propagation by incorporating not only varying free-space path losses and phase shifts for each element but also the influence of the projected aperture, resulting in superior accuracy compared to conventional NF models.
		\item Building upon the multiuser NFC framework, we derive closed-form expressions for the sum-rate channel capacity of NF MAC and BC under a two-user scenario, along with the corresponding capacity region. These results are then extended to more general scenarios with an arbitrary number of users. For comparison, we also analyze the MAC and BC sum-rates achieved by typical linear combiners and precoders. In the context of NF MC, we propose an optimal linear beamforming design for a two-user scenario and derive the corresponding multicast capacity. For scenarios with more users, we provide a closed-form expression for the upper bound of the NF multicast capacity.
		\item To gain deeper insights into system design, we explore three special cases: scenarios with infinitely large array apertures, co-directional users, and linear arrays. For each case, we revisit the corresponding NF capacity and compare it with its far-field (FF) counterpart. This analysis enables us to establish the power scaling law and optimal power allocation policy.    
		\item We present numerical results to demonstrate that, for MAC and BC, the asymptotic orthogonality of NFC in the range domain can enhance both the sum-rate capacity and the capacity region for users in the same direction. Conversely, under the same condition, the multicast capacity exhibits higher values under the FF model. Furthermore, we observe that as the number of array elements increases, the NF capacity converges to finite limits for all three multiuser channels, while its FF counterpart grows unlimitedly, potentially violating energy conservation laws.
	\end{itemize}
	
	The remainder of this article is organized as follows. Section \ref{system} presents the NF channel model and defines the capacity of the three multiuser channels. Then, Sections \ref{sec_MAC}, \ref{sec_BC} and \ref{sec_MC} analyzes the NF channel capacity of MAC, BC and MC, respectively. Section \ref{numerical} provides numerical results to validate the derived insights. Finally, Section \ref{conclusion} concludes the article.
	
	\subsubsection*{Notations}
	Throughout this paper, scalars, vectors, and matrices are denoted by non-bold, bold lower-case, and bold upper-case letters, respectively. For the matrix $\mathbf{A}$, ${\mathbf{A}}^{\mathsf{T}}$, ${\mathbf{A}}^{*}$, and ${\mathbf{A}}^{\mathsf{H}}$ denote the transpose, conjugate, and transpose conjugate of $\mathbf{A}$, respectively. For the square matrix $\mathbf{B}$, ${\mathsf{tr}}(\mathbf{B})$ and $\det(\mathbf{B})$ denote the trace and determinant of $\mathbf{B}$, respectively. The notations $\lvert a\rvert$ and $\lVert \mathbf{a} \rVert$ denote the magnitude and norm of scalar $a$ and vector $\mathbf{a}$, respectively. The identity matrix with dimensions $N\times N$ is represented by $\mathbf{I}_N$, and the zero matrix is denoted by $\mathbf{0}$. The matrix inequality ${\mathbf{A}}\succeq{\mathbf{0}}$ implies that $\mathbf{A}$ is positive semi-definite. The sets $\mathbbmss{R}$ and $\mathbbmss{C}$ stand for the real and complex spaces, respectively, and notation $\mathbbmss{E}\{\cdot\}$ represents mathematical expectation. The notation $f(x)=\mathcal{O}\left(g(x)\right)$ means that $\lim\sup_{x\rightarrow\infty}\frac{\lvert f(x)\rvert}{g(x)}<\infty$. Finally, ${\mathcal{CN}}({\bm\mu},\mathbf{X})$ is used to denote the circularly-symmetric complex Gaussian distribution with mean $\bm\mu$ and covariance matrix $\mathbf{X}$.
	
	\section{System Model}\label{system}
\begin{figure}[!t]
		\centering
		\setlength{\abovecaptionskip}{0pt}
		\includegraphics[height=0.27\textwidth]{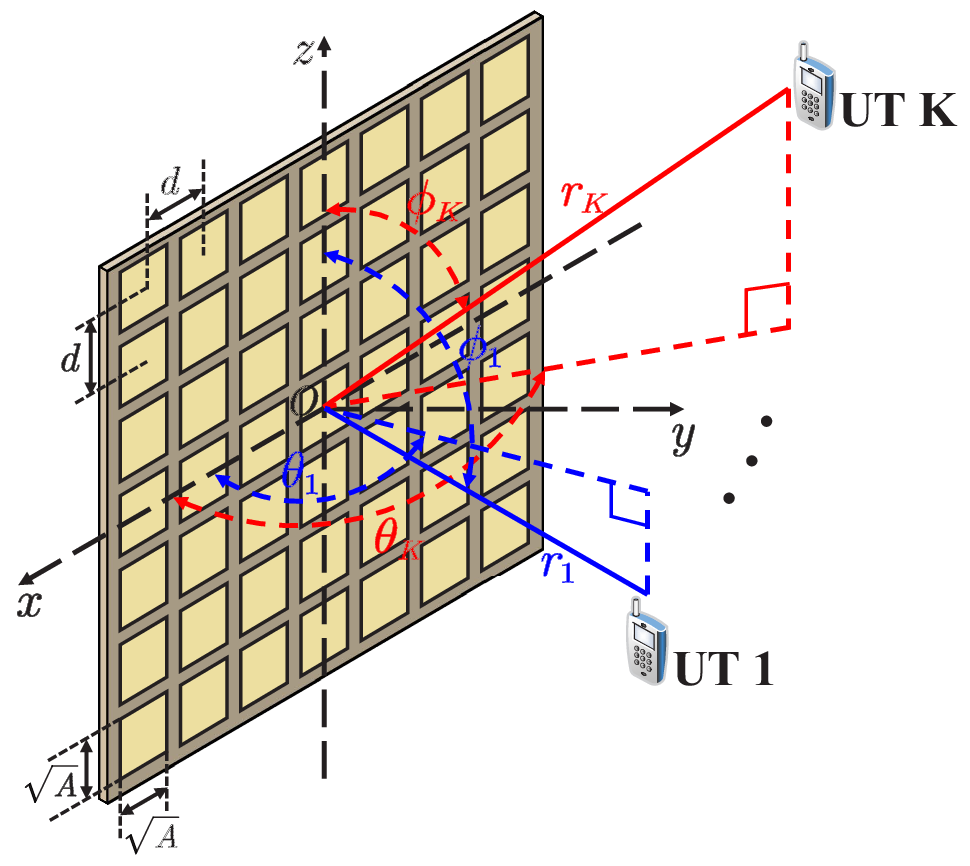}
		\caption{Illustration of the array geometry.}
		\label{LoS_3D_Model_MU}
		\vspace{-10pt}
	\end{figure}

	Consider a narrowband single-cell multiuser system where one base station (BS) simultaneously serves a set of $K$ user terminals (UTs), as depicted in {\figurename} {\ref{LoS_3D_Model_MU}}. Each UT $k\in{\mathcal{K}}\triangleq\{1,\ldots,K\}$ is a single-antenna device, while the BS is equipped with a large-aperture uniform planar array (UPA) containing $M\gg K$ antennas. The deployment of this massive array extends the NF region, which resides \emph{all the UTs within the near field}. Since NF channels are sparsely-scattered and dominated by LoS propagation \cite{liu2023near}, we consider pure-LoS propagation scenarios for a theoretical exploration of fundamental capacity limits. 
	
	As illustrated in {\figurename} {\ref{LoS_3D_Model_MU}}, the UPA is placed on the $x$-$z$ plane and centered at the origin. We set $M=M_{x}M_{z}$, where $M_{x}$ and $M_{z}$ denote the number of array elements along the $x$- and $z$-axes, respectively. Without loss of generality, we assume that $M_{x}$ and $M_{z}$ are odd numbers with $M_x=2\tilde{M}_x+1$ and $M_z=2\tilde{M}_z+1$. The physical dimensions of each BS array element along the $x$- and $z$-axes are denoted by $\sqrt{A}$, and the inter-element distance is $d$, where $d\geq\sqrt{A}$. The central location of the $(m_x,m_z)$th element is denoted by ${\mathbf{s}}_{m_x,m_z}=[m_xd,0,m_zd]^{\mathsf{T}}$, where $m_x\in{\mathcal{M}}_x\triangleq\{0,\pm1,\ldots,\pm\tilde{M}_x\}$ and $m_z\in{\mathcal{M}}_z\triangleq\{0,\pm1,\ldots,\pm\tilde{M}_z\}$. 
	
	Regarding each UT $k\in\mathcal{K}$, they are all assumed to be equipped with a single hypothetical isotropic array element to receive or transmit signals. Let $r_k$ denote the propagation distance from the center of the antenna array to UT $k$, and $\theta_k\in(0,\pi)$ and $\phi_k\in(0,\pi)$ denote the associated azimuth and elevation angles, respectively. Thus, the location of UT $k$ can be expressed as ${\mathbf{r}}_k=[r_k\Phi_k,r_k\Psi_k,r_k\Omega_k]^{\mathsf{T}}$, where $\Phi_k\triangleq\sin{\phi_k}\cos{\theta_k}$, $\Psi_k\triangleq\sin\phi_k\sin\theta_k$, and $\Omega_k\triangleq\cos{\phi_k}$. In particular, the distance between UT $k$ and the center of the $(m_x,m_z)$th array element is given by
	\begin{align}
		&r_{m_x,m_z,k}=\lVert{\mathbf{r}}_k-{\mathbf{s}}_{m_x,m_z}\rVert\notag\\
		&=r_k\sqrt{(m_x^2+m_z^2)\epsilon _k^2-2m_x\epsilon _k\Phi_k-2m_z\epsilon _k\Omega_k+1},\label{distance}   
	\end{align}
	where $\epsilon _k=\frac{d}{r_k}$. Note that $r_k=r_{0,0,k}$, and since the array element separation $d$ is typically on the order of a wavelength, in practice, we have $\epsilon _k\ll1$.
	\subsection{Channel Model}\label{Section: System Model: Near-Field Channel Model}
	The channel response from UT $k$ to the $(m_x,m_z)$th antenna element of the BS is given by $h_{m_x,m_z,k}=\sqrt{g_{m_x,m_z,k}}\mathrm{e}^{-\mathrm{j}\phi_{m_x,m_z,k}}$, where
	\begin{align}
		g_{m_x,m_z,k}=\int_{{\mathcal{S}}_{m_x,m_z}}{\mathsf{g}}(\mathbf{r}_k,{\mathbf{s}})
		{\frac{\lvert{\mathbf{e}}_y^{\mathsf{T}}({\mathbf{s}}-{\mathbf{r}}_k)\rvert}{\lVert{\mathbf{r}}_k-{\mathbf{s}}\rVert}}{\rm{d}}{\mathbf{s}}.
	\end{align}
	Specifically, ${\mathcal{S}}_{m_x,m_z}=[m_xd-{\sqrt{A}}/{2},m_xd+{\sqrt{A}}/{2}]\times[m_zd-{\sqrt{A}}/{2},m_zd+{\sqrt{A}}/{2}]$ denotes the aperture of the $(m_x,m_z)$th array element, and the term $\frac{\lvert{\mathbf{e}}_y^{\mathsf{T}}({\mathbf{s}}-{\mathbf{r}})\rvert}{\lVert{\mathbf{r}}-{\mathbf{s}}\rVert}$ captures the impact of the projected aperture of the UPA with ${\mathbf{e}}_y=[0,1,0]^{\mathsf{T}}$ being the UPA normal vector. Moreover, ${\mathsf{g}}(\mathbf{r},{\mathbf{s}})$ models the influence of free-space EM propagation, which is given by \cite{ouyang2024primer}
	\begin{equation}\label{Green_Function_Full_Version}
		{\mathsf{g}}(\mathbf{r},{\mathbf{s}})=\frac{1}{{4\pi} \lVert{\mathbf{r}}-{\mathbf{s}}\rVert^2}
		\bigg(1-\frac{1}{k_0^2\lVert{\mathbf{r}}-{\mathbf{s}}\rVert^2}+\frac{1}{k_0^4\lVert{\mathbf{r}}-{\mathbf{s}}\rVert^4}\bigg),
	\end{equation}
	where $k_0=\frac{2\pi}{\lambda}$ is the wavenumber with $\lambda$ denoting the wavelength. Due to the small antenna size, i.e., $\sqrt{A}$, compared to the propagation distance between the user and antenna elements, i.e., $\lVert{\mathbf{r}}_k-{\mathbf{s}}_{m_x,m_z}\rVert$, the variation of the channel response within an antenna element is negligible. This gives
	\begin{align}\label{SPD_Electric_Field_Antenna_Element}
		g_{m_x,m_z,k}\approx A {\mathsf{g}}(\mathbf{r}_k,{\mathbf{s}}_{m_x,m_z})
		\frac{\lvert{\mathbf{e}}_y^{\mathsf{T}}({\mathbf{s}}_{m_x,m_z}-{\mathbf{r}}_k)\rvert}{\lVert{\mathbf{r}}_k-{\mathbf{s}}_{m_x,m_z}\rVert}.
	\end{align}
	Furthermore, applying this approximation to the phase component, we obtain $\phi_{m_x,m_z,k}\approx\frac{2\pi}{\lambda}r_{m_x,m_z,k}$.
	
	The function ${\mathsf{g}}(\mathbf{r},{\mathbf{s}})$ comprises three terms: the first term corresponds to the radiating NF and FF regions, while the remaining two terms correspond to the reactive NF region. Note that $1-\frac{1}{k_0^2\lVert{\mathbf{r}}-{\mathbf{s}}\rVert^2}+\frac{1}{k_0^4\lVert{\mathbf{r}}-{\mathbf{s}}\rVert^4}\approx0.97$ at distance $\lVert{\mathbf{r}}-{\mathbf{s}}\rVert=\lambda$ \cite{ouyang2024impact}. Hence, when considering practical NFC systems with $r_k\gg \lambda$, the last two terms in \eqref{Green_Function_Full_Version} can be neglected. Consequently, the NF channel coefficient can be modeled as
	\begin{align}\label{channel_NF}
		h_{m_x,m_z,k}\approx \sqrt{{Ar_k\Psi_k}/{(4\pi r_{m_x,m_z,k}^3)}}\mathrm{e}^{-\mathrm{j}\frac{2\pi}{\lambda}r_{m_x,m_z,k}}.   
	\end{align}
	For clarity, we denote $\mathbf{h}_k=[h_{m_x,m_z,k}]_{\forall m_x,m_z}\in\mathbbmss{C}^{M\times1}$ as the channel vector from UT $k$ to the BS. 
	
	For comparison, we also present the planar-wave based FF channel model. In contrast to the NF model, the FF model assumes that the angles of the links between each array element and UT $k$ are approximated to be identical, which results in linearly varying phase shifts. Additionally, variations in channel power across the BS array are considered negligible. Thus, the FF channel coefficient satisfies
	\begin{align}\label{channel_FF}
		h_{m_x,m_z,k}\approx \sqrt{{A\Psi_k}/({4\pi r_k^2})}\mathrm{e}^{-\mathrm{j}\frac{2\pi r_k(1-m_x\epsilon _k\Phi_k-m_z\epsilon _k\Omega_k)}{\lambda}}.    
	\end{align}
	By comparing \eqref{channel_NF} with \eqref{channel_FF}, we can observe that, in contrast to FF channels, spherical wavefronts introduce additional range dimensions $\{r_{m_x,m_z,k}\}_{m_x,m_z}$ to NF channels. 
	\subsection{Signal Models for Multiuser Communications}
	The system layout illustrated in {\figurename} {\ref{LoS_3D_Model_MU}} establishes the foundational framework for multiuser NFC. We next refine this basic model into three models of significant research interest: \emph{MAC}, \emph{BC}, and \emph{MC}. Throughout this paper, the channel is assumed to be known perfectly at the transceivers.
	\subsubsection{MAC}
	MAC refers to the scenario where all UTs simultaneously send its own message to the BS. The received signal vector at the BS is given by
	\begin{align}\label{MAC_Channel}
		{\mathbf{y}}=\sum\nolimits_{k=1}^K{\mathbf{h}_kx_k}+{\mathbf{n}},
	\end{align}
	where $x_{k}\in{\mathbbmss{C}}$ is the signal sent by UT $k$ with mean zero and variance ${\mathbbmss{E}}\{\lvert x_k\rvert^2\}\triangleq p_k$, and ${\mathbf{n}}\sim{\mathcal{CN}}({\mathbf{0}},\sigma^2{\mathbf{I}})$ is the additive white Gaussian noise (AWGN) with power $\sigma^2$. 
	\subsubsection{BC}
	If we reverse the MAC and have one BS broadcasting simultaneously to all UTs, it becomes the BC. The received signal at each UT $k$ is given by
	\begin{align}\label{Downlink_Signal}
		y_k={\mathbf{h}}_k^{\mathsf{H}}{\mathbf{x}}+{{n}}_k,
	\end{align}
	where ${\mathbf{x}}\in{\mathbbmss{C}}^{M\times1}$ is the transmitted vector with mean zero and covariance matrix ${\mathbbmss{E}\{{\mathbf{x}}{\mathbf{x}}^{\mathsf{H}}\}}\triangleq{\bm\Sigma}$, and $n_k\sim{\mathcal{CN}}(0,\sigma_k^2)$ is the AWGN with power $\sigma_k^2$. After receiving $y_k$, UT $k$ will decode the private message dedicated for himself from $y_k$.
	\subsubsection{MC}
	MC refers to the scenario where the BS sends a common message to all UTs. In this case, the received signal at each UT $k$ can be still described as \eqref{Downlink_Signal}. 
	
	In the sequel, we will analyze the NF capacity of the above three channels, and compare them with their FF counterpart to gather insights.
	\section{Multiple Access Channel}\label{sec_MAC}
	In this section, we analyze the NF channel capacity of a MAC by deriving its sum-rate capacity and capacity region. The sum-rate capacity of the MAC in \eqref{MAC_Channel} is given by \cite{heath2018foundations}
	\begin{align}\label{MAC_K>2}
		{\mathsf{C}}_{\mathsf{MAC}}=\max_{0\leq p_k\leq P_k}\log_2\det\Big({\mathbf{I}}_M+\sum\nolimits_{k=1}^{K}\frac{p_k}{\sigma^2}{\mathbf{h}}_k{\mathbf{h}}_k^{\mathsf{H}}\Big),
	\end{align}
	where $P_k$ represents the power budget of UT $k$. This capacity is achieved using \emph{point-to-point Gaussian random coding} along with \emph{successive interference cancellation (SIC) decoding} in a specific message decoding order. It is evident that the MAC sum-rate capacity is maximized when $p_k=P_k$ for $k\in{\mathcal{K}}$. By defining $\gamma_k\triangleq\frac{P_k}{\sigma^2}$ as the transmit signal-to-noise ratio (SNR), we can rewrite the sum-rate capacity as follows:
	\begin{align}\label{MAC_K>2_2}
		{\mathsf{C}}_{\mathsf{MAC}}=\log_2\det\Big({\mathbf{I}}_M+\sum\nolimits_{k=1}^{K}\gamma_k{\mathbf{h}}_k{\mathbf{h}}_k^{\mathsf{H}}\Big).
	\end{align}
	Deriving an analytically tractable expression for ${\mathsf{C}}_{\mathsf{MAC}}$ under the NF model is a challenging task. As a compromise, we will focus on the two-user case to gain more insights, which will then be extended to scenarios with more than two UTs.
	\subsection{Near-Field Capacity of the Two-User Case}
	\subsubsection{Sum-Rate Capacity}\label{Discussion of the Two-User Case: Sum-Rate Capacity}
	By denoting ${\mathsf{g}}_k=\lVert{\mathbf{h}}_k\rVert^2$ as the channel gain of UT $k$ and defining $\rho\triangleq\frac{\lvert{\mathbf{h}}_1^{\mathsf{H}}{\mathbf{h}}_2\rvert^2}{\lVert{\mathbf{h}}_1\rVert^2\lVert{\mathbf{h}}_2\rVert^2}\in[0,1]$ as the channel correlation factor (CCF)\footnote{The CCF is an important metric that quantifies the correlation between the channels of two UTs. When $\rho= 0$, the spatial channels for the UTs are orthogonal; when $\rho=1$, the spatial channels are parallel.} between the two UTs, the rates of the UTs achieved by SIC with different decoding orders are calculated as follows.
		\vspace{-5pt} 
		\begin{lemma}\label{MAC_region_lem}
			When the SIC decoding order $1\rightarrow2$ is adopted, the rates of the UTs are, respectively, given by
			\begin{subequations}\label{MAC_corner}
				\begin{align}
					&{\mathsf{R}}_{1}^{1\rightarrow2}= \log _2\left( 1+\frac{\gamma_1{\mathsf{g}}_1+\gamma_1\gamma_2{\mathsf{g}}_1{\mathsf{g}}_2(1-\rho)}{1+\gamma_2{\mathsf{g}}_2} \right),\label{r12}\\
					&{\mathsf{R}}_{2}^{1\rightarrow2}= \log_2\left(1+\gamma_2{\mathsf{g}}_2\right).
				\end{align}
			\end{subequations}
			For the decoding order $2\rightarrow1$, we have
			\begin{subequations}
				\begin{align}
					&{\mathsf{R}}_{1}^{2\rightarrow1}= \log_2\left(1+\gamma_1{\mathsf{g}}_1\right), \\
					&{\mathsf{R}}_{2}^{2\rightarrow1}=\log _2\left( 1+\frac{\gamma_2{\mathsf{g}}_2+\gamma_1\gamma_2{\mathsf{g}}_1{\mathsf{g}}_2(1-\rho)}{1+\gamma_1{\mathsf{g}}_1}\right).
				\end{align}
			\end{subequations}
		\end{lemma}
		\vspace{-3pt} 
		\begin{IEEEproof}
			Please refer to Appendix~\ref{Appendix:D} for more details.    
		\end{IEEEproof}
		The sum-rate capacity of the two-user MAC is given by
		\begin{align}\label{MAC_capacity}
			\mathsf{C}_{\mathsf{MAC}}&={\mathsf{R}}_{1}^{1\rightarrow2}+{\mathsf{R}}_{2}^{1\rightarrow2}={\mathsf{R}}_{1}^{2\rightarrow1}+{\mathsf{R}}_{2}^{2\rightarrow1}\notag\\
			&=\log_2( 1+{\gamma_1{\mathsf{g}}_1+\gamma_2{\mathsf{g}}_2}+{\gamma_1\gamma_2{\mathsf{g}}_1{\mathsf{g}}_2(1-{\rho})}) ,
		\end{align} 
		which can also be obtained by directly simplifying \eqref{MAC_K>2_2} \cite[Eq. (14)]{sdma}. The results suggest that the sum-rate capacity is determined by the channel gains and the CCF. By incorporating the NF model \eqref{channel_NF} into the analysis, we derive a new expression for NF sum-rate capacity of the MAC as follows.
	\vspace{-5pt}
	\begin{theorem}\label{MAC_capacity_NF_the}
		The sum-rate capacity of the MAC under the NF model can be expressed as follows:
		\begin{align}\label{Channel_Capacity_Two_User}
			{\mathsf{C}}_{\mathsf{MAC}}=\log _2( 1+\gamma_1{{\mathsf{g}}^{\mathsf{n}}_1}+ \gamma_2{{\mathsf{g}}}^{\mathsf{n}}_2+{\gamma_1\gamma_2{{\mathsf{g}}^{\mathsf{n}}_1}{{\mathsf{g}}^{\mathsf{n}}_2}(1-{{\rho}_{\mathsf{n}}})}). 
		\end{align}
		For the terms appearing in \eqref{Channel_Capacity_Two_User}, 
		\begin{equation}\label{channelgain_NF} 
			\begin{split} {{\mathsf{g}}}^{\mathsf{n}}_k=\frac{\xi}{4\pi}\sum_{x\in{\mathcal{X}}_k}\sum_{z\in{\mathcal{Z}}_k}
				\arctan\bigg(\frac{xz/\Psi_k}{\sqrt{\Psi_k^2+x^2+z^2}}\bigg)
			\end{split}
		\end{equation}
		is the NF channel gain of UT $k$, where ${\mathcal{X}}_k=\{\frac{M_x}{2}\epsilon_k\pm\Phi_k\}$, ${\mathcal{Z}}_k=\{\frac{M_z}{2}\epsilon_k\pm\Psi_k\}$, and $\xi = \frac{A}{d^2}\in \left( 0,1 \right] $ denotes the array occupation ratio (AOR). Furthermore, the CCF under the NF model can be calculated as
		\begin{align}
			{\rho}_{\mathsf{n}}&=\prod\nolimits_{k=1}^2{\frac{\pi MA\Psi _k}{4r_{k}^{2}\overline{\mathsf{g}}_kT^2}}\left| \sum\nolimits_{t=1}^T\sum\nolimits_{t^{\prime}=1}^T{\sqrt{\left( 1-\delta _{t}^{2} \right) \left( 1-\delta _{t^{\prime}}^{2} \right)}}\right.\nonumber\\
			&\left.\times f_1\left( M_x\epsilon _1\delta _t,M_z\epsilon _1\delta _{t^{\prime}} \right) f_2\left( M_x\epsilon _1\delta _t,M_z\epsilon _1\delta _{t^{\prime}} \right) \right|^2, \label{rho_NF}  
		\end{align}
		where $T$ is a complexity-vs-accuracy tradeoff parameter, $\delta _t=\cos \left( \frac{2t-1}{2T}\pi \right) $ for $t\in\left\{ 1,\ldots,T \right\} $, and 
		\begin{subequations}
			\begin{align}
				f_1\left( x,z \right) &\triangleq\frac{\mathrm{e}^{\mathrm{j}\frac{2\pi}{\lambda}r_1\left( x^2+z^2-2\Phi _1x-2\Omega _1z+1 \right) ^{\frac{1}{2}}}}{\left( x^2+z^2-2\Phi _1x-2\Omega _1z+1 \right) ^{\frac{3}{2}}}   ,\\
				f_2\left( x,z \right) &\triangleq\frac{\mathrm{e}^{-\mathrm{j}\frac{2\pi}{\lambda}r_2\left( \upsilon ^2x^2+\upsilon ^2z^2-2\upsilon \Phi _2x-2\upsilon \Omega _2z+1 \right) ^{\frac{1}{2}}}}{\left( \upsilon ^2x^2+\upsilon ^2z^2-2\upsilon \Phi _2x-2\upsilon \Omega _2z+1 \right) ^{\frac{3}{2}}}
			\end{align}
		\end{subequations}
		with $\upsilon ={r_1}/{r_2}$. 
	\end{theorem}
	\vspace{-5pt}
	\begin{IEEEproof}
		Please refer to Appendix~\ref{Appendix:B} for more details.    
	\end{IEEEproof}
	We note the CCF reflects the level of IUI, and thus an ideal scenario occurs when $\rho=0$, indicating the absence of IUI. This establishes an upper bound for $\mathsf{C}_{\mathsf{MAC}}$, which can be expressed as follows:
	\begin{subequations}
		\begin{align}
			\mathsf{C}_{\mathsf{MAC}}&\leq\log _2( 1+{\gamma_1\mathsf{g}_1^{\mathsf{n}}+\gamma_2\mathsf{g}_2^{\mathsf{n}}}+{\gamma_1\gamma_2\mathsf{g}_1^{\mathsf{n}}\mathsf{g}_2^{\mathsf{n}}} )\\
			&=\sum\nolimits_{k=1}^{2} 
			\log _2( 1+\gamma_k\mathsf{g}_k^{\mathsf{n}}).\label{MAC_Sum_Capacity_UB}
		\end{align}
	\end{subequations}
	Next, we aim to gain further insights into NF MAC capacity by considering an infinitely large array aperture. Specifically, when $M_x,M_z \rightarrow \infty$, we have
	\begin{align}
		\lim_{M_x,M_z\rightarrow \infty} \mathsf{g}_k^{\mathsf{n}}&=\frac{4\xi}{4\pi}\lim_{x,z\rightarrow \infty} \arctan \bigg( \frac{xz/\Psi_k}{\sqrt{\Psi _k^{2}+x^2+z^2}} \bigg)\nonumber\\
		&=\frac{4\xi}{4\pi}\frac{\pi}{2}=\frac{\xi}{2},\label{gain_asy}
	\end{align}
	which, together with \eqref{MAC_Sum_Capacity_UB}, yields
	\begin{align}\label{MAC_Sum_Capacity_UB2}
		\lim\nolimits_{M_x,M_z\rightarrow \infty} \mathsf{C}_{\mathsf{MAC}}\leq\sum\nolimits_{k=1}^{2} 
		\log _2( 1+\xi\gamma_k/2),    
	\end{align}
	The results suggest that the NF MAC capacity is upeer bounded by a finite value as $M_x,M_z \rightarrow \infty$.
	
	In contrast to $\lim_{M_x,{M}_z\rightarrow \infty}\mathsf{g}_k^{\mathsf{n}}$, $\lim_{M_x,{M}_z\rightarrow \infty}\rho _{\mathsf{n}}$ is computationally intractable. However, numerical results presented in \cite{NFISAC_performance} demonstrate that $\lim_{M_x,{M}_z\rightarrow \infty}\rho _{\mathsf{n}}\ll 1$, which is also verified by the results in Section \ref{numerical}. This observation suggests that as the array aperture increases, NF channels of UTs positioned at different locations tend to become orthogonal, which presents an opportunity to mitigate IUI.
	\vspace{-5pt}
	\begin{corollary}\label{MAC_NF_M_cor}
		When $M_x,M_z \rightarrow \infty$, the asymptotic NF MAC capacity is given by
		\begin{align}\label{MAC_NF_M}
			\lim\nolimits_{M_x,M_z\rightarrow \infty} \mathsf{C}_{\mathsf{MAC}}\approx\sum\nolimits_{k=1}^{2} 
			\log _2( 1+\xi\gamma_k/2).    
		\end{align}    
	\end{corollary}
	\vspace{-5pt} 
	\begin{IEEEproof}
		Equation \eqref{MAC_NF_M} is derived using \eqref{gain_asy} and the fact that $\lim_{M_x,{M}_z\rightarrow \infty}\rho _{\mathsf{n}}\ll 1$.
	\end{IEEEproof}
	\vspace{-5pt} 
	\begin{remark}\label{MAC_constant}
		The results of \textbf{Corollary \ref{MAC_NF_M_cor}} suggest that, as $M_x,M_z \rightarrow \infty$, the NF MAC capacity converge to a finite value positively correlated to the AOR. 
	\end{remark}
	\vspace{-5pt} 
	\vspace{-5pt} 
	\begin{remark}\label{MAC_Upper_Bound}
		The asymptotic NF MAC capacity closely approximates its upper bound presented in \eqref{MAC_Sum_Capacity_UB2}, as setting $M_x,M_z\rightarrow\infty$ nearly removes the impact of IUI. 
	\end{remark}
	\vspace{-5pt} 
	
	\begin{figure}[!t]
		\centering
		\setlength{\abovecaptionskip}{0pt}
		\includegraphics[height=0.24\textwidth]{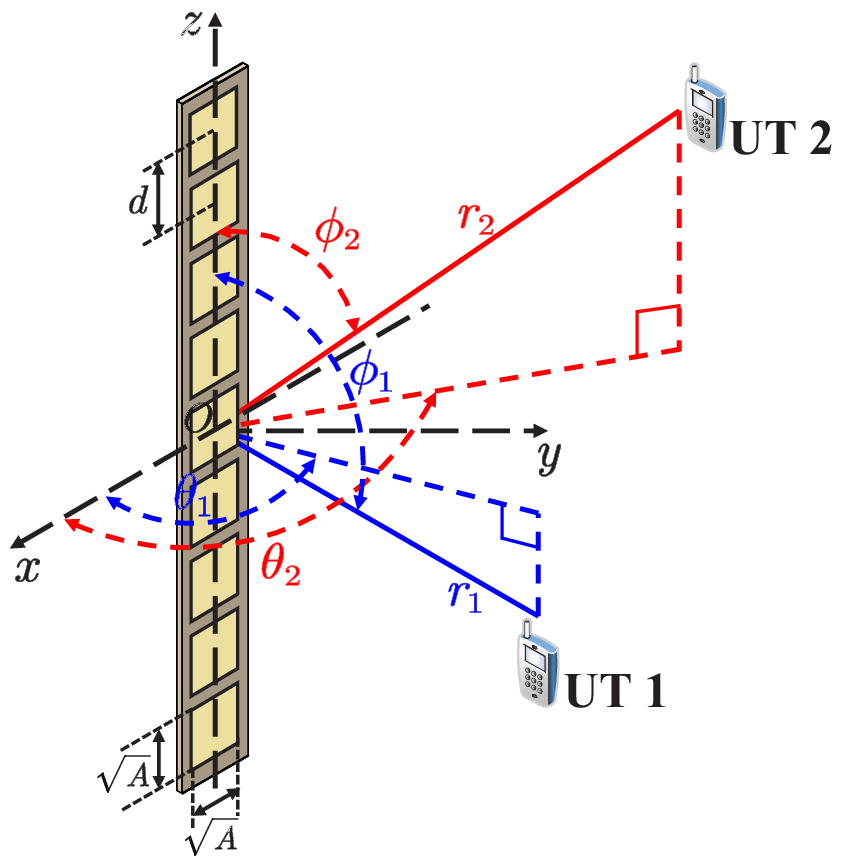}
		\caption{Illustration of the ULA.}
		\label{ula}
		\vspace{-7pt}
	\end{figure}

	Turn to a special case where the BS uses a uniform linear array (ULA), i.e., $M_x=1$ and $M_z=M$, as shown in Fig~\ref{ula}.
	\vspace{-5pt}
	\begin{corollary}\label{ula_mac_cor}
		When using a ULA, the NF channel gains satisfy
		\begin{align}\label{channelgain_ula}
			\mathsf{g}_k^{\mathsf{n}}=\frac{\xi \epsilon _k\sin \phi _k\Xi}{4\pi \sin \theta _k},
		\end{align}
		where $\Xi =\frac{M\epsilon _k-2\cos \theta _k}{\sqrt{M^2\epsilon _{k}^{2}-4M\cos \theta _k\epsilon _k+4}}+\frac{M\epsilon _k+2\cos \theta _k}{\sqrt{M^2\epsilon _{k}^{2}+4M\cos \theta _k\epsilon _k+4}}$. Accordingly, the MAC capacity satisfies
		\begin{align}\label{ula_mac}
				\lim\nolimits_{M\rightarrow \infty} \mathsf{C}_{\mathsf{MAC}}\approx \sum\nolimits_{k=1}^2{\log _2\left( 1+\frac{\gamma_kA\sin \phi _k\!}{2\pi dr_k\sin \theta _k} \right)},  
		\end{align}
		which is also a finite value.
	\end{corollary}
	\vspace{-5pt}
	\begin{IEEEproof}
		Please refer to Appendix~\ref{Appendix:C} for more details.
	\end{IEEEproof}
	We note from \eqref{ula_mac} that, different from the asymptotic capacity for the UPA given in \eqref{MAC_NF_M}, which is independent of the UTs' locations, the MAC capacity for the ULA is always affected by the locations of the UTs even as the antenna number approaches infinity. Specifically, it depends on the projected aperture $A\sin \phi _k$ and the projected distance to the ULA $r_k\sin \theta _k$ for each UT.
	
	\begin{figure*}[!b]
		\hrulefill
		\setcounter{equation}{27}
		\begin{align}\label{far_MAC}
			\mathsf{C}_{\mathsf{MAC}}\simeq \begin{cases}
				\log _2\left( \frac{MA}{4\pi}\left( \frac{\gamma_1\Psi _1}{r_{1}^{2}}+\frac{\gamma_2\Psi _2}{r_{2}^{2}} \right) \right) \triangleq \mathsf{F}_{\mathsf{MAC}}^{\mathsf{s}}&		\left( \theta _1,\phi _1 \right) =\left( \theta _2,\phi _2 \right)\\
				\log _2\left( \frac{MA}{4\pi }\left( \frac{\gamma_1\Psi _1}{r_{1}^{2}}+\frac{\gamma_2\Psi _2}{r_{2}^{2}} \right) +\frac{M^2A^2\gamma_1\gamma_2\Psi _1\Psi _2}{16\pi ^2r_{1}^{2}r_{2}^{2}} \right) \triangleq \mathsf{F}_{\mathsf{MAC}}^{\mathsf{d}}&		\left( \theta _1,\phi _1 \right) \ne \left( \theta _2,\phi _2 \right)\\
			\end{cases}    
		\end{align}
		\begin{align}\label{MAC_ratio}
			\mathsf{F}_{\mathsf{MAC}}^{\mathsf{d}}-\mathsf{F}_{\mathsf{MAC}}^{\mathsf{s}}=\log _2\left( 1+\frac{MA}{4\pi \left( \gamma_{1}^{-1}\Psi _{1}^{-1}r_{1}^{2}+\gamma_{2}^{-1}\Psi _{2}^{-1}r_{2}^{2} \right) } \right) >0
		\end{align}
		\setcounter{equation}{23}
	\end{figure*}
	
	\subsubsection{Capacity Region}
	Having obtained the sum-rate capacity, we now explore the capacity region. For a two-user NF MAC, its capacity region contains all the achievable rate pairs $({\mathsf{R}}_{1},{\mathsf{R}}_{2})$ such that \cite{el2011network}
	\begin{subequations}
		\begin{align}
			&{\mathsf{R}}_{1}\leq \log_2(1+\gamma_1\mathsf{g}_1^{\mathsf{n}}),\ {\mathsf{R}}_{2}\leq \log_2(1+{\gamma_2}\mathsf{g}_2^{\mathsf{n}}),\\
			&{\mathsf{R}}_{1}+{\mathsf{R}}_{2}\leq {\mathsf{C}}_{\mathsf{MAC}},
		\end{align}
	\end{subequations}
	The MAC capacity region forms a pentagon, with two corner points attained through SIC decoding, i.e., $({\mathsf{R}}_{1}^{1\rightarrow2},{\mathsf{R}}_{2}^{1\rightarrow2})$ and $({\mathsf{R}}_{1}^{2\rightarrow1},{\mathsf{R}}_{2}^{2\rightarrow1})$, and the line segment connecting these points achieved through time sharing, as seen in \cite[{\figurename} 7]{goldsmith}. 
	
	Time sharing means that applying the SIC order $1\rightarrow2$ with probability $\tau$, while applying $2\rightarrow1$ with probability $1-\tau$, where $\tau\in(0,1)$. Given $\tau$, by denoting the achievable rate pair as $( {\mathsf{R}}_{1}^{\tau},{\mathsf{R}}_{2}^{\tau}) $, we have ${\mathsf{R}}_{1}^{\tau}=\tau  {\mathsf{R}}_{1}^{1\rightarrow2}+( 1-\tau  ) {\mathsf{R}}_{1}^{2\rightarrow1}$ and ${\mathsf{R}}_{2}^{\tau}=\tau  {\mathsf{R}}_{2}^{1\rightarrow2}+( 1-\tau  ) {\mathsf{R}}_{2}^{2\rightarrow1}$. As a result, the MAC capacity region is given by
	\begin{align}
		\mathcal{R} _{\mathsf{MAC}}=\left\{ ( {\mathsf{R}}_{1},{\mathsf{R}}_{2} ) \left| {\mathsf{R}}_{1}\in \left[ 0,{\mathsf{R}}_{1}^{\tau} \right] ,{\mathsf{R}}_{2}\in [ 0,{\mathsf{R}}_{2}^{\tau} ] ,\tau \in [ 0,1 ] \right.  \right\} .
	\end{align}    
	Next, we consider the case where $M_x,M_z\rightarrow\infty$.
	\vspace{-5pt}
	\begin{corollary}\label{MAC_NF_region_M_cor}
		When $M_x,M_z \rightarrow \infty$, the corner points of the NF MAC capacity region satisfy
		\begin{align}
			&\lim\nolimits_{M_x,M_z\rightarrow \infty} ( \mathsf{R}_{1}^{1\rightarrow 2},\mathsf{R}_{2}^{1\rightarrow 2} )\approx\lim\nolimits_{M_x,M_z\rightarrow \infty} ( \mathsf{R}_{1}^{2\rightarrow 1},\mathsf{R}_{2}^{2\rightarrow 1} )  \nonumber\\
			&\approx\left( \log _2( 1+{\xi\gamma_1}/2),\log _2( 1+{\xi\gamma_2}/{2} ) \right).
		\end{align}
	\end{corollary} 
	\begin{IEEEproof}
		Similar to the proof of \textbf{Corollary~\ref{MAC_NF_M_cor}}.
	\end{IEEEproof}
	\vspace{-5pt}
	\begin{remark}\label{MAC_rectangle}
		The findings in \textbf{Corollary~\ref{MAC_NF_region_M_cor}} indicate that as the number of antennas increases, the two corner points tend to approach each other. This causes the NF MAC capacity region to transition from a pentagon to a finite rectangle, implying that the rates are no longer influenced by the SIC order.
	\end{remark}
	\vspace{-5pt} 
	\subsection{Comparison with the FF Capacity}
	For comparison, we next discuss the FF case.
	\vspace{-5pt} 
	\begin{lemma}\label{MAC_capacity_FF_the}
		Under the FF model, the channel gains are given by $\mathsf{g}_k^{\mathsf{f}}=\frac{MA\Psi _k}{4\pi r_{k}^{2}}$ for $k=1,2$, and the CCF satisfies
		\begin{equation}
			\rho _{\mathsf{f}} =\left\{ \!\begin{array}{ll}
				1&		\left( \theta _1,\phi _1 \right)\! =\!\left( \theta _2,\phi _2 \right)\\
				\frac{1-\cos \left( M_x\Delta _{\Phi} \right)}{M^2\left( 1-\cos \Delta _{\Phi} \right)}&		\Phi _1\!\ne\! \Phi _2,\Omega _1\!=\!\Omega _2\\
				\frac{1-\cos \left( M_z\Delta _{\Omega} \right)}{M^2\left( 1-\cos \Delta _{\Omega} \right)}&		\Phi _1\!=\!\Phi _2,\Omega _1\!\ne\! \Omega _2\\
				\frac{4\left( 1-\cos \left( M_x\Delta _{\Phi} \right) \right) \left( 1-\cos \left( M_z\Delta _{\Omega} \right) \right)}{M^2\left( 1-\cos \Delta _{\Phi} \right) \left( 1-\cos \Delta _{\Omega} \right)}&		\text{else}\\
			\end{array} \right. , \label{rho_FF}
		\end{equation}
		where $\Delta _{\Phi}=\frac{2\pi}{\lambda}d\left( \Phi _1-\Phi _2 \right)$, and $\Delta _{\Omega}=\frac{2\pi}{\lambda}d\left( \Omega _1-\Omega _2 \right)$. 
	\end{lemma}
	\vspace{-5pt}
	\begin{IEEEproof}
		The results can be derived from \eqref{channel_FF} by using the sum of the geometric series and trigonometric identities.  
	\end{IEEEproof}
	Using \textbf{Lemma \ref{MAC_capacity_FF_the}}, the closed-form expressions for the FF MAC capacity and the capacity region follow immediately, which are omitted due to space limitations.
	\vspace{-5pt} 
	\begin{corollary}\label{MAC_FF_M}
		Under the FF model, when $M\rightarrow\infty$, the asymptotic MAC capacity is given in \eqref{far_MAC}, shown at the bottom of this page, which yields $\mathsf{C}_{\mathsf{MAC}}\simeq\mathcal{O} \left( \log M \right) $.
	\end{corollary}
	\vspace{-5pt}
	\begin{IEEEproof}
		The results can be obtained using the fact that $\lim _{M\rightarrow \infty}\rho _{\mathsf{f}}=\begin{cases}
			1&		\left( \theta _1,\phi _1 \right) =\left( \theta _2,\phi _2 \right)\\
			0&		\left( \theta _1,\phi _1 \right) \ne \left( \theta _2,\phi _2 \right)\\
		\end{cases}$, along with the approximation $\log_2(1+x)\approx\log_2x$ for large $x$.
	\end{IEEEproof}
	\vspace{-5pt}
	\begin{remark}\label{MAC_unbound}
		Rather than converging to a finite bound as under the NF model, the FF MAC sum-rate capacity grows unboundedly with the number of the BS antennas, theoretically achieving any desired level, which poses a contradiction to the energy-conservation laws.    
	\end{remark}
	\vspace{-5pt} 
	This contradiction stems from the assumption of uniform channel power across each antenna element, which becomes increasingly inaccurate as the number of the elements grows. Furthermore, comparing the FF MAC capacity for co-directional UTs and UTs with differing directions, as indicated in \eqref{MAC_ratio}, we find that the FF MAC capacity significantly diminishes when UTs are oriented in the same direction. This contrast with the NF model arises from the high channel correlation ($\rho _{\mathsf{f}}=1$) among UTs in the same direction, leading to notable IUI.
	
	We then characterize the asymptotic MAC capacity region under the FF model. When $M\rightarrow\infty$, if UTs are in the same direction, the corner points of the MAC capacity region satisfy\setcounter{equation}{29}\begin{subequations}
		\begin{align}
			(\mathsf{R}_{1}^{1\rightarrow 2},\mathsf{R}_{2}^{1\rightarrow 2})&\simeq\! (\log _2( 1\!+\!{\gamma_1r_{2}^{2}}/{(\gamma_2r_{1}^{2})} ),\mathcal{O}(\log M)),\\
			(\mathsf{R}_{1}^{2\rightarrow 1},\mathsf{R}_{2}^{2\rightarrow 1})&\simeq\! (\mathcal{O}(\log M),\log _2( 1\!+\!{\gamma_2r_{1}^{2}}/{(\gamma_1r_{2}^{2})} )).
	\end{align}\end{subequations}
	If UTs are positioned in different directions, we have
	\begin{subequations}
		\begin{align}
			(\mathsf{R}_{1}^{1\rightarrow 2},\mathsf{R}_{2}^{1\rightarrow 2})&\approx(\mathsf{R}_{1}^{2\rightarrow 1},\mathsf{R}_{2}^{2\rightarrow 1})\\
			&\simeq (\mathcal{O}(\log M) ,\mathcal{O}(\log M)).
		\end{align}
	\end{subequations}
	The above results indicate that the MAC capacity region under the FF model can expand unboundedly with the number of array elements, rendering it impractical. Additionally, when UTs face different directions, the FF MAC capacity region roughly forms a rectangle and is more extensive than that when UTs are in the same direction.
	
	By comparing the NF capacity with its FF counterpart, we have the following observations:
	\begin{itemize}
		\item  The adoption of the FF channel model for a large antenna array may lead to outcomes that conflict with the fundamental principle of energy conservation. This issue arises because the NF region expands as the array size increases, which challenges the FF assumption of uniform plane-wave propagation. On the other hand, the NF model is more sustainable under energy considerations. The above facts underscore the necessity for channel modeling within the NF region using spherical-wave propagation to capture the physical reality more accurately.    
		\item Different from the FF MAC capacity that degenerates for co-directional UTs due to the significant IUI, the NF capacity can be preserved when UTs are located at different locations (different directions or distances), which underscores superior flexibility and robust interference management capabilities inherent to NFC. This additional resolution in the range domain enables \textit{space-division multiple access} \cite{sdma} from FF \textit{angle-division multiple access} to NF \textit{range-division multiple access}, demonstrating the NFC's potential as a promising approach for efficient and interference-free communication systems.  
	\end{itemize}
	\subsection{Comparison with Linear Combiners}
		Achieving the MAC sum-rate capacity in \eqref{MAC_K>2_2} requires the utilization of a non-linear, SIC-based combiner. However, to reduce decoding complexity, linear combiners are often preferred in practical systems. In the following analysis, we aim to evaluate the sum-rate performance of several typical linear combiners and examine the sum-rate gap between these linear combiners and the capacity-achieving non-linear combiners. Specifically, let $\mathbf{v}_k\in{\mathbbmss{C}}^{M\times1}$ denote the linear combiner used to decode $x_k$, and the achievable rate is given by
		\begin{align}\label{Linear_Combiner_MAC_Per_User_Rate}
			{\mathsf{R}}_k=\log_2\bigg(1+\frac{\lvert{\mathbf{v}}_k^{\mathsf{H}}{\mathbf{h}}_k\rvert^2P_k}
			{\sigma^2\lVert{\mathbf{v}}_k\rVert^2+\sum_{k'\ne k}\lvert{\mathbf{v}}_k^{\mathsf{H}}{\mathbf{h}}_{k'}\rvert^2P_{k'}}\bigg).
		\end{align}
		For the sake of brevity, we continue to consider the two-user case with $K=2$.
		\subsubsection{Optimal Linear Combining}
		We begin with the optimal linear combiner, which is designed by first whitening the interference-plus-noise term and then applying a maximal-ratio combiner to the resulting effective channel. According to \cite{heath2018foundations}, the optimal linear combiner for UT $k$ is given by:
		\begin{align}\label{Optimal_Combiner_MAC}
			{\mathbf{v}}_k=(\sigma^2{\mathbf{I}}_M+P_{k'}{\mathbf{h}}_{k'}{\mathbf{h}}_{k'}^{\mathsf{H}})^{-1}{\mathbf{h}}_k,
		\end{align}
		where $k'\ne k$. Substituting \eqref{Optimal_Combiner_MAC} into \eqref{Linear_Combiner_MAC_Per_User_Rate} gives
		\begin{align}
			{\mathsf{R}}_k=\log_2(1+P_k{\mathbf{h}}_k^{\mathsf{H}}(\sigma^2{\mathbf{I}}_M+P_{k'}{\mathbf{h}}_{k'}{\mathbf{h}}_{k'}^{\mathsf{H}})^{-1}{\mathbf{h}}_k),
		\end{align}
		which, together with the Woodbury matrix identity, yields
		\begin{equation}
			{\mathsf{R}}_1+{\mathsf{R}}_2=\sum_{k=1}^{2}
			\log _2\Big( 1+\gamma_k{\mathsf{g}}_k\Big(1-\frac{\gamma_{k'}{\mathsf{g}}_{k'}}{1+\gamma_{k'}{\mathsf{g}}_{k'}}\rho\Big) \Big)\triangleq {\mathsf{R}}_{{\mathsf{opt}}}.
		\end{equation}
		\subsubsection{Maximal-Ratio Combining}
		Next, we consider maximal-ratio combining (MRC), which aims to maximize the channel gain for each UT. This yields $\mathbf{v}_k=\mathbf{h}_k$. Substituting  $\mathbf{v}_k=\mathbf{h}_k$ into \eqref{Linear_Combiner_MAC_Per_User_Rate} and performing some basic mathematical manipulations, we express the sum-rate as follows:
		\begin{equation}
			{\mathsf{R}}_1+{\mathsf{R}}_2=\sum_{k=1}^{2}
			\log _2\Big( 1+\gamma_k{\mathsf{g}}_k\Big(1-\frac{\gamma_{k'}{\mathsf{g}}_{k'}\rho}{1+\gamma_{k'}{\mathsf{g}}_{k'}\rho}\Big) \Big)\triangleq {\mathsf{R}}_{{\mathsf{mrc}}}.
		\end{equation}
		\subsubsection{Zero-Forcing Combining}
		Finally, we consider zero-forcing (ZF) combining, which aims to nullify the IUI. In this case, ${\mathbf{v}}_k^{\mathsf{H}}{\mathbf{h}}_{k'}=0$ for $k'\ne k$. To further maximize the channel gain of UT $k$, we set ${\mathbf{v}}_k$ as the projection of $\mathbf{h}_k$ onto the space orthogonal to $\mathbf{h}_{k'}$, yielding \cite{heath2018foundations}
		\begin{align}\label{Zero-Forcing_Combiner_MAC}
			{\mathbf{v}}_k=( \mathbf{I}_M-\mathbf{h}_{k'}\mathbf{h}_{k'}^{\mathsf{H}}/\| \mathbf{h}_{k'} \| ^2 ) \mathbf{h}_k.
		\end{align}
		Substituting \eqref{Zero-Forcing_Combiner_MAC} into \eqref{Linear_Combiner_MAC_Per_User_Rate} and performing some simple manipulations, we obtain
		\begin{align}
			{\mathsf{R}}_1+{\mathsf{R}}_2=\sum_{k=1}^{2}
			\log _2(1+\gamma _k\mathsf{g}_k(1-\rho))\triangleq {\mathsf{R}}_{{\mathsf{zf}}}.
		\end{align}
		\subsubsection{Summary and Discussion}
		Through the derivations above, we observe that the sum-rate achieved by the considered linear combining schemes can be expressed as a function of the channel gains and the CCF, i.e., 
		\begin{align}\label{Sum_Rate_Linear_General_up}
			{\mathsf{R}}_{\ell}=\sum_{k=1}^{2} \log_2(1+\gamma _k\mathsf{g}_k (1-f_{\ell}(\gamma_{k'}\mathsf{g}_{k'},\rho))),
		\end{align}
		where $\ell\in\{{\mathsf{opt}},{\mathsf{mrc}},{\mathsf{zf}}\}$, $f_{{\mathsf{opt}}}(x,z)\triangleq\frac{xz}{1+x}$, $f_{{\mathsf{mrc}}}(x,z)\triangleq\frac{xz}{1+xz}$, and $f_{{\mathsf{zf}}}(x,z)\triangleq z$. By substituting \eqref{channelgain_NF} and \eqref{rho_NF} into \eqref{Sum_Rate_Linear_General_up}, we obtain the NF sum-rate expressions under the linear combiners, which can be written as follows:
		\begin{align}
			{\mathsf{R}}_{\ell}=\sum_{k=1}^{2} \log_2(1+\gamma _k\mathsf{g}_k (1-f_{\ell}(\gamma_{k'}\mathsf{g}_{k'}^{\mathsf{n}},\rho_{\mathsf{n}}))).
		\end{align}
		Moreover, it can be observed that $\mathsf{C}_{\mathsf{MAC}}\geq {\mathsf{R}}_{{\mathsf{opt}}}\geq {\mathsf{R}}_{{\mathsf{zf}}}$ and $\mathsf{C}_{\mathsf{MAC}}\geq {\mathsf{R}}_{{\mathsf{opt}}}\geq {\mathsf{R}}_{{\mathsf{mrc}}}$. However, the relationship between ${\mathsf{R}}_{{\mathsf{mrc}}}$ and ${\mathsf{R}}_{{\mathsf{zf}}}$ depends on the values of $\rho$ and ${\mathbf{g}}_k$.
		
		As discussed in Section \ref{Discussion of the Two-User Case: Sum-Rate Capacity}, we have $\lim_{M_x,M_z\rightarrow \infty}\rho_{\mathsf{n}}\approx 0$ under the NF channel model. This suggests that
		\begin{align}
			\lim\nolimits_{M_x,M_z\rightarrow \infty}{\mathsf{R}}_{\ell}\approx \sum_{k=1}^2{\log _2\left( 1+\gamma _k\mathsf{g}_k^{\mathsf{n}} \right)}
		\end{align}
		holds for $\ell\in\{{\mathsf{opt}},{\mathsf{mrc}},{\mathsf{zf}}\}$.
		\vspace{-5pt}
		\begin{remark}\label{rem_linear_up}
			The above discussion implies that in NF MAC, when the number of array elements is sufficiently large, the considered linear combining schemes can approach the upper bound of the sum-rate capacity, similar to the capacity-achieving schemes. This occurs because the NF effect can make the UTs' channels orthogonal to each other, resulting in zero IUI.
		\end{remark}
		
		\begin{figure}[!t]
			\centering
			\subfigbottomskip=5pt
			\subfigcapskip=0pt
			\setlength{\abovecaptionskip}{0pt}
			\subfigure[$(\theta_2,\phi_2)=(\frac{\pi}{3},\frac{2\pi}{3})$.]
			{
				\includegraphics[height=0.175\textwidth]{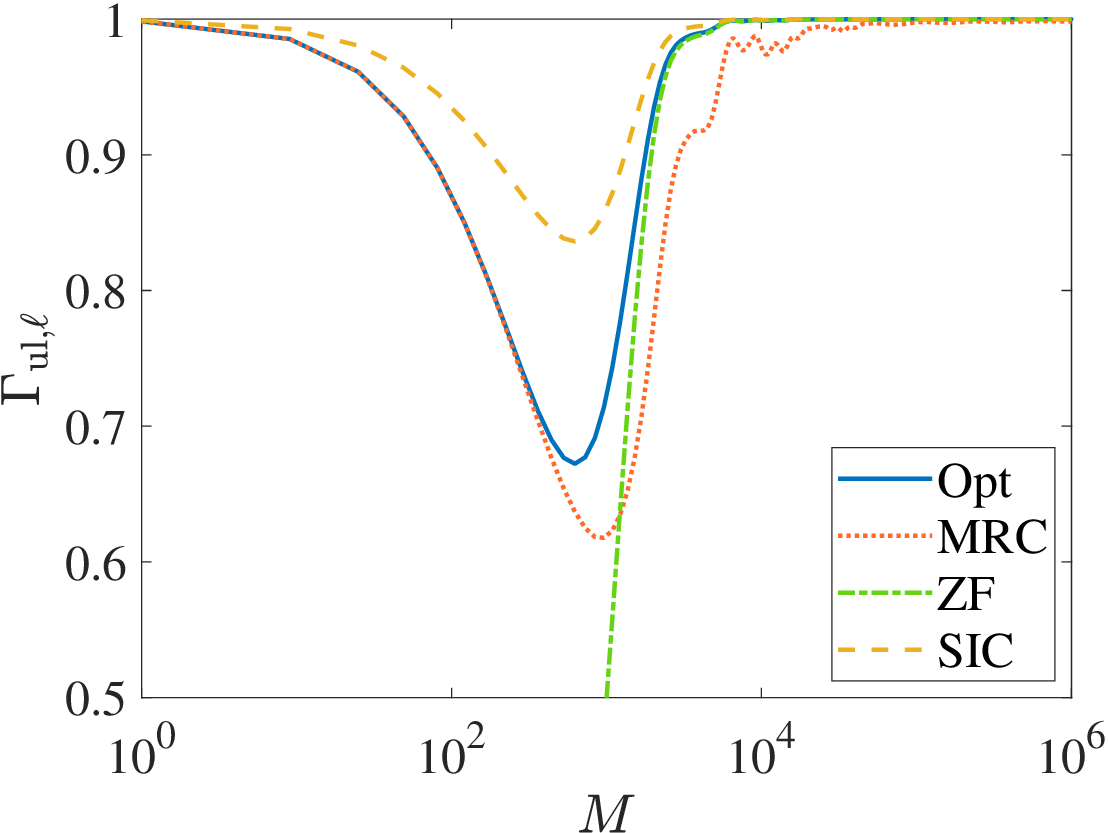}
			}%
			\subfigure[$(\theta_2,\phi_2)=(\frac{\pi}{3}\!+\!\frac{\pi}{60},\frac{2\pi}{3}\!+\!\frac{\pi}{60})$.]
			{
				\includegraphics[height=0.175\textwidth]{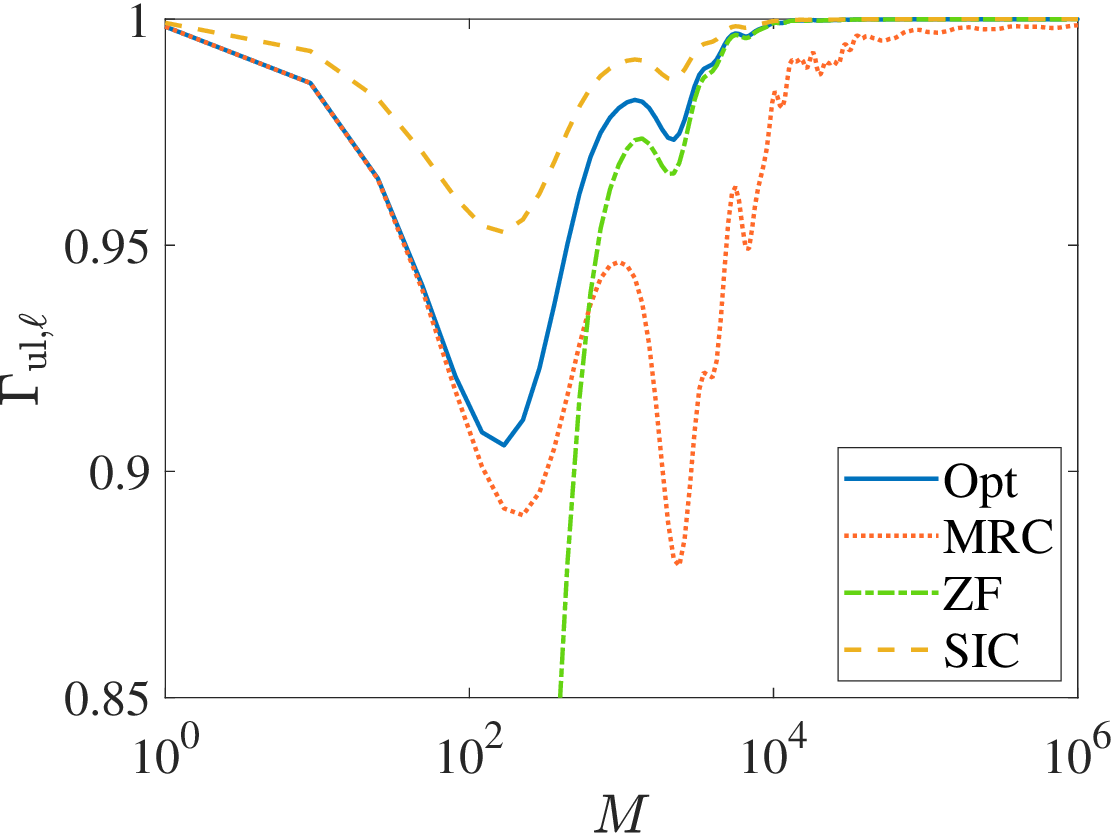}
			}
			\caption{$\Gamma_{{\mathsf{ul}},\ell}$ versus $M$ for different combiners with $(\theta_1,\phi_1)=(\frac{\pi}{3},\frac{2\pi}{3})$ and the other parameter settings given in Section~\ref{numerical}.}
			\label{linear_MAC}
			\vspace{-8pt}
		\end{figure}
		
		However, it is important to note that the conclusion in \textbf{Remark \ref{rem_linear_up}} only holds when $M_x,M_z\rightarrow \infty$. In practical NFC systems with large yet finite values of $M_x,M_z$, there will be a gap between the achievable sum-rate and its upper bound $\sum_{k=1}^{2}\log_2(1+\gamma_k\mathsf{g}_k^{\mathsf{n}})$, which is evaluated as follows:
		\begin{align}
			\Gamma_{{\mathsf{ul}},\ell}\triangleq\frac{\sum\nolimits_{k=1}^{2} \log_2(1+\gamma _k\mathsf{g}_k^{\mathsf{n}} (1-f_{\ell}(\gamma_{k'}\mathsf{g}_{k'}^{\mathsf{n}},\rho_{\mathsf{n}})))}{\sum\nolimits_{k=1}^2{\log _2\left( 1+\gamma _k\mathsf{g}_k^{\mathsf{n}} \right)}}.
		\end{align}
		Due to the complexity involved in $\rho_{\mathsf{n}}$ (Eq. \eqref{rho_NF}), a quantitative comparison of $\Gamma_{{\mathsf{ul}},\ell}$ for different $\ell$ is challenging. Therefore, we use numerical results to explore the relationship between $\Gamma_{{\mathsf{ul}},\ell}$ and the antenna number $M$, as shown in {\figurename} {\ref{linear_MAC}}. For comparison, the ratio between the capacity achieved by SIC and the upper bound is also presented. As demonstrated, with the increasing number of antennas, all of the discussed combining schemes achieve a sum-rate close to their upper bound. Among the discussed linear combining schemes, the optimal linear combining scheme yields the best sum-rate performance. This indicates that the optimal combining scheme can leverage the weak IUI arising from the NF effect to the greatest extent among the three linear combining schemes. For the other two schemes, we observe that the MRC scheme is preferred over ZF combining with fewer antennas, while ZF combiner outperforms MRC combiner when the antenna number exceeds a certain threshold, demonstrating that eliminating interference is more beneficial than maximizing channel gain in larger antenna arrays. Furthermore, the sum-rates achieved by linear combiners are lower than the capacity when $M$ is small, but nearly the same as the capacity after $M$ reaching about $10^5$, which suggests that in NFC, linear combiners are able to provide adequate performance compared to optimal SIC for large-scale arrays.
	
	\subsection{Extension to Cases of $K>2$}\label{Section: MAC: General Case}
	We then consider the case with $K>2$. By defining ${\mathbf{H}}\triangleq [\sqrt{\gamma_1}{\mathbf{h}}_1 \sqrt{\gamma_2}{\mathbf{h}}_2,\ldots,\sqrt{\gamma_K}{\mathbf{h}}_K]\in{\mathbbmss{C}}^{M\times K}$, we can rewrite \eqref{MAC_K>2} as follows:
		\begin{equation}
			{\mathsf{C}}_{\mathsf{MAC}}=\log_2\det({\mathbf{I}}_M
			+{\mathbf{H}}{\mathbf{H}}^{\mathsf{H}})=\log_2\det({\mathbf{I}}_K+{\mathbf{H}}^{\mathsf{H}}{\mathbf{H}}),
		\end{equation}
		where the last equality follows from Sylvester's identity. For clarity, we denote ${\mathbf{A}}_K={\mathbf{I}}_K+{\mathbf{H}}^{\mathsf{H}}{\mathbf{H}}\in{\mathbbmss{C}}^{K\times K}$, which yields
		\begin{align}
			{\mathbf{A}}_K
			=\left[ \begin{matrix}
				1+\gamma_1\| \mathbf{h}_1 \| ^2&		\cdots&		{\sqrt{\gamma_1\gamma_K}}\mathbf{h}_{1}^{\mathsf{H}}\mathbf{h}_K\\
				\vdots&		\ddots&		\vdots\\
				{\sqrt{\gamma_1\gamma_K}}\mathbf{h}_{K}^{\mathsf{H}}\mathbf{h}_1&		\cdots&		1+\gamma_K\| \mathbf{h}_K \| ^2\\
			\end{matrix} \right].
		\end{align}
		Note that the diagonal elements of ${\mathbf{A}}_K$ are determined by the channel gains ${\mathsf{g}}_k\triangleq \lVert{\mathbf{h}}_k\rVert^2$. Furthermore, the off-diagonal elements, say the $(i,j)$th element $[{\mathbf{A}}_K]_{i,j}$ with $i\ne j$, can be written as follows:
		\begin{align}
			[{\mathbf{A}}_K]_{i,j}={\sqrt{\gamma_i\gamma_j}}\mathbf{h}_{i}^{\mathsf{H}}\mathbf{h}_j=
			{\sqrt{\gamma_i\gamma_j}}\sqrt{{\mathsf{g}}_i{\mathsf{g}}_j}\frac{\mathbf{h}_{i}^{\mathsf{H}}\mathbf{h}_j}
			{\lVert{\mathbf{h}}_i\rVert\lVert{\mathbf{h}}_j\rVert}.
		\end{align}
		This fact suggests that the sum-rate capacity when $K>2$ remains a function of the channel gains and the CCF. Therefore, the NF sum-rate capacity can be analyzed in a similar way as in the two-user case. Particularly, as the BS antenna number approaches infinity, it can be concluded that the NF sum-rate capacity approaches its upper limit $\sum_{k=1}^{K}\log_2(1+\gamma_k{\mathsf{g}}_k^{\mathsf{n}})$, which satisfies
		\begin{align}
			\lim\nolimits_{M_x,M_z\rightarrow \infty} \mathsf{C}_{\mathsf{MAC}}\approx\sum\nolimits_{k=1}^{K} 
			\log _2( 1+\xi\gamma_k/2).   
		\end{align}
		
		The capacity region of the MAC exhibits a polymatroidal structure in $K$-dimensional space, and each of the $K!$ corner points can be attained by employing SIC decoding in a specific message decoding order. The remaining capacity region is realized through time-sharing among these corner points. Specifically, when considering the SIC decoding order $\varpi  (K)\rightarrow\varpi  (K-1)\rightarrow\ldots\rightarrow\varpi  (1)$ with $\{\varpi  (k)\}_{k=1}^{K}={\mathcal{K}}$, the achievable rate tuple $({\mathsf{R}}_{1},\ldots,{\mathsf{R}}_{K})$ at the corner point of the capacity region is given by
		\begin{equation}\label{R_K>2}
			\begin{split}
				\mathsf{R}_{\varpi \left( k \right)}=&\log _2\det\big(\mathbf{I}_M+\gamma_{\varpi \left( k \right)}\mathbf{h}_{\varpi \left( k \right)}\mathbf{h}_{\varpi \left( k \right)}^{\mathsf{H}}\\
				&\!\!\times\Big(\mathbf{I}+\sum\nolimits_{k^{\prime}<k}^{}{\gamma_{\varpi \left( k^{\prime} \right)}\mathbf{h}_{\varpi \left( k^{\prime} \right)}\mathbf{h}_{\varpi \left( k^{\prime} \right)}^{\mathsf{H}}}\Big)^{\!-1}\Big).     
			\end{split}
		\end{equation}
		By following steps similar to those in Appendix~\ref{Appendix:D}, \eqref{R_K>2} can be calculated as follows:
		\begin{equation}\label{R_K>2_2}
			\mathsf{R}_{\varpi \left( k \right)}=\log _2( 1+\gamma_{\varpi \left( k \right)}(\| \mathbf{h}_{\varpi \left( k \right)} \| ^2-\mathbf{a}_{k}^{\mathsf{H}}\mathbf{B}_{k}^{-1}\mathbf{a}_k) ), 
		\end{equation}
		where $\mathbf{a}_k=[ \mathbf{h}_{\varpi \left( k \right)}^{\mathsf{H}}\mathbf{h}_{\varpi ( k-1 )},\ldots,\mathbf{h}_{\varpi ( k )}^{\mathsf{H}}\mathbf{h}_{\varpi ( 1 )} ] ^{\mathsf{H}}$, and \scalebox{0.78}{$\mathbf{B}_k=\left[ \begin{matrix}
				1+{\gamma_{\varpi \left( k-1 \right)}}\left\| \mathbf{h}_{\varpi \left( k-1 \right)} \right\| ^2&		\cdots&		{\sqrt{\gamma_{\varpi \left( k-1 \right)}\gamma_{\varpi \left( 1 \right)}}}\mathbf{h}_{\varpi \left( 1 \right)}^{\mathsf{H}}\mathbf{h}_{\varpi \left( k-1 \right)}\\
				\vdots&		\ddots&		\vdots\\
				{\sqrt{\gamma_{\varpi \left( k-1 \right)}\gamma_{\varpi \left( 1 \right)}}}\mathbf{h}_{\varpi \left( k-1 \right)}^{\mathsf{H}}\mathbf{h}_{\varpi \left( 1 \right)}&		\cdots&		1+{\gamma_{\varpi \left( 1 \right)}}\left\| \mathbf{h}_{\varpi \left( 1 \right)} \right\| ^2\\
			\end{matrix} \right] $}. Eq. \eqref{R_K>2_2} can be also expressed in terms of a function of the channel gains and CCFs. Thus, the results derived in the case of $K=2$ can be trivially extended to the case of $K>2$.
	
	\section{Broadcast Channel}\label{sec_BC}
	In this section, we investigate the sum-rate capacity and the capacity region of the BC. 
	
	The capacity of the multiantenna Gaussian BC is achieved by successive dirty-paper encoding with Gaussian codebooks \cite{costa1983writing,weingarten2006capacity}. Let ${\bm\Sigma}_k\in{\mathbbmss{C}}^{M\times M}$ denote the covariance matrix in generating the dirty-paper Gaussian codebook for UT $k$. The transmitted signal is subject to the power budget ${\mathsf{tr}}({\bm\Sigma})=\sum_{k=1}^{K}{\mathsf{tr}}({\bm\Sigma}_k)\leq P$. Without loss of generality, we consider the encoding order $\varepsilon (K)\rightarrow\varepsilon (K-1)\rightarrow\ldots\rightarrow\varepsilon (1)$ with $\{\varepsilon (k)\}_{k=1}^{K}={\mathcal{K}}$. Concerning UT $\varepsilon (k)$, the dirty-paper encoder considers the interference signal caused by UTs $\varepsilon (k')$ for $k'>k$ as known non-causally and his decoder treats the interference signal caused by UTs $\varepsilon (k')$ for $k'<k$ as additional noise. By applying dirty-paper coding and by using minimum Euclidean distance decoding at each UT, it follows that the achieved rate of UT $\varepsilon (k)$ is given by \cite{vishwanath2003duality,jindal2005sum}
	\setlength\abovedisplayskip{5pt}
	\setlength\belowdisplayskip{5pt}
	\begin{align}
		{\mathsf{R}}_{\varepsilon (k)}=\log_2\left(1\!+\!\frac{{\mathbf{h}}_{\varepsilon (k)}^{\mathsf{H}}{\bm\Sigma}_{\varepsilon (k)}{\mathbf{h}}_{\varepsilon (k)}}
		{\sigma_{\varepsilon (k)}^2\!+\!\sum_{k'<k}{\mathbf{h}}_{\varepsilon (k)}^{\mathsf{H}}{\bm\Sigma}_{\varepsilon (k')}{\mathbf{h}}_{\varepsilon (k)}}\right).
	\end{align}
	Consequently, the sum-rate capacity can be written in terms of the following maximization:
	\begin{align}\label{Sum_Rate_Capacity_BC}
		{\mathsf{C}}_{\mathsf{BC}}=\max_{\{{\bm\Sigma}_k\}_{k=1}^{K}:{\bm\Sigma}_k\succeq{\mathbf{0}},\sum_{k=1}^{K}{\mathsf{tr}}({\bm\Sigma}_k)\leq P}\sum\nolimits_{k=1}^{K}{{\mathsf{R}}_{\varepsilon (k)}}.
	\end{align}
	Deriving a closed-form expression for ${\mathsf{C}}_{\mathsf{BC}}$ under the NF model is a challenging task. As a compromise, we continue to consider the two-user case to gain more insights.
	\subsection{Near-Field Capacity of the Two-User Case}
	\subsubsection{Sum-Rate Capacity}
	Based on \eqref{Sum_Rate_Capacity_BC}, the sum-rate capacity of BS with two UTs is written as follows:
	\begin{equation}\label{Sum_Rate_Capacity_BC_k=2}
		\begin{split}
			{\mathsf{C}}_{\mathsf{BC}}=&\max_{{\bm\Sigma}_k\succeq{\mathbf{0}},\sum_{k=1}^{2}{\mathsf{tr}}({\bm\Sigma}_k)\leq P}\log_2\left(1+\frac{{\mathbf{h}}_1^{\mathsf{H}}{\bm\Sigma}_1{\mathbf{h}}_1}{\sigma_1^2}\right)\\
			&\qquad\qquad+\log_2\left(1+\frac{{\mathbf{h}}_2^{\mathsf{H}}{\bm\Sigma}_2{\mathbf{h}}_2}
			{\sigma_2^2+{\mathbf{h}}_2^{\mathsf{H}}{\bm\Sigma}_1{\mathbf{h}}_2}\right).
		\end{split}
	\end{equation}
	The above expression is derived under the encoding order $2\rightarrow1$. However, as indicated in \cite{vishwanath2003duality,jindal2005sum}, the capacity is always the same regardless of the encoding order.
	
	According to \eqref{Sum_Rate_Capacity_BC_k=2}, solving the broadcast capacity is a non-convex problem, where numerically finding the maximum is nontrivial. However, in \cite{vishwanath2003duality}, a duality is shown to exist between the uplink and downlink which establishes that the dirty-paper capacity region for the BC is equal to the capacity region of the dual MAC (described in \eqref{MAC_Channel}). Particularly, in the dual MAC, the UTs in the system have a joint power constraint ($\sum_{k=1}^{K}p_k\leq P$) instead of individual constraints ($p_k\leq P_k$) as in the conventional MAC. Consequently, the BC sum-rate capacity in \eqref{Sum_Rate_Capacity_BC_k=2} is equivalent to the following dual MAC capacity:
	\begin{align}\label{Sum_Rate_Capacity_BC2}
		&{\mathsf{C}}_{\mathsf{BC}}=\max_{0\leq p_k,\sum_{k}p_k\leq P}\log_2\det\Big({\mathbf{I}}_M+\sum\nolimits_{k=1}^{K}\tilde{\gamma}_k{\mathbf{h}}_k{\mathbf{h}}_k^{\mathsf{H}}\Big)\notag\\
		&=\max_{0\leq p_k,\sum_{k}p_k\leq P}\log_2(1+\tilde{\gamma}_1{\mathsf{g}}_1+\tilde{\gamma}_2{\mathsf{g}}_2
		+\tilde{\gamma}_1{\mathsf{g}}_1\tilde{\gamma}_2{\mathsf{g}}_2(1-\rho )).
	\end{align}
	where $\tilde{\gamma}_k=\frac{p_k}{\sigma_k^2}$ denotes the SNR. In this case, the capacity can be found through a convex problem that can be optimally solved. A novel expression for the NF sum-rate capacity of BC is derived in the following theorem.
	\vspace{-5pt}
	\begin{theorem}\label{BC_capacity_the}
		The optimal power allocation for the dual MAC under NFC is given by
		\begin{equation}
			\left( p_{1}^{\star},p_{2}^{\star} \right) =\begin{cases}
				\left( P,0 \right)&		\kappa _1\leq 0\\
				\left( 0,P \right)&		\kappa _2\leq 0\\
				\left( \frac{\kappa  _1}{\sigma _{2}^{-2}\mathsf{g}_2^{\mathsf{n}}},\frac{\kappa  _2}{\sigma _{1}^{-2}\mathsf{g}_1^{\mathsf{n}}} \right)&		\mathrm{else}\\
			\end{cases},
		\end{equation}
		where $\kappa _1=\frac{P\sigma _{1}^{-2}\sigma _{2}^{-2}\mathsf{g}_1^{\mathsf{n}}\mathsf{g}_2^{\mathsf{n}}(1-\rho_{\mathsf{n}})-\sigma _{1}^{-2}\mathsf{g}_1^{\mathsf{n}}+\sigma _{2}^{-2}\mathsf{g}_2^{\mathsf{n}}}{2\sigma _{1}^{-2}\mathsf{g}_1^{\mathsf{n}}(1-\rho_{\mathsf{n}})} $, and $\kappa _2=\frac{P\sigma _{1}^{-2}\sigma _{2}^{-2}\mathsf{g}_1^{\mathsf{n}}\mathsf{g}_2^{\mathsf{n}}(1-\rho_{\mathsf{n}})+\sigma _{1}^{-2}\mathsf{g}_1^{\mathsf{n}}-\sigma _{2}^{-2}\mathsf{g}_2^{\mathsf{n}}}{2\sigma _{2}^{-2}\mathsf{g}_2^{\mathsf{n}}(1-\rho_{\mathsf{n}})}$. Accordingly, the NF sum-rate capacity of the BC is given by 
		\begin{equation}\label{BC_capacity}
			\mathsf{C}_{\mathsf{BC}}=\begin{cases}
				\log _2\big( 1+P\sigma _{1}^{-2}\mathsf{g}_1 \big)&		\kappa _1\leq 0\\
				\log _2\big( 1+P\sigma _{2}^{-2}\mathsf{g}_2 \big)&		\kappa _2\leq 0\\
				\log _2\big( 1+{\kappa  _1+\kappa _2+\kappa  _1\kappa  _2(1-\rho_{\mathsf{n}})} \big) &		\text{else}\\
			\end{cases}.
		\end{equation} 
	\end{theorem}
	\vspace{-5pt}
	\begin{IEEEproof}
		The results can be derived by applying the Karush–Kuhn–Tucker (KKT) conditions to the right-hand side of \eqref{Sum_Rate_Capacity_BC2}.
	\end{IEEEproof}
	After obtaining the optimized power allocation $\{p_k\}_{k=1}^{2}$ for the dual MAC, we can employ the transformation in \cite[Eqs. (8)–(10)]{vishwanath2003duality} to recover the corresponding downlink covariance matrices that achieve the same rates and the sum-rate. Specifically, for the encoding order $2\rightarrow1$, the optimal covariance matrices for UT 1 and 2 are, respectively, given by
		\begin{align}
			\mathbf{\Sigma }_{1}^{\star}=\frac{p_{1}^{\star}\mathbf{\Lambda h}_1\mathbf{h}_{1}^{\mathsf{H}}\mathbf{\Lambda }}{\mathbf{h}_{1}^{\mathsf{H}}\mathbf{\Lambda h}_1}, \
			\mathbf{\Sigma }_{2}^{\star}=\frac{p_{2}^{\star}\left( 1+\mathbf{h}_{2}^{\mathsf{H}}\mathbf{\Sigma }_{1}^{\star}\mathbf{h}_2 \right)}{\mathsf{g}_2^{\mathsf{n}}}\mathbf{h}_2\mathbf{h}_{2}^{\mathsf{H}},
		\end{align}
		where $\mathbf{\Lambda }=\mathbf{I}-\frac{p_{2}^{\star}\mathbf{h}_2\mathbf{h}_{2}^{\mathsf{H}}}{1+p_{2}^{\star}\mathsf{g}_2^{\mathsf{n}}}$.
	
We then study the asymptotic NF BC capacity for $M_x,M_z\rightarrow \infty$ 
	\vspace{-5pt}
	\begin{corollary}\label{BC_NF_M_cor}
		When $M_x,M_z\rightarrow \infty$, the optimal power allocation policy degenerates into
		\begin{align}
			\lim_{M_x,M_z\rightarrow \infty} \left( p_{1}^{\star},p_{2}^{\star} \right) =\begin{cases}
				\left( P,0 \right)&		\tilde{\kappa}_1\leq 0\\
				\left( 0,P \right)&		\tilde{\kappa}_2\leq 0\\
				\left( \frac{2\tilde{\kappa}_1}{\sigma _{2}^{-2}\xi},\frac{2\tilde{\kappa}_2}{\sigma _{1}^{-2}\xi} \right)&		\mathrm{else}\\
			\end{cases},
		\end{align}
		where $\tilde{\kappa}_1=\frac{P\sigma _{1}^{-2}\sigma _{2}^{-2}\frac{\xi}{2}-\sigma _{1}^{-2}+\sigma _{2}^{-2}}{2\sigma _{1}^{-2}}$, and $\tilde{\kappa}_2=\frac{P\sigma _{1}^{-2}\sigma _{2}^{-2}\frac{\xi}{2}+\sigma _{1}^{-2}-\sigma _{2}^{-2}}{2\sigma _{2}^{-2}}$. Accordingly, the asymptotic BC capacity under the NF model is given by
		\begin{align}
			\!\lim_{M_x,M_z\rightarrow \infty} \mathsf{C}_{\mathsf{BC}}\approx \!\begin{cases}
				\log _2\left( 1\!+\!\frac{P\xi}{2\sigma _{1}^{2}} \right)&		\!\tilde{\kappa}_1\leq 0\\
				\log _2\left( 1\!+\!\frac{P\xi}{2\sigma _{2}^{2}} \right)&		\!\tilde{\kappa}_2\leq 0\\
				\log _2\left( 1\!+\!\tilde{\kappa}_1\!+\!\tilde{\kappa}_2\!+\!\tilde{\kappa}_1\tilde{\kappa}_2 \right)&		\mathrm{else}\\
			\end{cases}.    
		\end{align}
	\end{corollary}
	\vspace{-5pt}
	\begin{IEEEproof}
		Similar to the proof of \textbf{Corollary~\ref{MAC_NF_M_cor}}.
	\end{IEEEproof}
	\vspace{-5pt}
	\begin{remark}\label{BC_constant}
		The results in \textbf{Corollary~\ref{BC_NF_M_cor}} suggest that, as $M_x,M_z \rightarrow \infty$, the NF BC capacity converge to a finite value positively correlated to the AOR. 
	\end{remark}
	\vspace{-5pt}
	
	\begin{figure}[!t]
		\centering
		\setlength{\abovecaptionskip}{0pt}
		\includegraphics[height=0.22\textwidth]{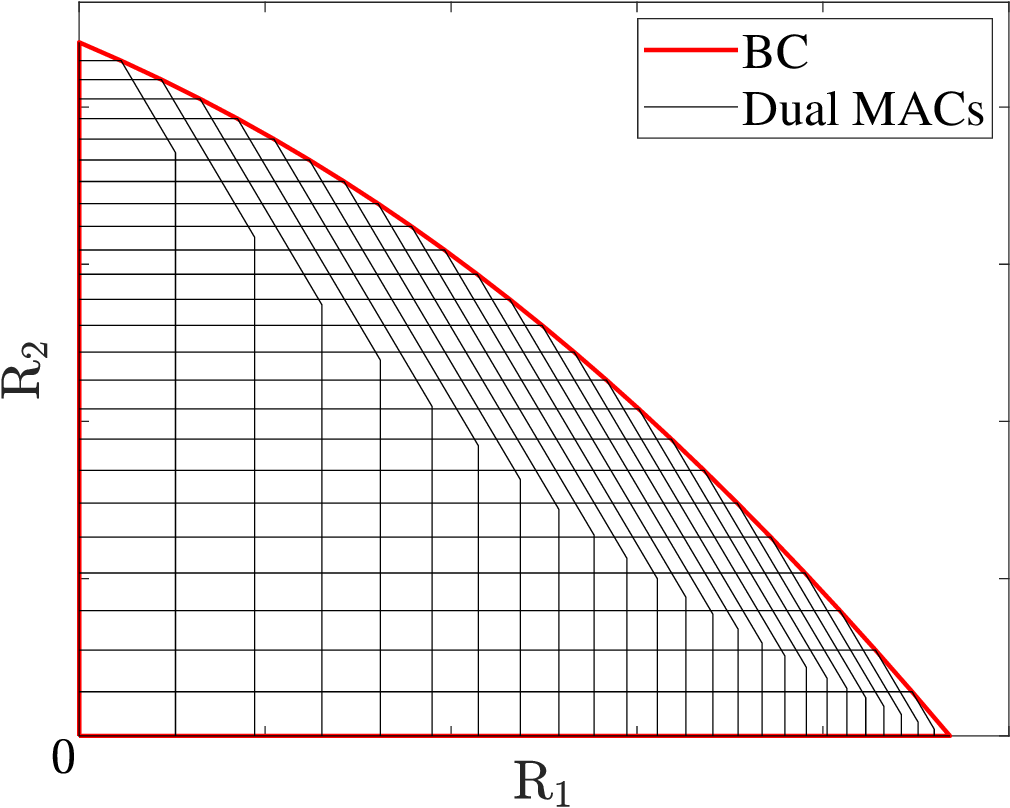}
		\caption{Illustration of the BC and dual MAC capacity regions.}
		\label{BC_illu}
		\vspace{-8pt}
	\end{figure}
	
	Next, we consider the ULA case.
	\vspace{-5pt}
	\begin{corollary}
		When the BS is equipped with a ULA, we have 
		\begin{align}
			\lim_{M\rightarrow \infty} \left( p_{1}^{\star},p_{2}^{\star} \right) =\begin{cases}
				\left( P,0 \right)&		\tilde{\kappa}_{1}^{\prime}\le 0\\
				\left( 0,P \right)&		\tilde{\kappa}_{2}^{\prime}\le 0\\
				\left( \frac{2\pi \tilde{\kappa}_{1}^{\prime}}{\nu _2\xi},\frac{2\pi \tilde{\kappa}_{1}^{\prime}}{\nu _1\xi} \right)&		\mathrm{else}\\
			\end{cases},
		\end{align}
		where $\nu _1=\sigma _{1}^{-2}\epsilon _1\sin \phi _1\sin \theta _{1}^{-1}$, $\nu _2=\sigma _{2}^{-2}\epsilon _2\sin \phi _2\sin \theta _{2}^{-1}$, $\tilde{\kappa}_{1}^{\prime}=\frac{\xi \left( 2\pi \right) ^{-1}P\nu _1\nu _2-\nu _1+\nu _2}{2\nu _1}$, and $\tilde{\kappa}_{2}^{\prime}=\frac{\xi \left( 2\pi \right) ^{-1}P\nu _1\nu _2+\nu _1-\nu _2}{2\nu _2}$. Subsequently, the sum-rate capacity satisfies
		\begin{align}
			\lim_{M\rightarrow \infty} \mathsf{C}_{\mathsf{BC}}=\begin{cases}
				\log _2\left( 1+\frac{P\nu _1\xi}{2\pi} \right)&		\tilde{\kappa}_{1}^{\prime}\le 0\\
				\log _2\left( 1+\frac{P\nu _2\xi}{2\pi} \right)&		\tilde{\kappa}_{2}^{\prime}\le 0\\
				\log _2\left( 1+\tilde{\kappa}_{1}^{\prime}+\tilde{\kappa}_{2}^{\prime}+\tilde{\kappa}_{1}^{\prime}\tilde{\kappa}_{2}^{\prime} \right)&		\mathrm{else}\\
			\end{cases},    
		\end{align}
		which is a finite value.
	\end{corollary}
	\vspace{-5pt}
	\begin{IEEEproof}
		Similar to the proof of \textbf{Corollary~\ref{ula_mac_cor}}.
	\end{IEEEproof}
	
	Notably, if the noise power at each UTs is same, for the UPA case, we can deduce that $\lim_{M_x,M_z\rightarrow \infty} \left( p_{1}^{\star},p_{2}^{\star} \right) =\left(\frac{P}{2},\frac{P}{2}\right)$, which indicates that the optimal power allocation of the dual MAC tends to an equal power allocation when the number of array element becomes extremely large. On the other hand, in the ULA scenario, the optimal power allocation always depends on the UT's positions even when $M\rightarrow\infty$.
	
	\begin{figure*}[!b]
		\hrulefill
		\setcounter{equation}{62}
		\begin{align}\label{far_BC}
			\mathsf{C}_{\mathsf{BC}}\simeq 
			\begin{cases}
				\log _2\!\left( \frac{MPA}{4\pi}\max_{k=1,2} \frac{\Psi _k}{r_{k}^{2}\sigma _{k}^{2}} \right) \triangleq \mathsf{F}_{\mathsf{BC}}^{\mathsf{s}}&		\left( \theta _1,\phi _1 \right) =\left( \theta _2,\phi _2 \right)\\
				\log _2\left( \frac{\left( MPA\left( 4\pi \right) ^{-1}+\Psi _{1}^{-1}r_{1}^{2}\sigma _{1}^{2}+\Psi _{2}^{-1}r_{2}^{2}\sigma _{2}^{2} \right) ^2}{4\Psi _{1}^{-1}\Psi _{2}^{-1}r_{1}^{2}r_{2}^{2}\sigma _{1}^{2}\sigma _{2}^{2}}-1 \right) \triangleq \mathsf{F}_{\mathsf{BC}}^{\mathsf{d}}&		\left( \theta _1,\phi _1 \right) \ne \left( \theta _2,\phi _2 \right)\\
			\end{cases}   
		\end{align}
		\begin{align}\label{BC_ratio}
			\mathsf{F}_{\mathsf{BC}}^{\mathsf{d}}-\mathsf{F}_{\mathsf{BC}}^{\mathsf{s}}=\log _2\left( 1+\frac{\left( MPA\left( 4\pi \right) ^{-1}-\left| \Psi _{1}^{-1}r_{1}^{2}\sigma _{1}^{2}-\Psi _{2}^{-1}r_{2}^{2}\sigma _{2}^{2} \right| \right) ^2}{MPA\pi ^{-1}\max_{k=1,2} \Psi _{k}^{-1}r_{k}^{2}\sigma _{k}^{2}} \right) >0    
		\end{align}
		\setcounter{equation}{59}
	\end{figure*}
	
	\subsubsection{Capacity Region}
	Given a power allocation scheme $\{p_k\}_{k=1}^{2}$ with $p_1+p_2= P$, we can obtain the capacity region of a dual MAC. The capacity region of the BC is the convex hull of all these dual MAC capacity regions \cite{vishwanath2003duality}, which means that the boundary of capacity region of the BC is formed by the corner points of the dual MAC capacity regions, as illustrated in Fig~\ref{BC_illu}. For a given power allocation scheme $\{p_k\}_{k=1}^{2}$, the two corner points of the NF capacity region of the dual MAC, $(\hat{\mathsf{R}}_{1}^{1\rightarrow2},\hat{\mathsf{R}}_{2}^{1\rightarrow2})$ and $(\hat{\mathsf{R}}_{1}^{2\rightarrow1},\hat{\mathsf{R}}_{2}^{2\rightarrow1})$, are given as follows. When the dirty-paper encoding order is $2\rightarrow1$, we have
	\begin{subequations}\label{BC_corner}
		\begin{align}
			&\hat{\mathsf{R}}_{1}^{2\rightarrow1}= \log_2(1+\tilde{\gamma}_1\mathsf{g}_1^{\mathsf{n}}), \\
			&\hat{\mathsf{R}}_{2}^{2\rightarrow1}= \log_2\Big( 1+\frac{\tilde{\gamma}_2\mathsf{g}_2^{\mathsf{n}}+\tilde{\gamma}_1\mathsf{g}_1^{\mathsf{n}}\tilde{\gamma}_2\mathsf{g}_2^{\mathsf{n}}(1-\rho_{\mathsf{n}})}{1+\tilde{\gamma}_1\mathsf{g}_1^{\mathsf{n}}} \Big).
		\end{align}
	\end{subequations}
	When the dirty-paper encoding order is $1\rightarrow2$, we have
	\begin{subequations}
		\begin{align}
			&\hat{\mathsf{R}}_{1}^{1\rightarrow2}= \log _2\Big( 1+\frac{\tilde{\gamma}_2\mathsf{g}_2^{\mathsf{n}}+\tilde{\gamma}_1\mathsf{g}_1^{\mathsf{n}}\tilde{\gamma}_2\mathsf{g}_2^{\mathsf{n}}(1-\rho_{\mathsf{n}})}{1+\tilde{\gamma}_2\mathsf{g}_2^{\mathsf{n}}} \Big),\\
			&\hat{\mathsf{R}}_{2}^{1\rightarrow2}= \log_2\left(1+\tilde{\gamma}_2\mathsf{g}_2^{\mathsf{n}}\right).
		\end{align}
	\end{subequations}
	For any given power allocation scheme $\{p_k\}_{k=1}^{2}$ with $p_1+p_2= P$, we can obtain the closed-form expressions for the corner points of the dual MAC regions. The NF BC capacity region can be then characterized by the convex hull of these corner points.
	\vspace{-5pt}
	\begin{corollary}\label{BC_NF_region_M_cor}
		When $M_x,M_z \rightarrow \infty$, the corner points of the NF dual MAC capacity region with a given power allocation $\{p_k\}_{k=1}^{2}$ satisfy
		\begin{align}
			&\lim_{M_x,M_z\rightarrow \infty} \left( \hat{\mathsf{R}}_{1}^{1\rightarrow 2},\hat{\mathsf{R}}_{2}^{1\rightarrow 2} \right)\approx\lim_{M_x,M_z\rightarrow \infty} \left( \hat{\mathsf{R}}_{1}^{2\rightarrow 1},\hat{\mathsf{R}}_{2}^{2\rightarrow 1} \right)  \notag\\
			&~~~\approx\left( \log _2( 1+{\tilde{\gamma}_1\xi}/{2} ),\log _2( 1+{\tilde{\gamma}_2\xi}/{2}) \right).
		\end{align}    
	\end{corollary}
	\begin{IEEEproof}
		Similar to the proof of \textbf{Corollary~\ref{MAC_NF_M_cor}}.
	\end{IEEEproof}
	\vspace{-5pt}
	\begin{remark}\label{BC_square}
		The results in \textbf{Corollary~\ref{BC_NF_region_M_cor}} suggest that, as the number of the array elements increases, the dual MAC capacity region approaches a finite rectangle, which implies that the NF capacity region of the BC is bounded.
	\end{remark}

	\subsection{Comparison with the FF Capacity}
	Next, we discuss the FF BC capacity for comparison.
	\vspace{-5pt}
	\begin{corollary}\label{BC_FF_M}
		For the FF case, the asymptotic BC capacity for $M\rightarrow\infty$ is given in \eqref{far_BC}, shown at the bottom of this page, which yields $\mathsf{C}_{\mathsf{BC}}\simeq \mathcal{O}(\log M)$.
	\end{corollary}
	\vspace{-5pt}
	\begin{IEEEproof}
		Similar to the proof of \textbf{Corollary~\ref{MAC_FF_M}}.
	\end{IEEEproof}
	\vspace{-5pt}
	\begin{remark}\label{BC_unbounded}
		Rather than converging to a finite value as under the NF model, the BC sum-rate capacity under the FF model grows unboundedly with the number of the array elements, which potentially violates the energy-conservation laws.
	\end{remark}
	
	Further, according to the results of \eqref{BC_ratio}, we find that, in contrast to the NF case, the FF BC capacity with co-directional UTs is smaller than that when the UTs are in different directions.

	The capacity region of the BC in the FF scenario, when considering an infinitely large array, is influenced by the relative directions of the UTs. Specifically, when $M\rightarrow\infty$, if the UTs are located in the same direction, the corner points of the dual MAC capacity region satisfy
	\setcounter{equation}{64} 
	\begin{subequations}
		\begin{align}
			(\hat{\mathsf{R}}_{1}^{1\rightarrow 2},\hat{\mathsf{R}}_{2}^{1\rightarrow 2})&\simeq (\log _2( 1+{\tilde{\gamma}_1r_{2}^{2}}/{(\tilde{\gamma}_2r_{1}^{2})} ),\mathcal{O}(\log M)),\\
			(\hat{\mathsf{R}}_{1}^{2\rightarrow 1},\hat{\mathsf{R}}_{2}^{2\rightarrow 1})&\simeq (\mathcal{O}(\log M),\log _2( 1\!+{\tilde{\gamma}_2r_{1}^{2}}/{(\tilde{\gamma}_1r_{2}^{2})} )).
		\end{align}
	\end{subequations}
	If the UTs are positioned in different directions, we have
	\begin{align}
		\!\!(\hat{\mathsf{R}}_{1}^{1\rightarrow 2}\!,\hat{\mathsf{R}}_{2}^{1\rightarrow 2})\!\approx\!(\hat{\mathsf{R}}_{1}^{2\rightarrow 1}\!,\hat{\mathsf{R}}_{2}^{2\rightarrow 1})\!\simeq\! (\mathcal{O}(\log M) ,\mathcal{O}(\log M)).\!\!
	\end{align}
	The above results suggest that the BC capacity region under the FF model can extend unboundedly with the number of the array elements, which is impractical. Additionally, the FF BC capacity region of cases when the UTs are in different directions is more extensive than that of cases when the UTs are in a same direction.   
	
	The comparison between BC capacity under the NF and FF models, similar to the MAC, highlights the importance of incorporating NF modeling in asymptotic evaluations, which provides a more accurate and feasible framework in terms of energy sustainability. Additionally, the introduction of range dimensions by spherical-wave propagation in the NF model enhances the BC capacity for co-directional UTs.
	
	\subsection{Comparison with Linear Precoders}
		Although DPC is a potent capacity-achieving scheme, its practical implementation proves challenging. Consequently, non-DPC linear downlink precoding schemes hold practical significance. In the following analysis, we aim to evaluate the sum-rate performance of several typical linear precoders and examine the sum-rate gap between these linear precoders and the capacity-achieving DPC precoders in the context of NFC. Specifically, let $\mathbf{w}_k\in{\mathbbmss{C}}^{M\times1}$ denote the normalized linear precoder used to deliver $x_k$ with $\lVert{\mathbf{w}}_k\rVert^2=1$. The achievable downlink per-user rate is given by
		\begin{align}\label{Linear_Precoder_BC_Per_User_Rate}
			\hat{\mathsf{R}}_k=\log_2\bigg(1+\frac{\lvert{\mathbf{w}}_k^{\mathsf{H}}{\mathbf{h}}_k\rvert^2\hat{p}_k}
			{\sigma_k^2+\sum_{k'\ne k}\lvert{\mathbf{w}}_{k'}^{\mathsf{H}}{\mathbf{h}}_{k}\rvert^2\hat{p}_{k'}}\bigg),
		\end{align}
		where $\hat{p}_k$ is the transmit power. For the sake of brevity, we continue to consider the two-user case with $K=2$.
		\subsubsection{Maximal-Ratio Transmission Precoding}
		We begin with the maximal-ratio transmission (MRT) precoder, which is designed to maximize the channel gain for each UT. This yields ${\mathbf{w}}_k=\frac{{\mathbf{h}}_k}{\lVert{\mathbf{h}}_k\rVert}$ for $k\in{\mathcal{K}}$. Substituting ${\mathbf{w}}_k=\frac{{\mathbf{h}}_k}{\lVert{\mathbf{h}}_k\rVert}$ into \eqref{Linear_Precoder_BC_Per_User_Rate} and performing some basic mathematical manipulations, we express the sum-rate as follows:
		\begin{equation}
			\hat{\mathsf{R}}_1+\hat{\mathsf{R}}_2=\sum_{k=1}^{2}
			\log _2\Big( 1+{\hat{\gamma}}_k{\mathsf{g}}_k\Big(1-\frac{\hat{\gamma}_{k'}{\mathsf{g}}_{k}\rho}{1+\hat{\gamma}_{k'}{\mathsf{g}}_{k}\rho}\Big) \Big)\triangleq \hat{\mathsf{R}}_{{\mathsf{mrt}}},
		\end{equation}
		where ${\hat{\gamma}}_k= \frac{\hat{p}_{k}}{\sigma_k^2}$ denotes the SNR.
		\subsubsection{Zero-Forcing Precoding}
		Next, we consider ZF precoding, which aims to nullify the IUI. In this case, ${\mathbf{w}}_k^{\mathsf{H}}{\mathbf{h}}_{k'}=0$ for $k'\ne k$. Following the steps to obtain \eqref{Zero-Forcing_Combiner_MAC}, we write the ZF precoder for UT $k$ as follows:
		\begin{align}\label{Zero-Forcing_Precoder_BC}
			{\mathbf{w}}_k=\frac{( \mathbf{I}_M-\mathbf{h}_{k'}\mathbf{h}_{k'}^{\mathsf{H}}/\| \mathbf{h}_{k'} \| ^2 ) \mathbf{h}_k}
			{\lVert( \mathbf{I}_M-\mathbf{h}_{k'}\mathbf{h}_{k'}^{\mathsf{H}}/\| \mathbf{h}_{k'} \| ^2 ) \mathbf{h}_k\rVert}.
		\end{align}
		Substituting \eqref{Zero-Forcing_Precoder_BC} into \eqref{Linear_Precoder_BC_Per_User_Rate} and performing some simple manipulations, we obtain
		\begin{align}
			\hat{\mathsf{R}}_1+\hat{\mathsf{R}}_2=\sum\nolimits_{k=1}^{2}
			\log _2(1+\hat{\gamma}_k\mathsf{g}_k(1-\rho))\triangleq {\hat{\mathsf{R}}}_{{\mathsf{zf}}}.
		\end{align}
		\subsubsection{Summary and Discussion}
		Through the derivations above, we observe that the sum-rate achieved by the considered linear precoding schemes can be expressed as a function of the channel gains and the CCF, i.e., 
		\begin{align}\label{Sum_Rate_Linear_General}
			\hat{\mathsf{R}}_{\ell}=\sum\nolimits_{k=1}^{2} \log_2(1+\hat{\gamma}_k\mathsf{g}_k (1-\hat{f}_{\ell}(\hat{\gamma}_{k'}\mathsf{g}_{k},\rho))),
		\end{align}
		where $\ell\in\{{\mathsf{mrt}},{\mathsf{zf}}\}$, $\hat{f}_{{\mathsf{mrt}}}(x,z)\triangleq\frac{xz}{1+xz}$, and $\hat{f}_{{\mathsf{zf}}}(x,z)\triangleq z$. By substituting \eqref{channelgain_NF} and \eqref{rho_NF} into \eqref{Sum_Rate_Linear_General}, we obtain the NF sum-rate expressions under the linear precoders, which can be written as follows:
		\begin{align}
			\hat{\mathsf{R}}_{\ell}=\sum\nolimits_{k=1}^{2} \log_2(1+\hat{\gamma}_k\mathsf{g}_k^{\mathsf{n}} (1-\hat{f}_{\ell}(\hat{\gamma}_{k'}\mathsf{g}_{k}^{\mathsf{n}},\rho_{\mathsf{n}}))).
		\end{align}
		
		As previously mentioned, we have $\lim_{M_x,M_z\rightarrow \infty}\rho\approx 0$ under the NF channel model. This means that
		\begin{align}
			\lim_{M_x,M_z\rightarrow \infty}\hat{\mathsf{R}}_{\ell}\approx \sum\nolimits_{k=1}^2{\log _2\left( 1+\hat{\gamma}_k\mathsf{g}_k^{\mathsf{n}} \right)}
		\end{align}
		holds for $\ell\in\{{\mathsf{mrt}},{\mathsf{zf}}\}$.
		\vspace{-5pt}
		\begin{remark}\label{rem_linear_down}
			The above discussion implies that in NF BC, when the BS antenna number is sufficiently large, the considered linear precoding schemes can approach the upper bound of the sum-rate capacity, akin to the capacity-achieving DPC.
		\end{remark}
		
		\begin{figure}[!t]
			\centering
			\subfigbottomskip=5pt
			\subfigcapskip=0pt
			\setlength{\abovecaptionskip}{0pt}
			\subfigure[$(\theta_2,\phi_2)=(\frac{\pi}{3},\frac{2\pi}{3})$.]
			{
				\includegraphics[height=0.175\textwidth]{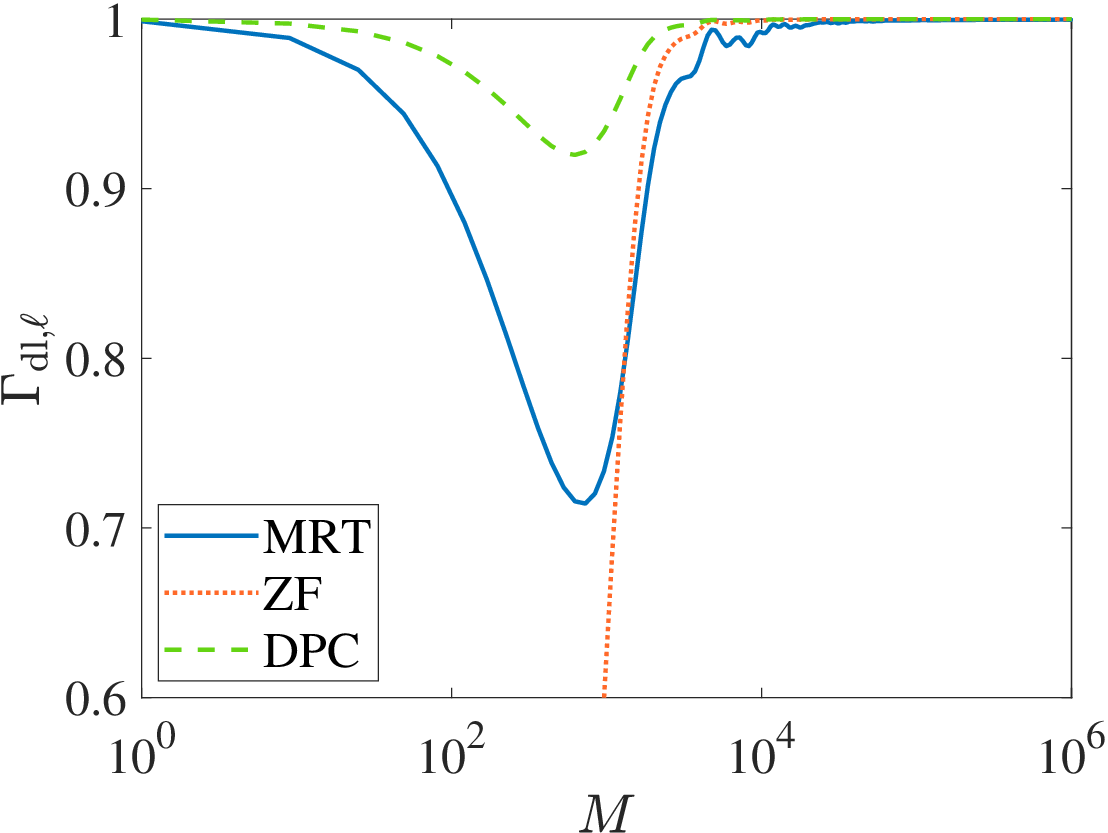}
			}%
			\subfigure[$(\theta_2,\phi_2)=(\frac{\pi}{3}\!+\!\frac{\pi}{60},\frac{2\pi}{3}\!+\!\frac{\pi}{60})$.]
			{
				\includegraphics[height=0.175\textwidth]{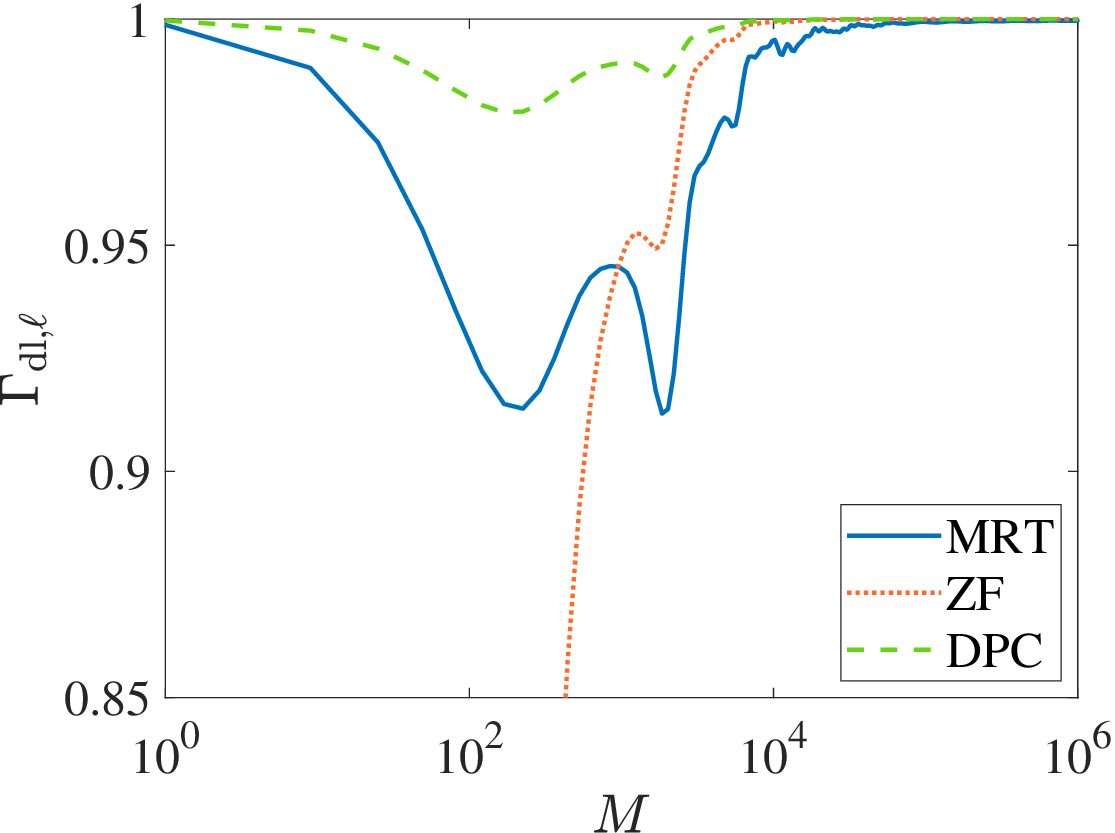}
			}
			\caption{$\Gamma_{{\mathsf{dl}},\ell}$ versus $M$ for different precoders with $(\theta_1,\phi_1)=(\frac{\pi}{3},\frac{2\pi}{3})$ and the other parameter settings given in Section~\ref{numerical}.}
			\label{linear_BC}
			\vspace{-7pt}
		\end{figure}

		This conclusion in \textbf{Remark \ref{rem_linear_down}} holds when $M_x,M_z\rightarrow \infty$. However, in practical NFC systems with large yet finite values of $M_x,M_z$, there will be a gap between the achievable sum-rate and its bound $\sum_{k=1}^2{\log _2\left( 1+\hat{\gamma}_k\mathsf{g}_k^{\mathsf{n}} \right)}$, which is given by
		\begin{align}
			\Gamma_{{\mathsf{dl}},\ell}\triangleq\frac{\sum\nolimits_{k=1}^{2} \log_2(1+\hat{\gamma}_k\mathsf{g}_k^{\mathsf{n}} (1-\hat{f}_{\ell}(\hat{\gamma}_{k'}\mathsf{g}_{k}^{\mathsf{n}},\rho_{\mathsf{n}})))}
			{\sum_{k=1}^2{\log _2\left( 1+\hat{\gamma}_k\mathsf{g}_k^{\mathsf{n}} \right)}}.
		\end{align}
		{\figurename} {\ref{linear_BC}} plots $\Gamma_{{\mathsf{dl}},\ell}$ versus the BS antenna number $M$ under equal power allocation, i.e., $\hat{p}_k=\frac{P}{K}$ for $k=1,2$. For comparison, the ratio between the rate achieved by DPC precoder and the upper bound is also presented. As shown, with the increasing number of antennas, the considered precoding schemes achieve a sum-rate close to their upper bound. Among the linear schemes, the MRT precoding scheme achieves superior sum-rate performance for smaller $M$, while NF precoder is more effective with large-scale array. Additionally, we can observed that linear precoders exhibit comparable performance to the capacity-achieving DPC precoders for larger $M$ in the NF scenario.

	\subsection{Extension to Cases of $K>2$}
	When $K>2$, the channel capacity of the BC can be still analyzed by studying its dual MAC, which yields
		\begin{align}
			&{\mathsf{C}}_{\mathsf{BC}}=\max_{0\leq p_k,\sum_{k=1}^{K}p_k= P}\log_2\det\left({\mathbf{I}}+\sum\nolimits_{k=1}^{K}\frac{p_k}{\sigma_k^2}{\mathbf{h}}_k{\mathbf{h}}_k^{\mathsf{H}}\right)\nonumber\\
			&=\!\max_{0\leq p_k,\sum_{k=1}^{K}p_k= P}\!\log_2\det\!\left[\scalebox{0.85}{ \begin{matrix}
					1\!+\!\frac{p_1}{\sigma_1^2}\| \mathbf{h}_1 \| ^2&		\cdots&		\frac{{\sqrt{p_1p_K}}}{\sigma_1^2\sigma_K^2}\mathbf{h}_{1}^{\mathsf{H}}\mathbf{h}_K\\
					\vdots&		\ddots&		\vdots\\
					\frac{{\sqrt{p_1p_K}}}{\sigma_1^2\sigma_K^2}\mathbf{h}_{K}^{\mathsf{H}}\mathbf{h}_1&		\cdots&		1\!+\!\frac{p_K}{\sigma_K^2}\| \mathbf{h}_K \| ^2\\
			\end{matrix}} \right].\label{BC_Multiuser_Sum_Rate_Capacity}
		\end{align}
		Note that calculating the BC capacity ${\mathsf{C}}_{\mathsf{BC}}$ involves solving the convex optimization problem formulated on the right-hand side of \eqref{BC_Multiuser_Sum_Rate_Capacity}, which can be efficiently addressed using the sum-power iterative water-filling method, as introduced in \cite{jindal2005sum}. It is clearly shown that the results can be expressed in terms of the channel gains and CCFs. Furthermore, given a power allocation scheme $\{p_k\}_{k=1}^{K}$ with $\sum_{k=1}^{K}p_k= P$, we can obtain the capacity region of a dual MAC. The capacity region of the BC is the convex hull of all these dual MAC capacity regions, which means that the boundary of the BC capacity region is formed by the corner points of the dual MAC capacity regions.
	
	\section{Multicast Channel}\label{sec_MC}
	In this section, we investigate the NF multicast capacity. 
	\subsection{NF Multicast Capacity}
	Given ${\mathsf{tr}}({\bm\Sigma})\leq P$, the transmission rate of the MC from the BS to UT $k$ is given by
	\begin{align}
		{\mathsf{R}}_{\mathsf{MC},k}=\log_2\left(1+\sigma_k^{-2}{\mathbf{h}}_k^{\mathsf{H}}{\bm\Sigma}{\mathbf{h}}_k\right).
	\end{align}
	Since a common message is delivered in the MC, the multicast rate is limited by the minimum of the maximum transmission rate of all UTs \cite{jindal2006capacity}, which is given by
	\begin{align}
		{\mathsf{C}}_{\mathsf{MC}}&=\max_{{\bm\Sigma}\succeq{\mathbf{0}},{\mathsf{tr}}({\bm\Sigma})\leq P}\min_{k=1,\ldots,K}{\mathsf{R}}_{\mathsf{MC},k}\notag\\
		&=\max_{{\bm\Sigma}\succeq{\mathbf{0}},{\mathsf{tr}}({\bm\Sigma})\leq P}\log_2\Big(1+\min_{k=1,\ldots,K}\sigma_k^{-2}{\mathbf{h}}_k^{\mathsf{H}}{\bm\Sigma}{\mathbf{h}}_k\Big).\label{Multicast_Capacity}
	\end{align} 
	This suggests that the multicast capacity can be obtained by solving a convex problem, as shown on the right-hand side of \eqref{Multicast_Capacity}. However, no closed-form or water-filling-based solution is known to exist for this problem. 

For practical purposes, the covariance matrix $\bm\Sigma$ is often constrained. When transmit beamforming is employed, it holds that $\bm\Sigma=P\mathbf{w}\mathbf{w}^{\mathsf{H}}$, which constrains $\bm\Sigma$ to be of unit rank \cite{jindal2006capacity,beamforming}. In this case, the multicast capacity in \eqref{Multicast_Capacity} can be reformulated as follows:
	\begin{align}
		{\mathsf{C}}_{\mathsf{MC}}=\max_{\lVert\mathbf{w}\rVert\leq 1}\log_2\Big(1+P\min_{k=1,\ldots,K}\sigma_k^{-2}\lvert{\mathbf{h}}_k^{\mathsf{H}}\mathbf{w}\rvert^2\Big).\label{Multicast_Capacity_new}
	\end{align}
	
	Based on the monotonicity of $\sigma_k^{-2}\lvert{\mathbf{h}}_k^{\mathsf{H}}\mathbf{w}\rvert^2$ with respect to $\lVert{\mathbf{w}}\rVert$, it is easily shown that the optimal $\lVert{\mathbf{w}}\rVert$ satisfies $\lVert{\mathbf{w}}\rVert^2=1$. Therefore, according to \eqref{Multicast_Capacity_new}, the beamforming vector that achieves the multicast capacity for $K=2$ can be determined from the following problem:
	\begin{subequations}\label{Multicast_Capacity_Problem1}
		\begin{align}
			&\min\nolimits_{{\mathbf{w}},t}~-t\label{Multicast_Capacity_Problem_Obj1}\\
			&~{\rm{s.t.}}~\sigma_1^{-2}\lvert{\mathbf{h}}_1^{\mathsf{H}}\mathbf{w}\rvert^2\geq t, ~\sigma_2^{-2}\lvert{\mathbf{h}}_2^{\mathsf{H}}\mathbf{w}\rvert^2\geq t,~\lVert{\mathbf{w}}\rVert^2= 1.\label{Power_Min_Problem_Instantaneous_Cons}
		\end{align}
	\end{subequations}
	By using the KKT conditions, we obtain the optimal solution to problem \eqref{Multicast_Capacity_Problem1} as follows.
	\vspace{-5pt}
	\begin{theorem}\label{MC_capacity_the}
		The NF optimal beamforming vector that achieves the multicast capacity is given by
		\begin{equation}\label{Optimal_Beamformer_Solution}
			{\mathbf{w}}^{\star}=
			\begin{cases}
				{\mathsf{g}_1^{\mathsf{n}}}^{-\frac{1}{2}}{{\mathbf{h}}_1}             & {\frac{\mathsf{g}_1^{\mathsf{n}}}{\sigma_1^2}\leq\rho_{\mathsf{n}}\frac{\mathsf{g}_2^{\mathsf{n}}}{\sigma_2^2}}\\
				{\mathsf{g}_2^{\mathsf{n}}}^{-\frac{1}{2}}{{\mathbf{h}}_2}            & {\frac{\mathsf{g}_2^{\mathsf{n}}}{\sigma_2^2}\leq\rho_{\mathsf{n}}\frac{\mathsf{g}_1^{\mathsf{n}}}{\sigma_1^2}}\\
				\frac{\mu_1}{\sigma_1\sqrt{\eta}}{\mathbf{h}}_{1}+
				\frac{\mu_2}{\sigma_2\sqrt{\eta}} {\mathbf{h}}_{2}{\rm{e}}^{-{\rm{j}}\angle({\mathbf{h}}_{1}^{\mathsf{H}}{\mathbf{h}}_{2})}          & {\text{else}}    
			\end{cases},
		\end{equation}
		where $\mu _1=\frac{\sigma _{1}^{2}\mathsf{g}_2^{\mathsf{n}}-\sigma _1\sigma _2 \sqrt{\mathsf{g}_1^{\mathsf{n}}\mathsf{g}_2^{\mathsf{n}}\rho_{\mathsf{n}}}}{\chi}$, $\mu _2=\frac{\sigma _{2}^{2}\mathsf{g}_1^{\mathsf{n}}-\sigma _1\sigma _2\sqrt{\mathsf{g}_1^{\mathsf{n}}\mathsf{g}_2^{\mathsf{n}}\rho_{\mathsf{n}}}}{\chi}$, $\eta =\frac{\mathsf{g}_1^{\mathsf{n}}\mathsf{g}_2^{\mathsf{n}}(1-\rho_{\mathsf{n}})}{\chi}$, and $\chi =\sigma _{2}^{2}\mathsf{g}_1^{\mathsf{n}}+\sigma _{1}^{2}\mathsf{g}_2^{\mathsf{n}}-2\sigma _1\sigma _2 \sqrt{\mathsf{g}_1^{\mathsf{n}}\mathsf{g}_2^{\mathsf{n}}\rho_{\mathsf{n}}}$. Accordingly, the multicast capacity is given by
		\begin{equation}\label{MC_capacity}
			{\mathsf{C}}_{\mathsf{MC}}=
			\begin{cases}
				\log_2(1\!+\!P\sigma_1^{-2}\mathsf{g}_1^{\mathsf{n}})           & {{\sigma_1^{-2}}{\mathsf{g}_1^{\mathsf{n}}}\leq\rho_{\mathsf{n}}{\sigma_2^{-2}}{\mathsf{g}_2^{\mathsf{n}}}}\\
				\log_2(1\!+\!P\sigma_2^{-2}\mathsf{g}_2^{\mathsf{n}})           & {{\sigma_2^{-2}}{\mathsf{g}_2^{\mathsf{n}}}\leq\rho_{\mathsf{n}}{\sigma_1^{-2}}{\mathsf{g}_1^{\mathsf{n}}}}\\
				\log_2\left(1+P\eta\right)             & {\text{else}}
			\end{cases} .
		\end{equation}
	\end{theorem}
	\begin{IEEEproof}
		Please refer to Appendix~\ref{Appendix:F} for more details.
	\end{IEEEproof}
	\vspace{-5pt}
	\begin{corollary}\label{MC_NF_M_cor}
		When $M_x,M_z\rightarrow \infty$, the NF optimal beamforming vector satisfies
		\begin{align}
			\lim_{M_x,M_z\rightarrow \infty}\!\! \mathbf{w}^{\star}\!=\!\sqrt{\frac{2\sigma _{1}^{2}}{\xi \!\left( \sigma _{1}^{2}\!+\!\sigma _{2}^{2} \right)}}\mathbf{h}_1\!+\!\sqrt{\frac{2\sigma _{2}^{2}}{\xi\! \left( \sigma _{1}^{2}\!+\!\sigma _{2}^{2} \right)}}\mathbf{h}_2\mathrm{e}^{-\mathrm{j}\angle (\mathbf{h}_{1}^{\mathsf{H}}\mathbf{h}_2)}.
		\end{align}
		Accordingly, the asymptotic NF multicast capacity is given by
		\begin{align}
			\lim_{M_x,M_z\rightarrow \infty} \mathsf{C}_{\mathsf{MC}}=\log _2\left( 1+\frac{P\xi}{2\left( \sigma _{1}^{2}+\sigma _{2}^{2} \right)} \right) .    
		\end{align}
	\end{corollary}
	\begin{IEEEproof}
		Similar to the proof of \textbf{Corollary~\ref{MAC_NF_M_cor}}.
	\end{IEEEproof}
	\vspace{-3pt}
	\begin{remark}\label{MC_constant}
		The results in \textbf{Corollary~\ref{MC_NF_M_cor}} suggest that, as $M_x,M_z \rightarrow \infty$, the NF MC capacity converge to a finite value that is proportional to the AOR. 
	\end{remark}
	\vspace{-5pt}
	\vspace{-5pt}
	\begin{corollary}
		When the BS is equipped with a ULA, we have 
		\begin{align}
			\lim_{M\rightarrow \infty} \!\mathsf{C}_{\mathsf{MC}}\!=\!\log _2\!\Big( 1\!+\!\frac{P\xi \left( 2\pi \right) ^{-1}\epsilon _1\epsilon _2\sin \phi _1\sin \phi _2}{\sigma _{2}^{2}\epsilon _1\sin \phi _1\sin \theta _2\!+\!\sigma _{1}^{2}\epsilon _2\sin \phi _2\sin \theta _1} \Big),
		\end{align}
		which is a finite value.
	\end{corollary}
	\vspace{-3pt}
	\begin{IEEEproof}
		Similar to the proof of \textbf{Corollary~\ref{ula_mac_cor}}.
	\end{IEEEproof}
	
	\begin{figure*}[!t]
		\centering
\subfigbottomskip=0pt
	\subfigcapskip=-3pt
\setlength{\abovecaptionskip}{2pt}
		\subfigure[NF $\rho _{\mathsf{n}}$.]
		{
			\includegraphics[height=0.29\textwidth]{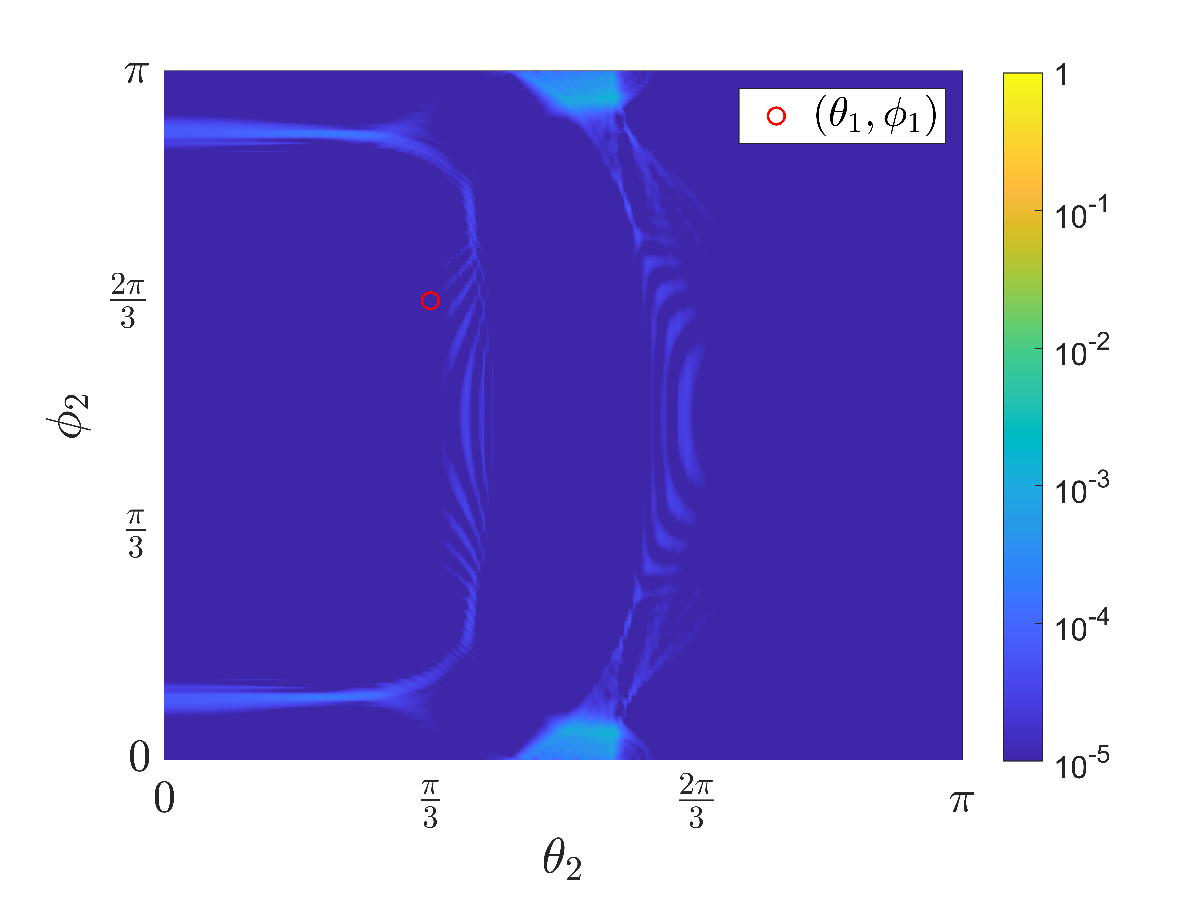}
			\label{rho_nf}	
		}
		\subfigure[FF $\rho _{\mathsf{f}}$.]
		{
			\includegraphics[height=0.29\textwidth]{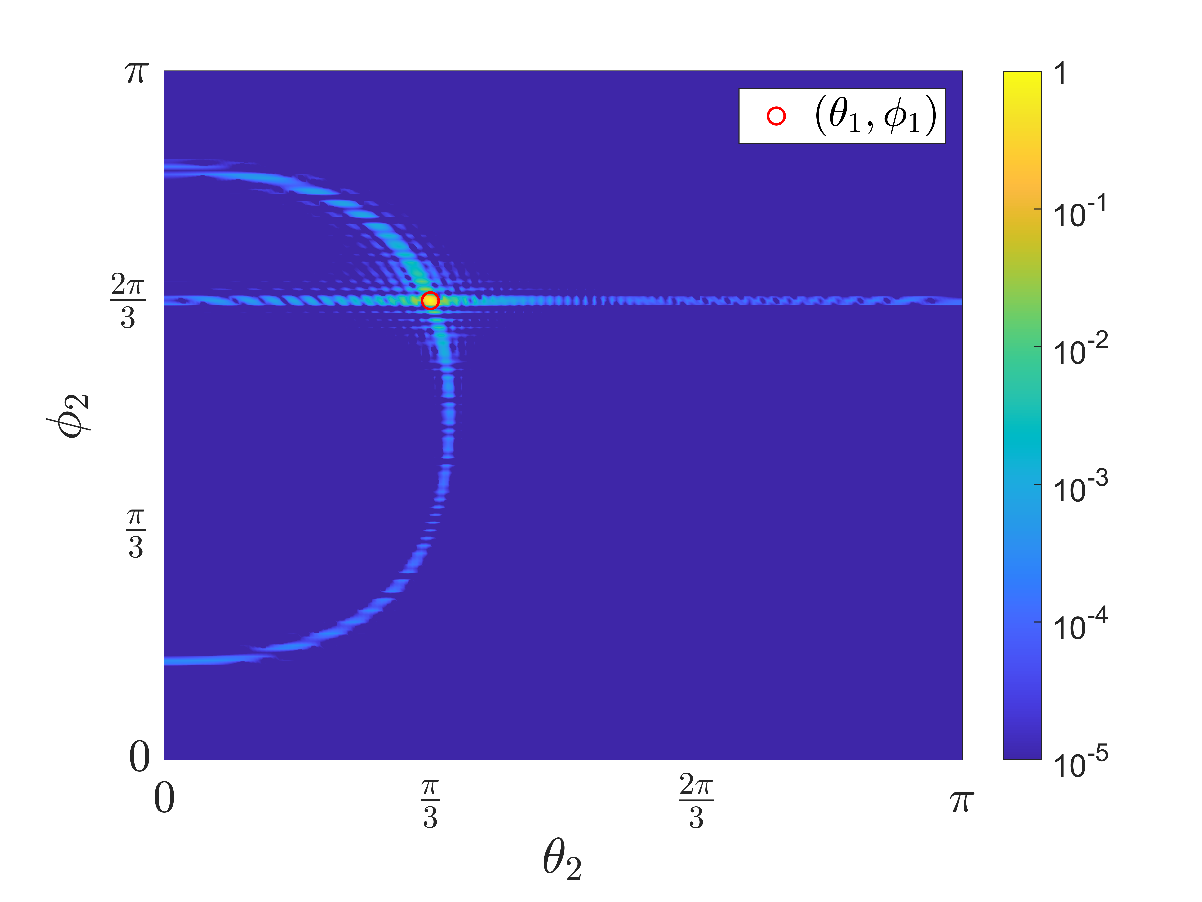}
			\label{rho_ff}	
		}
		\caption{Illustration of the CCF with respect to $(\theta_2, \phi_2)$, with $M_x=M_z=65$.}
		\label{rho}
		\vspace{-9pt}
	\end{figure*}
	
	\begin{figure*}[!t]
		\begin{minipage}[t]{0.5\linewidth}
			\centering
			\setlength{\abovecaptionskip}{2pt}
			\includegraphics[height=0.51\textwidth]{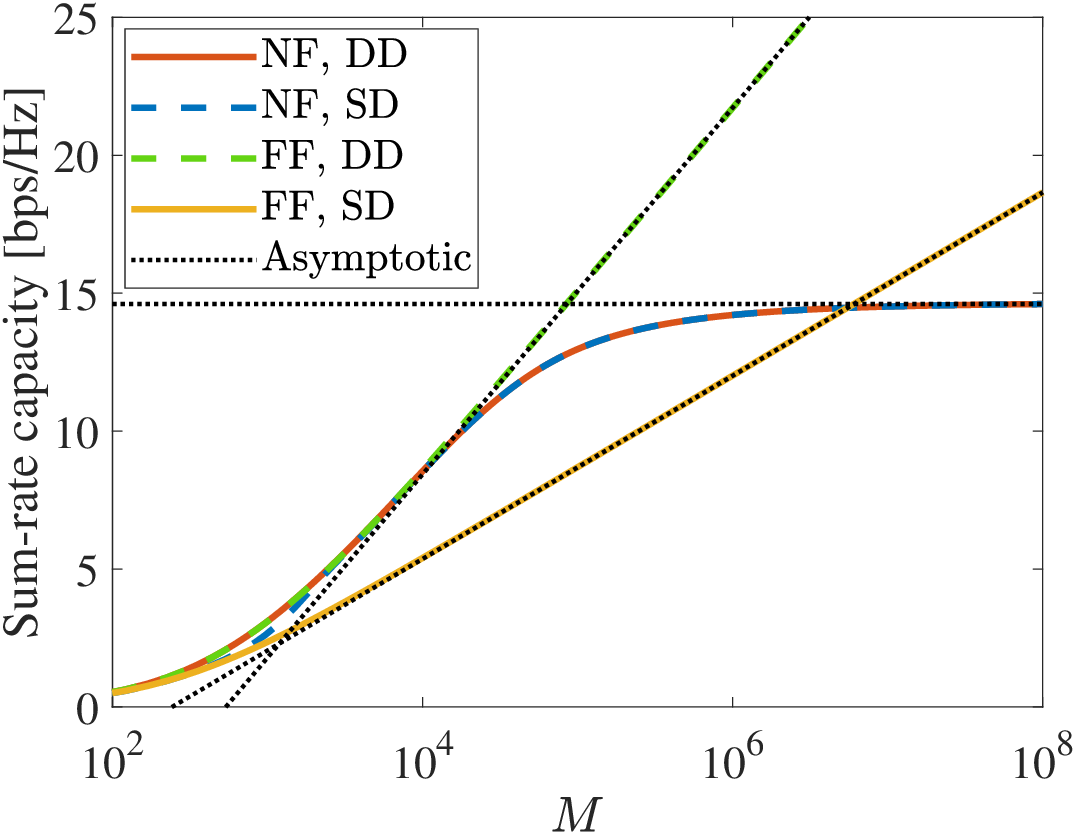}
			\caption{MAC sum-rate capacity versus $M$.}
			\vspace{-10pt}
			\label{MAC}
		\end{minipage}
		\begin{minipage}[t]{0.5\linewidth}
			\centering
			\setlength{\abovecaptionskip}{2pt}
			\includegraphics[height=0.52\textwidth]{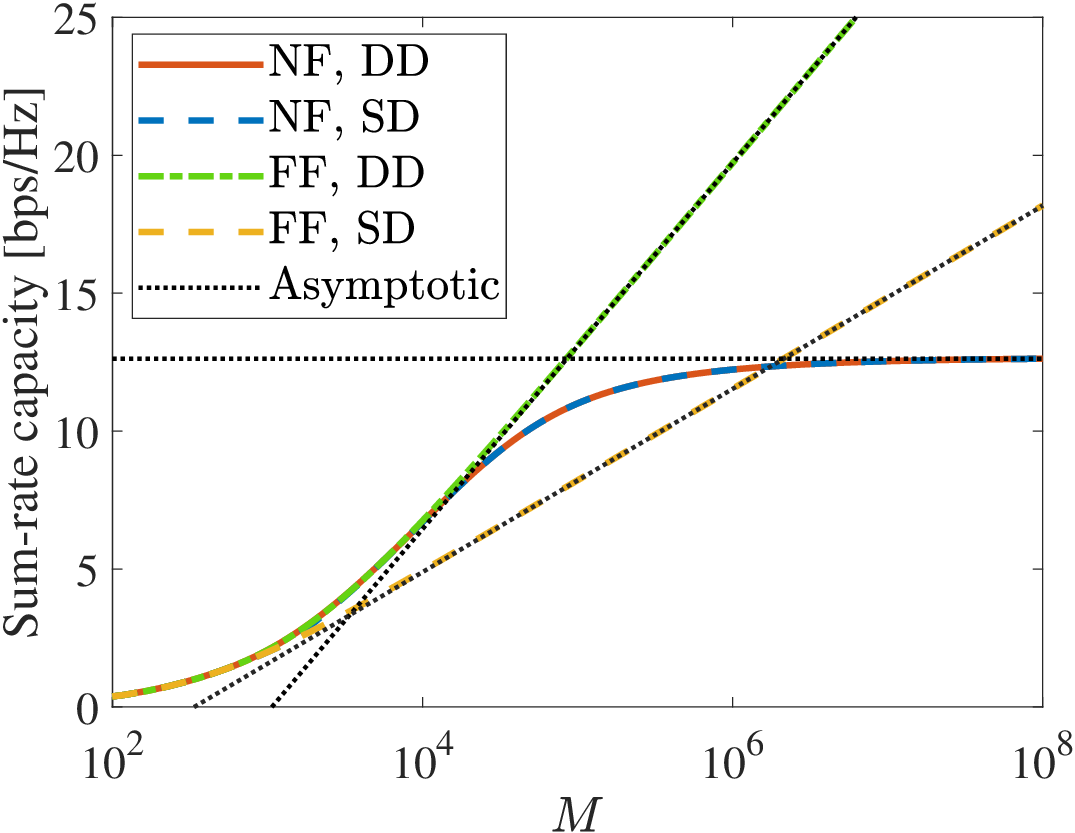}
			\caption{BC sum-rate capacity versus $M$.}
			\vspace{-10pt}
			\label{BC}
		\end{minipage}
	\end{figure*}

	\subsection{Comparison with the FF Capacity}
	We next calculate the FF MC capacity.
	\vspace{-5pt}
	\begin{corollary}\label{MC_FF_M}
		Under the FF model, the asymptotic multicast capacity for $M\rightarrow\infty$ is given by
		\begin{equation}\label{far_MC}
			\mathsf{C}_{\mathsf{MC}}\!\simeq\! 
			\begin{cases}
				\!\!\log _2\!\left( \frac{MPA}{4\pi}\underset{k=1,2}{\min}  \,\frac{\Psi _k}{r_{k}^{2}\sigma _{k}^{2}} \right) \!\triangleq\! \mathsf{F}_{\mathsf{MC}}^{\mathsf{s}}&		\!\left( \theta _1,\phi _1 \right)\! =\!\left( \theta _2,\phi _2 \right)\\
				\!\!\log _2\!\left( \frac{MPA/(4\pi)}{  \Psi _{1}^{-1}r_{1}^{2}\sigma _{1}^{2}+\Psi _{2}^{-1}r_{2}^{2}\sigma _{2}^{2} } \right) \!\triangleq\! \mathsf{F}_{\mathsf{MC}}^{\mathsf{d}}&		\!\left( \theta _1,\phi _1 \right) \!\ne\! \left( \theta _2,\phi _2 \right)\\
			\end{cases}\!,    
		\end{equation}
		which yields $\mathsf{C}_{\mathsf{MC}}\simeq \mathcal{O}(\log M)$.
		
	\end{corollary}
	\vspace{-3pt}
	\begin{IEEEproof}
		Similar to the proof of \textbf{Corollary~\ref{MAC_FF_M}}.
	\end{IEEEproof}
	\vspace{-3pt}
	\begin{remark}\label{MC_unbound}
		Rather than converging to a finite value as under the NF model, the multicast capacity under the FF model can grow unboundedly with the number of the array elements, which potentially violates the energy-conservation laws.
	\end{remark}
	\vspace{-5pt}
	By calculating $\mathsf{F}_{\mathsf{MC}}^{\mathsf{s}}-\mathsf{F}_{\mathsf{MC}}^{\mathsf{d}}$, we have
	\begin{align}\label{MC_ratio}
		\mathsf{F}_{\mathsf{MC}}^{\mathsf{s}}-\mathsf{F}_{\mathsf{MC}}^{\mathsf{d}}=\log _2\left( 1+\Upsilon \right)  \in(0,1],
	\end{align}
	where $\Upsilon =\frac{\min_{k=1,2} r_{k}^{2}\sigma _{k}^{2}}{\max_{k=1,2} r_{k}^{2}\sigma _{k}^{2}}\in(0,1]$. The above results suggest that, within the FF context, the multicast capacity is higher when the UTs are oriented in the same direction compared to when they are in different directions, while the discrepancy converges to a constant not greater than one as $M$ increases.    
	
	This particular observation stems from the absence of the IUI under the MC, given that all UTs are intended to receive the same message. In this case, a high level of channel correlation is actually beneficial. Therefore, it can be concluded that unlike in the MAC and BC, the effect of NFC, i.e., the added range dimension, is not favorable for co-directional UTs in the multicast setting.
	
	\vspace{-5pt}
	\subsection{Extension to Cases of $K>2$}
	The multicast capacity under the scenario of $K>2$ is still an open problem. However, we can derive an upper bound for multicast capacity as follows.
	\vspace{-3pt}
	\begin{lemma}\label{MC_K>2}
		The multicast capacity is upper bounded as
		\begin{align}
			\mathsf{C}_{\mathsf{MC}}\le \log _2\left(1+\frac{P}{K}\sum\nolimits_{k=1}^K{\sigma _{k}^{-2}}\mathsf{g}_k^{\mathsf{n}}\right).
		\end{align}
	\end{lemma}
	\vspace{-5pt}
	\begin{IEEEproof}
		Please refer to Appendix~\ref{Appendix:G}.
	\end{IEEEproof}
	Given that the upper bound is articulated as a function of the channel gains, closed-form expressions for this bound can be derived for both NF and FF models. Further, we can find that when $M_x,M_z \rightarrow \infty$, the NF multicast capacity has a finite upper bound $\log _2(1+\frac{\xi}{2K}\sum_{k=1}^K{\sigma _{k}^{-2}})$, while the upper bound for the FF capacity tends toward infinity.

	\section{Numerical Results}\label{numerical}
	In this section, numerical results for the capacities of the three channels are presented. Without otherwise specification, the simulation parameter settings are defined as follows: the frequency is set as $2.4$ GHz, $d=\frac{\lambda}{2}$ m, $A=\frac{\lambda ^2}{4\pi}$, $\gamma_1=\gamma_2=30$ dB, $\frac{P}{\sigma_1^2}=\frac{P}{\sigma_2^2}=30$ dB, $r_1=10$ m, and $r_2=5$ m. UT 1 is located in the direction $(\theta_1,\phi_1)=(\frac{\pi}{3},\frac{2\pi}{3})$; and UT 2 is located in the same direction or the different direction, $(\theta_2,\phi_2)=(\frac{2\pi}{3},\frac{\pi}{3})$, with respect to UT 1, denoted as ``SD'' and ``DD'' in the figures, respectively.
	\vspace{-5pt}
	\subsection{Channel Correlation Factor}
	Fig.~\ref{rho} illustrates the values of the CCF $\rho$ when UT 2 is in various directions for both NF and FF
	scenarios. It can be observed that, with an array size of $65\times65$, the CCF $\rho_{\mathsf{n}}$ in the NF model is very close to zero regardless of UT 2's direction, verifying that $\lim_{M_x,{M}_z\rightarrow \infty}\rho _{\mathsf{n}}\approx0$, which demonstrates the asymptotic orthogonality of NFC for UTs in different locations. In the FF context, the value of the CCF $\rho_{_{\mathsf{F}}}$ is also approximately zero when UT 2 is in a distinct direction from UT 1. However, when they are located in the same direction, their channels are fully correlated, which yields $\rho_{_{\mathsf{F}}}=1$.

	\begin{figure*}[!t]
		\centering
\subfigbottomskip=0pt
	\subfigcapskip=-2pt
\setlength{\abovecaptionskip}{3pt}
		\subfigure[NF MAC sum-rate capacity with respect to $(\Delta\theta,\Delta\phi)$.]
		{
			\includegraphics[height=0.29\textwidth]{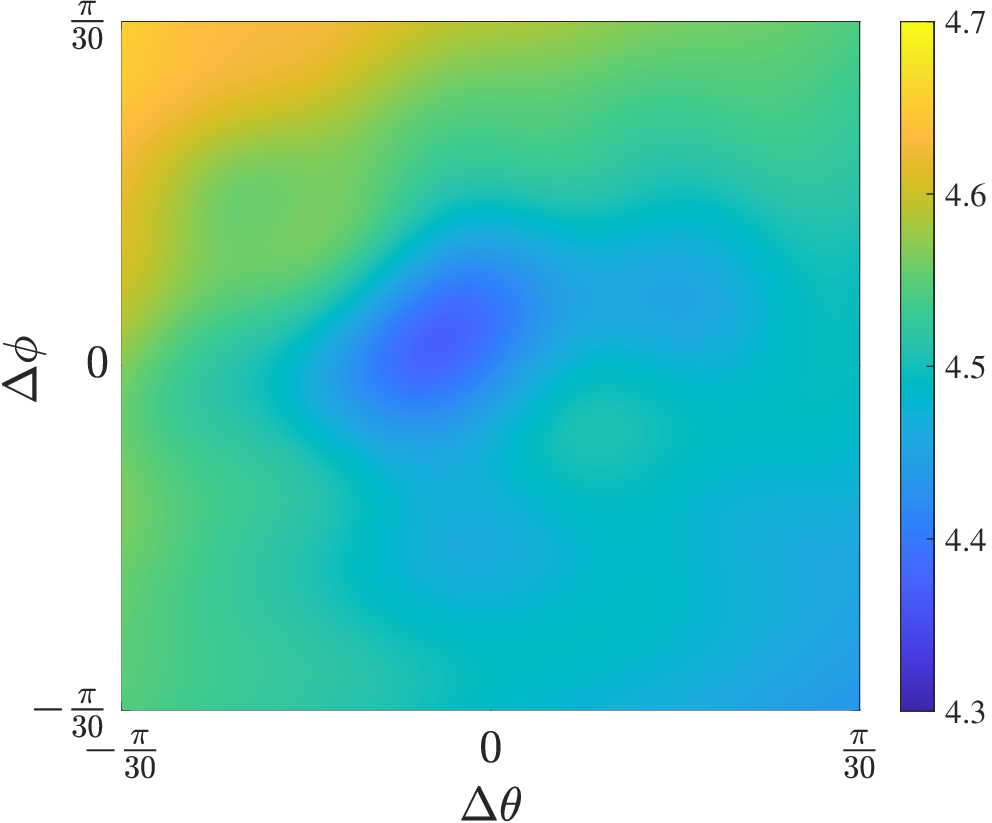}
			\label{per_nf}	
		}
		\quad
		\subfigure[FF MAC sum-rate capacity with respect to $(\Delta\theta,\Delta\phi)$.]
		{
			\includegraphics[height=0.29\textwidth]{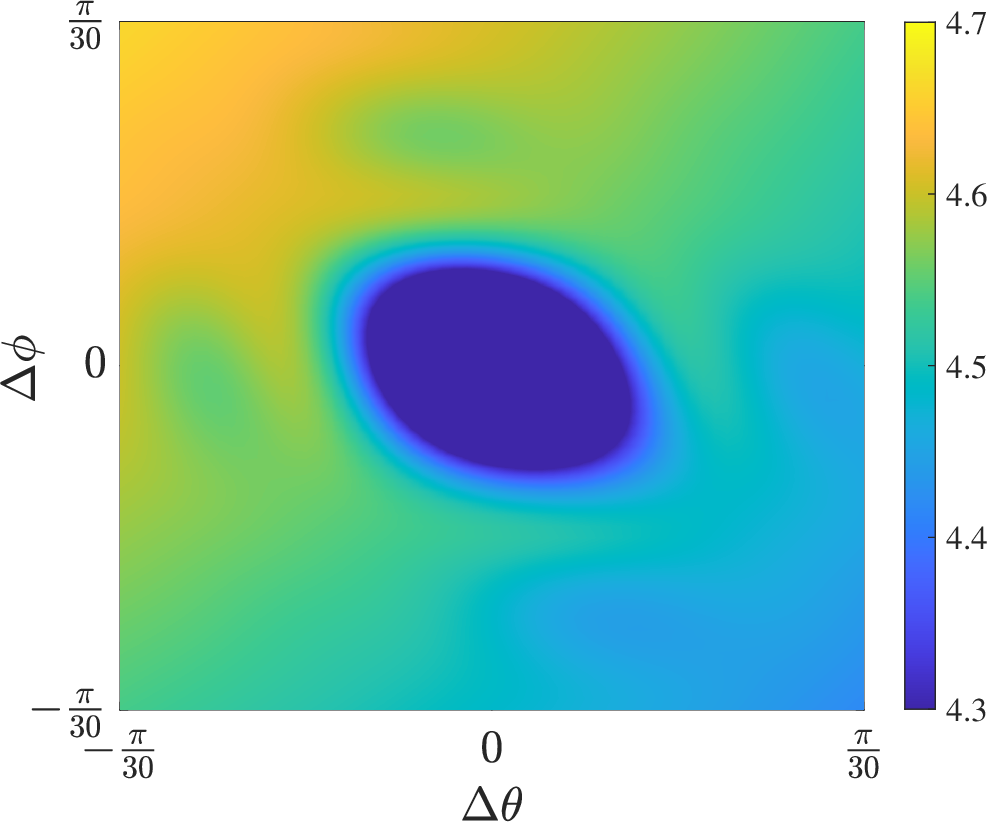}
			\label{per_ff}	
		}
		\caption{Illustration of impacts of small perturbations in angle difference for co-directional UTs on capacity, with $M_x=M_z=45$.}
		\vspace{-5pt}
		\label{per}
	\end{figure*}
	
	\begin{figure*}[!t]
		\begin{minipage}[t]{0.5\linewidth}
			\centering
			\setlength{\abovecaptionskip}{2pt}
			\includegraphics[height=0.5\textwidth]{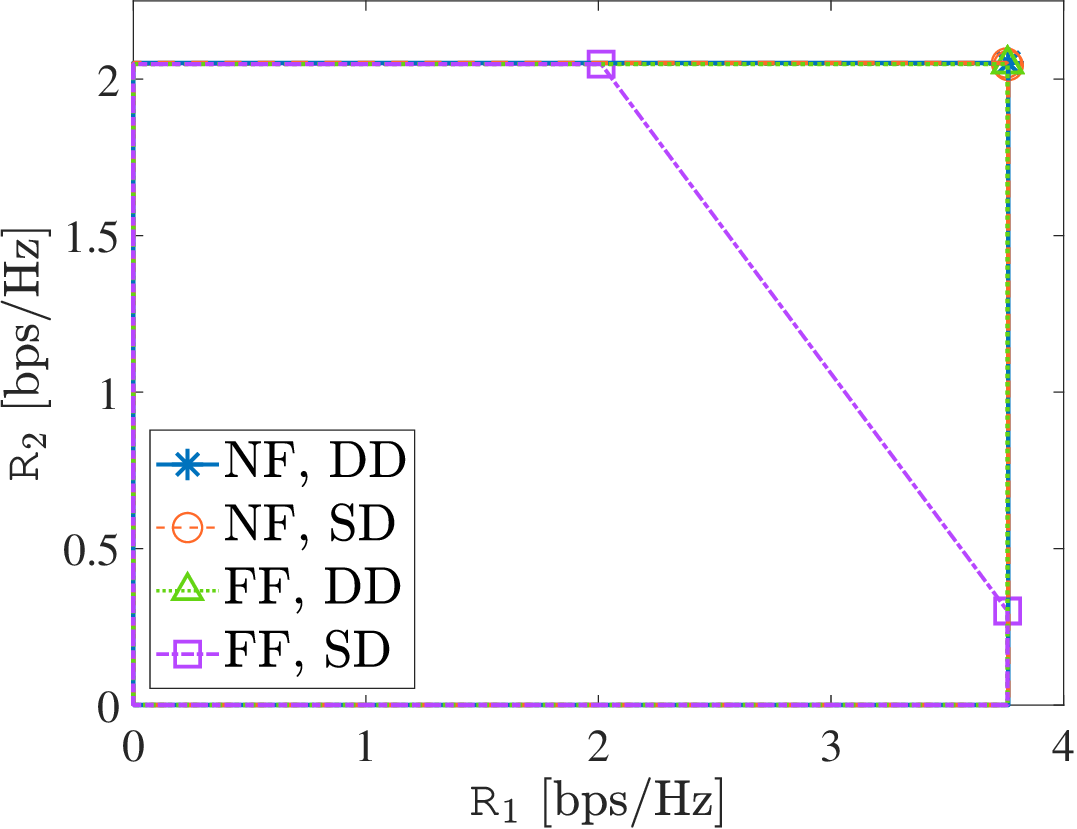}
			\caption{MAC capacity regions with $M_x=M_z=65$.}
			\vspace{-10pt}
			\label{MAC_2d}
		\end{minipage}
		\begin{minipage}[t]{0.5\linewidth}
			\centering
			\setlength{\abovecaptionskip}{2pt}
			\includegraphics[height=0.5\textwidth]{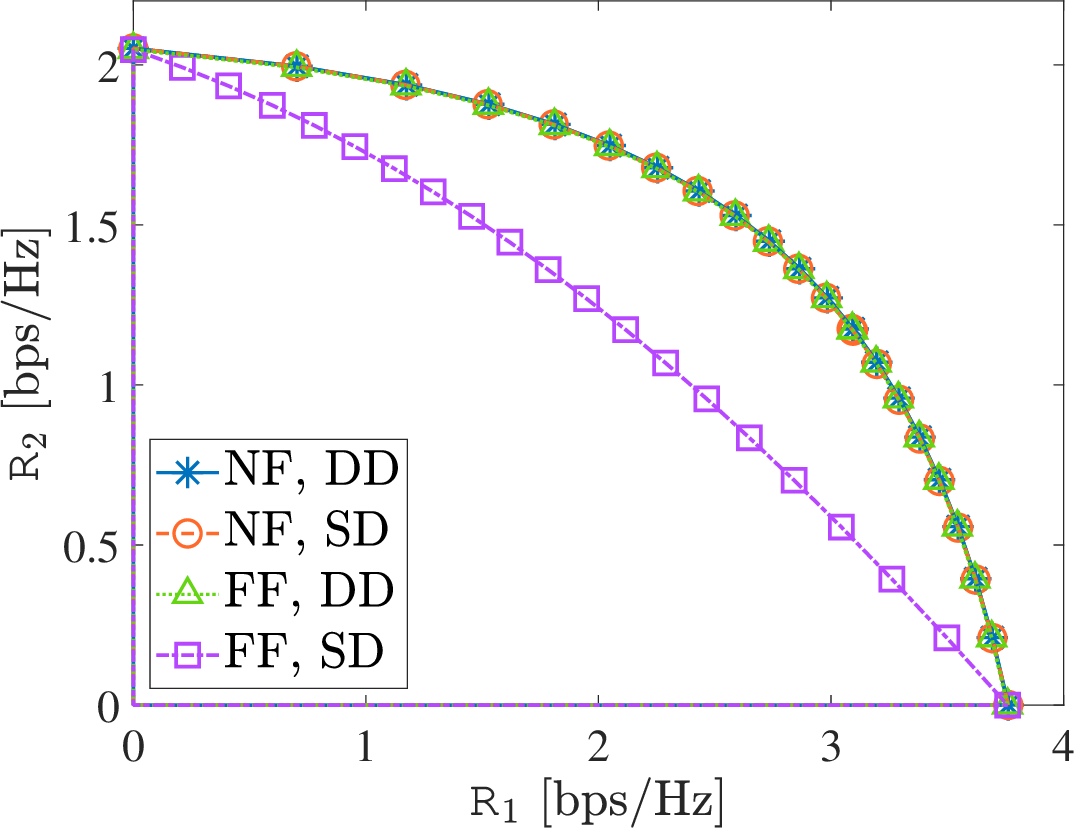}
			\caption{BC capacity regions with $M_x=M_z=65$.}
			\vspace{-10pt}
			\label{BC_2d}
		\end{minipage}
	\end{figure*}

	\vspace{-5pt}
	\subsection{Multiple Access Channel \& Broadcast Channel}
	\subsubsection{Sum-Rate Capacity}
	Fig.~\ref{MAC} and Fig.~\ref{BC} show the sum-rate capacity of the MAC and the BC concerning the number of array elements, respectively. We can observe that under both the MAC and BC, as $M$ increases, the NF capacity and the FF capacity follow distinctly different scaling patterns. Specifically, for relatively small values of $M$, when UT 2 is in the different direction with UT 1, the sum-rate capacity under the NF and FF models are similar and exhibit a linear increase with $\log M$. This similarity is attributed to the small array size, where the differences between NF and FF propagation effects are minimal and both models are accurate. However, when $M$ is sufficiently large, the disparity in channel powers and phases across the array becomes significant. In this case, the capacity under the FF model is overestimated due to the ignorance of such variations in wave propagation. As a result, the FF capacity will grow unboundedly with $M$, potentially breaking the energy-conservation laws, which is aligned with the statements in \textbf{Remark~\ref{MAC_unbound}} and \textbf{\ref{BC_unbounded}}. Conversely, as $M$ increases, the NF sum-rate capacity approach finite upper limits accurately tracked by our derived asymptotic results. This behavior is consistent with \textbf{Remark~\ref{MAC_constant}} and \textbf{\ref{BC_constant}}, revealing the superior accuracy for the NF channel model.

	Additionally, when UT 2 is in the same direction with UT 1, for both the MAC and BC, the capacity under the FF model are lower than in the case where the UTs are in the different directions, exhibiting a smaller scaling law with $M$. These observations corroborate our previous analysis. By contrast, in the NF model, the capacity for UTs in the same direction are nearly equivalent to those when they are located in different directions, surpassing the FF counterparts.

	While we find that the capacity of the MAC and BC with co-directional UTs can be improved under NFC compared to the FF case, it is nearly impossible for UTs to be positioned in the exact same direction in real-word scenarios, typically only similarly. Therefore, it would be crucial to explore how small perturbations in the angle difference between the UTs affect capacity. By letting $(\theta_2,\phi_2)=(\theta_1+\Delta\theta,\phi_1+\Delta\phi)$ with $(\Delta\theta,\Delta\phi)$ denoting the small angle perturbations, Fig.~\ref{per} illustrates how MAC capacity fluctuates with changes in $(\Delta\theta,\Delta\phi)$, noting that similar outcomes are expected for the BC, which are not shown for for brevity. It is clear that the NF capacity exceeds the FF capacity when the angle perturbations $\Delta\theta$ and $\Delta\phi$ are very small. More importantly, as the perturbations increase, the NF capacity grows more rapidly than the FF capacity, meaning that the NF capacity is more sensitive to the angle perturbations. In particular, compared to cases where the UTs are in the same direction, even slight angle differences can significantly enhance capacity, underscoring NFC's better angular resolution, which confirms the superior performance of NFC for UTs in the same direction or with small angle perturbations.

	\subsubsection{Capacity Region}
	In Fig.~\ref{MAC_2d} and Fig.~\ref{BC_2d}, the capacity regions for the MAC and the BC are presented, respectively. It is worth to note that the capacity regions under the NF model remain almost unchanged regardless of UT 2's direction relative to UT 1. These regions are characterized by rectangular shapes and resemble the FF capacity regions achieved when the UTs are in the different directions. On the other hand, the FF capacity regions for the UTs located in the same direction are shaped as pentagons, diverging from the rectangular regions observed under the NF model and the FF model with UTs in different directions. In particular, the rectangular capacity regions of the NF model envelop the pentagonal regions observed in the FF model for co-directional UTs. This encapsulation underscores the superior performance of NFC in managing signal correlation and interference, offering higher channel capacity for co-directional UTs.

\begin{figure*}[!t]
\centering
    \subfigbottomskip=0pt
	\subfigcapskip=-2pt
    \subfigure[MAC capacity region (SD).]
    {
        \includegraphics[height=0.187\textwidth]{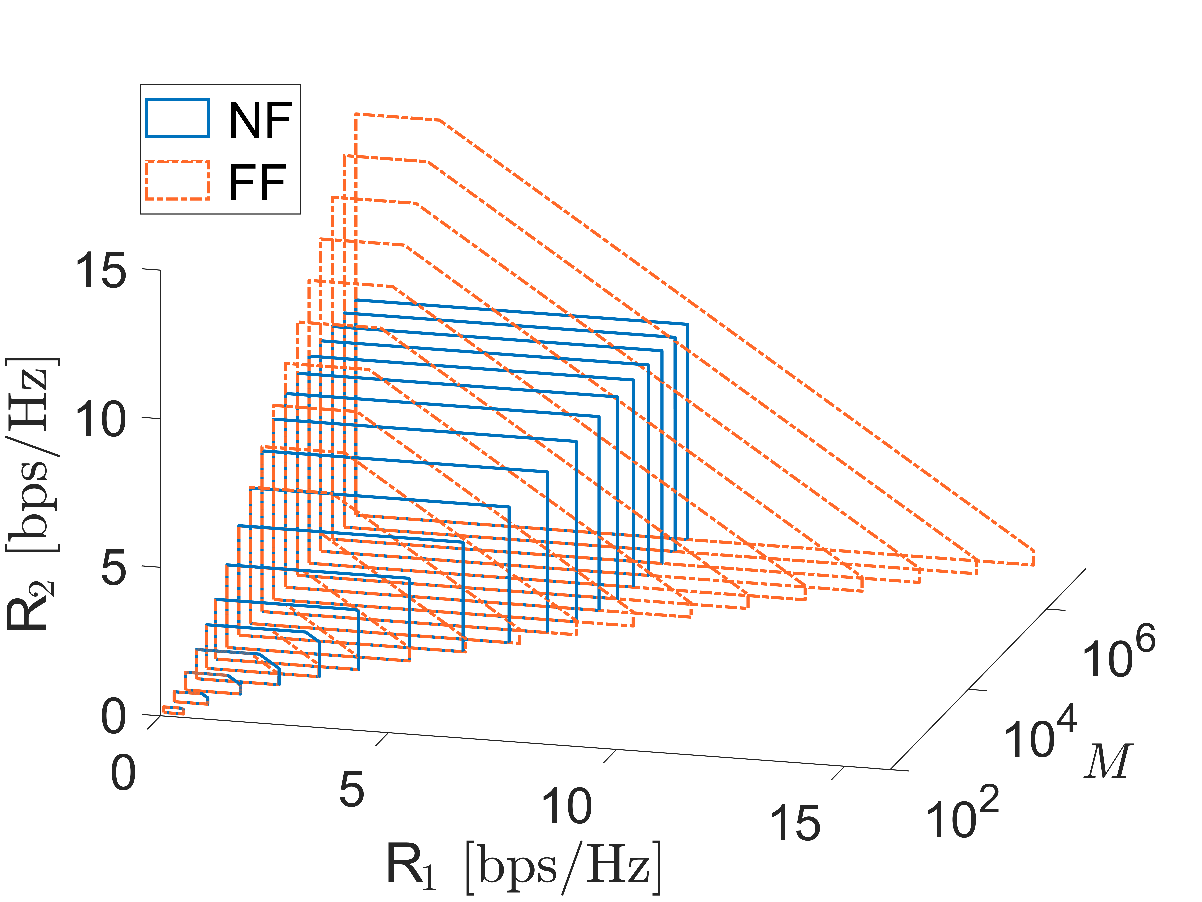}
	   \label{MAC_region_same}	
    }\hspace{-12pt}
   \subfigure[MAC capacity region (DD).]
    {
        \includegraphics[height=0.187\textwidth]{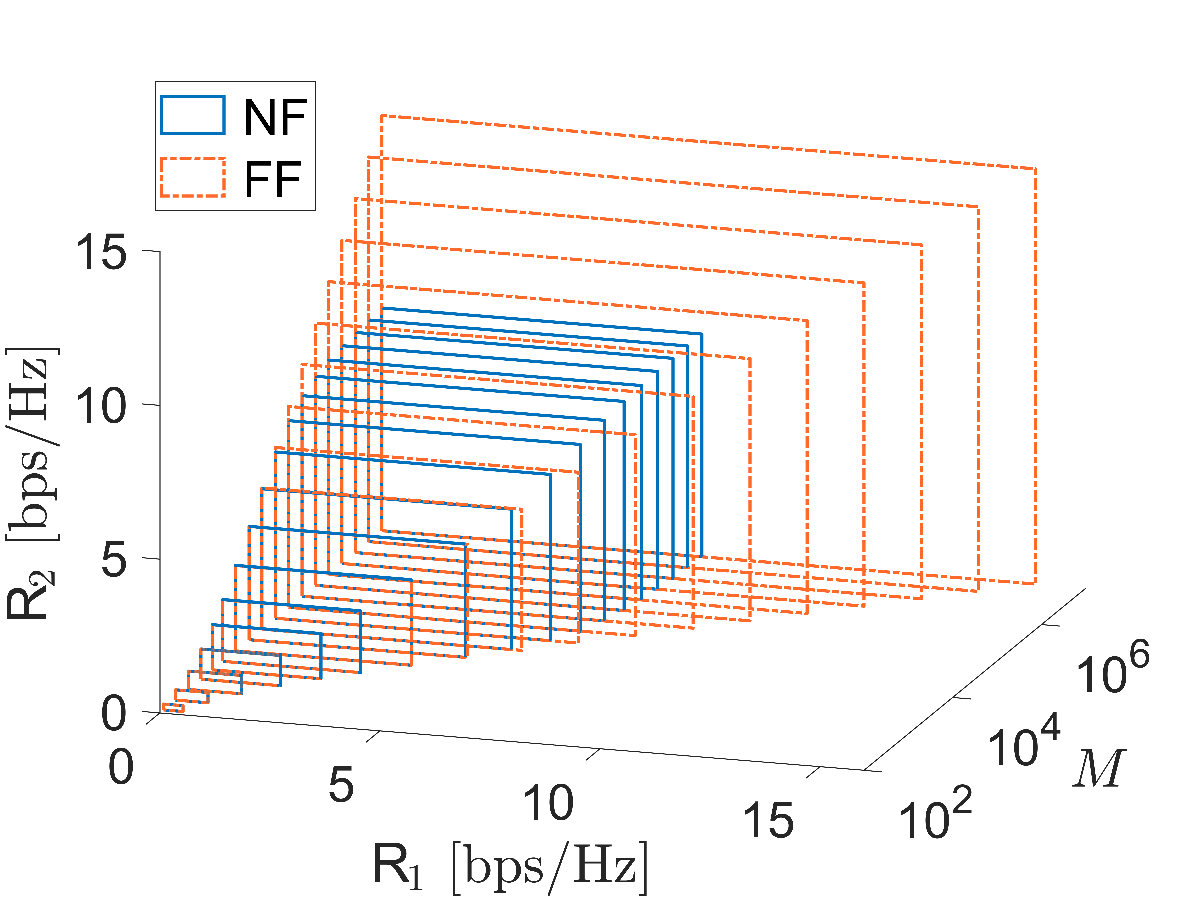}
	   \label{MAC_regions}	
    }\hspace{-12pt}
    \subfigure[BC capacity region (SD).]
    {
        \includegraphics[height=0.187\textwidth]{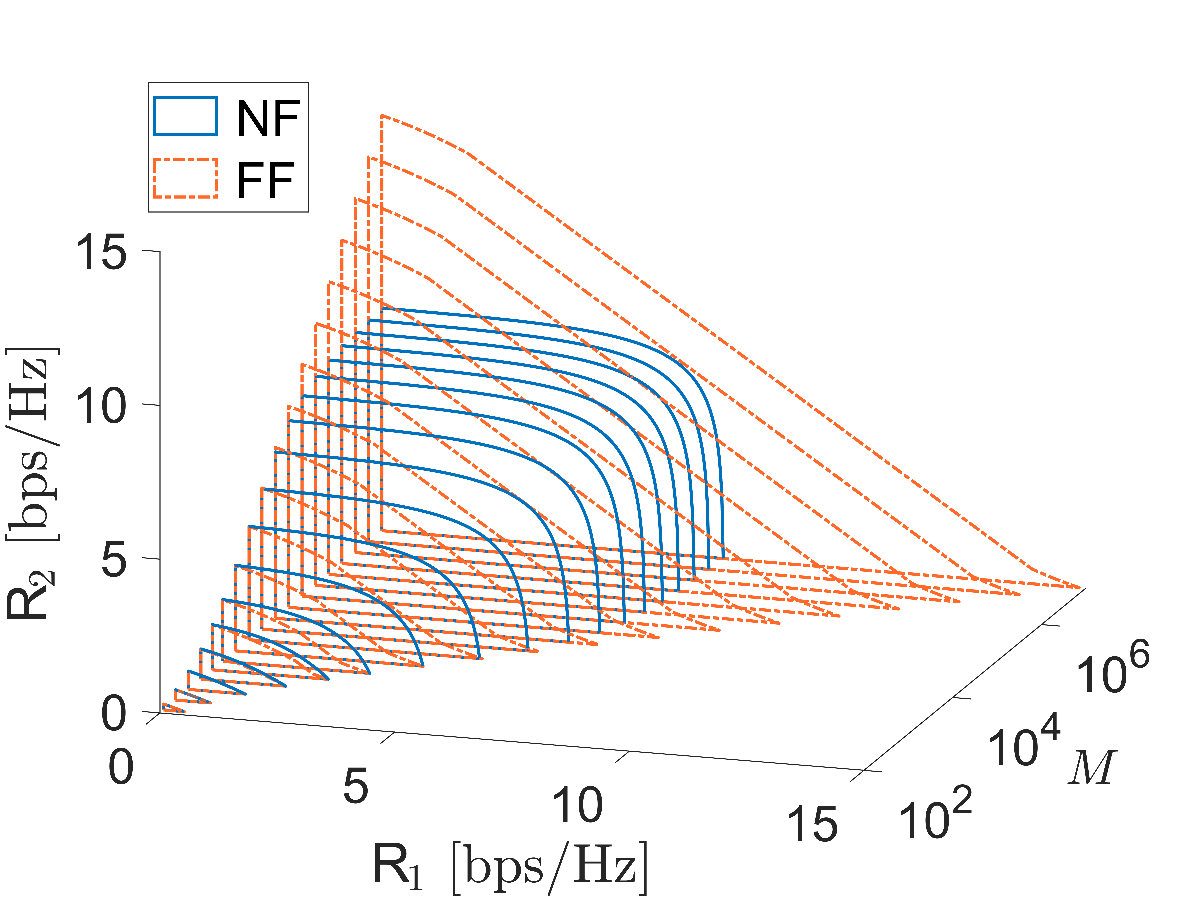}
	   \label{BC_region_same}	
    }\hspace{-12pt}
   \subfigure[BC capacity region (DD).]
    {
        \includegraphics[height=0.187\textwidth]{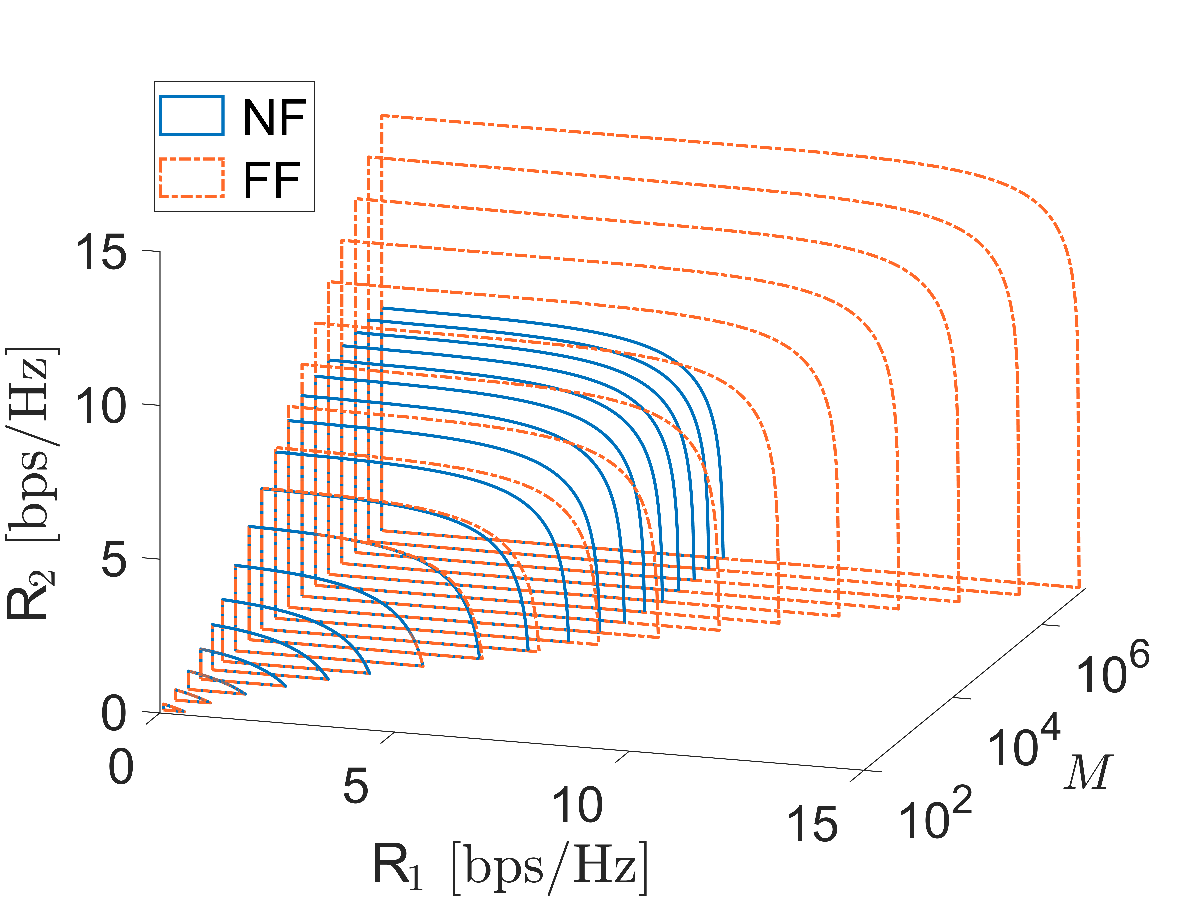}
	   \label{BC_region_diff}	
    }
\caption{MAC channel capacity and BC channel capacity versus the antenna number $M$.}
\label{3d}
\vspace{-10pt}
\end{figure*}
	
	Moreover, Fig.~\ref{3d} provides insightful visualizations of how the capacity regions for the MAC and BC evolve with changes in the number of array elements, contrasting the behaviors of NF and FF models. The visualizations demonstrate that the capacity regions under the NF model stabilize to finite areas as the array size increases, verifying the correctness of \textbf{Remark~\ref{MAC_rectangle}} and \textbf{\ref{BC_square}}. In contrast, the capacity regions in the FF context can expand without limitation with the number of array elements, which is unreasonable under energy consideration.

	\begin{figure*}[!t]
\centering
\subfigbottomskip=0pt
	\subfigcapskip=-2pt
\setlength{\abovecaptionskip}{2pt}
		\subfigure[Multicast Capacity versus $M$ with different $\xi$.]
		{
			\includegraphics[height=0.265\textwidth]{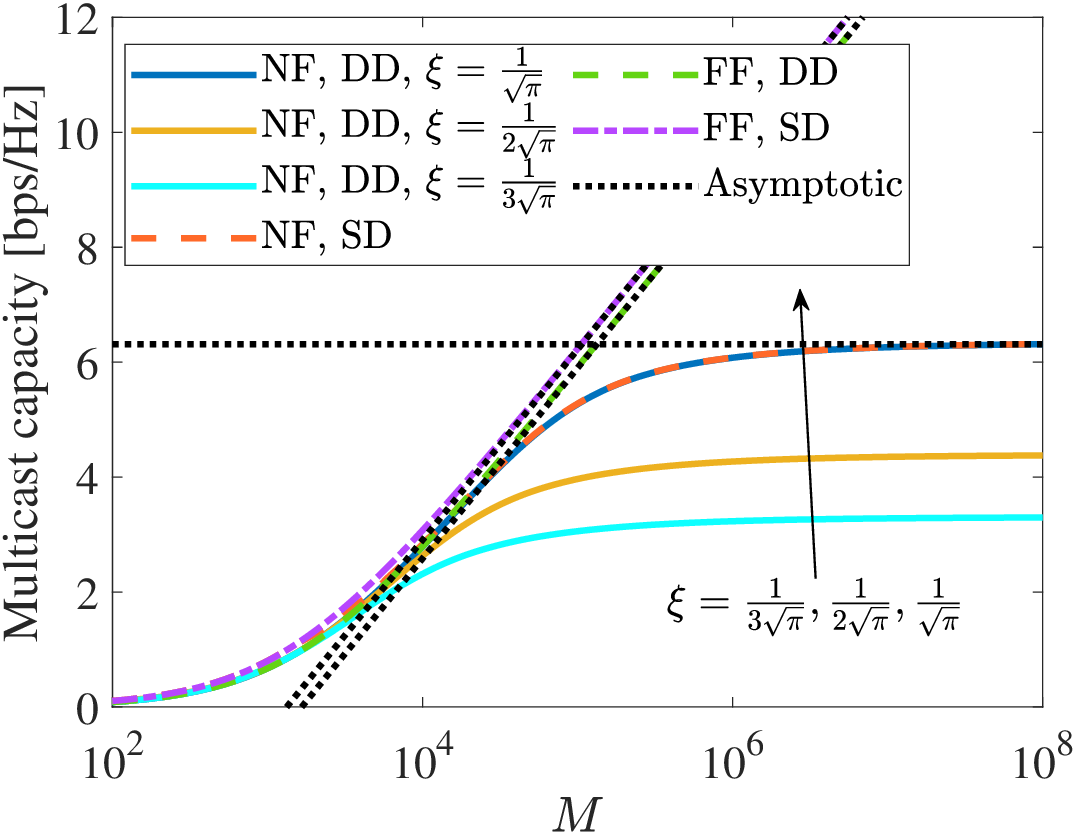}
			\label{MC_M}	
		}
		\quad\quad
		\subfigure[Multicast Capacity versus $r_2$ with $M_x\!=\!M_z\!=\!551$.]
		{
			\includegraphics[height=0.265\textwidth]{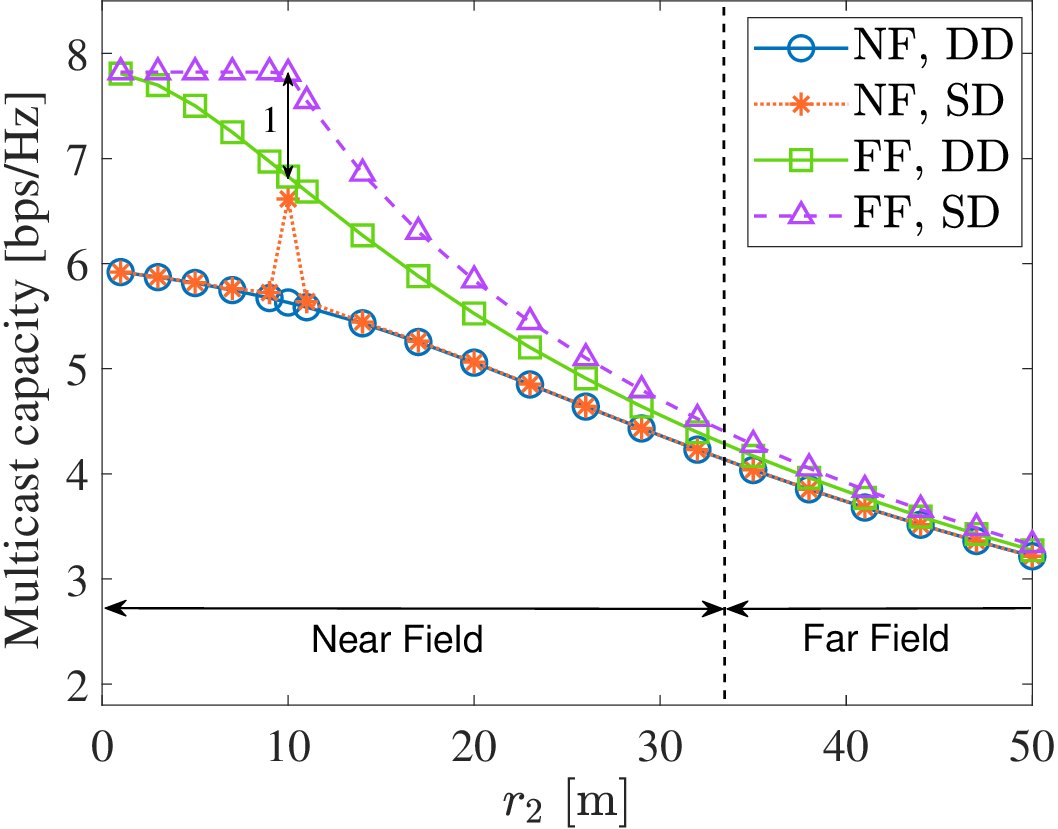}
			\label{MC_r}	
		}
		\caption{Multicast capacity.}
		\label{MC}
		\vspace{-10pt}
	\end{figure*}
	
	\subsection{Multicast Channel}
	Fig.~\ref{MC_M} demonstrates the multicast capacity versus $M$, with different values of the AOR $\xi$. For the NF case, the musticast capacity converge to finite upper limits accurately tracked by our derived asymptotic results, which are positively correlated to the values of $\xi$. This behavior validates the conclusions in \textbf{Remark~\ref{MC_constant}}. Additionally, contrary to the patterns observed in the MAC and BC, the FF multicast capacity for co-directional UTs exceeds that for UTs in different directions by a small gap which tend to be constant as $M$ becomes large, verifying our theoretical analysis. This distinct observation is attributed to the unique nature of the MC, where transmitting the same message to all UTs benefits from a high level of channel correlation.

	In Fig.~\ref{MC_r}, we present the multicast capacity as a function of the UT 2's distance $r_2$ with $r_1$ fixed at $10$ m. We observe that when the UTs are located near the BS, the multicast capacity under the FF model is significantly overestimated. However, this overestimation diminishes as the distance from the UT to the BS increases. This occurs because the NF effect is more pronounced at short distances and becomes less significant as the UT moves toward the far field. Further, the two lines under NF models are almost overlapping, with a notable exception at $r_2 = 10$ m. At this point, the scenario where UT 2 is in the same location as UT 1 exhibits a markedly higher capacity, highlighting the preference for high channel correlation in multicast settings. Regarding the FF case, the gap between the two lines initially increases with $r_2$, reaching a peak when UTs are equidistant from the BS ($\Upsilon = 1$). Beyond this point, as $r_2$ continues to increase, the gap decreases in accordance with $\Upsilon$, as demonstrated in \eqref{MC_ratio}.

	\section{Conclusion}\label{conclusion}
	This article has investigated the channel capacity of NF multiuser communications, by specializing the multiuser channels into the MAC, BC and MC. We derived the sum-rates capacity and capacity regions for the MAC and BC, and the multicast capacity for the MC. By comparing the results with the FF counterparts, we have drawn the following conclusions: 1) the additional asymptotic orthogonality of NFC in the distance domain can improve capacity of the MAC and BC for co-directional users, but it is not beneficial to the MC; 2) rather than growing unboundedly, the NF capacity of the three channels converge to finite values as the number of array elements increases, underscoring the importance of accurate channel modeling for NFC.
	
	\begin{appendix}   
		\setcounter{equation}{0}
		\renewcommand\theequation{A\arabic{equation}}
		\subsection{Proof of Lemma~\ref{MAC_region_lem}}\label{Appendix:D}
		When the SIC decoding order $1\rightarrow2$ is employed, the message of UT 1 is decoded firstly by treating the message of UT 2 as interference. In this case, the rate of UT 1 reads
		\begin{align}
			{\mathsf{R}}_{1}^{1\rightarrow2}&= \log_2\det({\mathbf{I}}_M+{\gamma_1}{\mathbf{h}}_1{\mathbf{h}}_1^{\mathsf{H}}({\mathbf{I}}_M+{\gamma_2}{\mathbf{h}}_2{\mathbf{h}}_2^{\mathsf{H}})^{-1})\notag\\
			&\overset{( a )}{=}\log _2\left(1+{\gamma_1}\mathbf{h}_1^{\mathsf{H}}\left(\mathbf{I}_M+{\gamma_2}\mathbf{h}_2\mathbf{h}_{2}^{\mathsf{H}}\right)^{-1}\mathbf{h}_{1}\right)\notag\\
			&\overset{( b )}{=}\log _2\bigg( 1+{\gamma_1}\mathbf{h}_1^{\mathsf{H}}\bigg( \mathbf{I}_M-\frac{{\gamma_2}\mathbf{h}_2\mathbf{h}_{2}^{\mathsf{H}}}{1+{\gamma_2}\left\| \mathbf{h}_2 \right\| ^2} \bigg) \mathbf{h}_{1} \bigg), \label{b1}
		\end{align}
		where equality $(a)$ is attained by using the Sylvester’s identity, and $(b)$ is attained by using the Woodbury matrix identity. The results of \eqref{r12} can be obtained by performing some manipulations on \eqref{b1}. After the message of UT 1 is decoded, it will be removed from the received signal, and the message of UT 2 will be decoded without interference. Consequently, the rate of UT 2 is given by 
		\begin{align}
			{\mathsf{R}}_{2}^{1\rightarrow2}=\log _2\det(\mathbf{I}_M+\gamma_2\mathbf{h}_2\mathbf{h}_{2}^{\mathsf{H}})
			\overset{\left( a \right)}{=} \log_2\big(1+\gamma_2\left\| \mathbf{h}_2 \right\| ^2\big),  
		\end{align}
		where $(a)$ follows from the Sylvester's identity. ${\mathsf{R}}_{1}^{2\rightarrow1}$ and ${\mathsf{R}}_{2}^{2\rightarrow1}$ can be derived following the similar steps.
		\subsection{Proof of Theorem~\ref{MAC_capacity_NF_the}}\label{Appendix:B}
		Based on \eqref{distance} and \eqref{channel_NF}, the channel gain under the NF model is calculated as 
		\begin{equation} \label{d1}
			\begin{split}
				&\mathsf{g}_k^{\mathsf{n}}=\frac{A\Psi _k}{4\pi r_k^{2}}\sum\nolimits_{m_x\in \mathcal{M}_x}\sum\nolimits_{m_z\in\mathcal{M}_z}\\
				&~~\times \big((m_x^2+m_z^2)\epsilon _k^2-2m_x\epsilon _k\Phi_k-2m_z\epsilon _k\Omega_k+1\big)^{-\frac{3}{2}}. 
			\end{split}
		\end{equation}
		We define the function 
		\begin{align}
			f\left( x,z \right) \triangleq (x^2+z^2-2\Phi_kx-2\Omega_kz+1)^{-\frac{3}{2}}    
		\end{align}
		in the rectangular area $\mathcal{H} =\left\{ \left( x,z \right) \mid -\frac{M_x\epsilon _k}{2}\leq x\leq \frac{M_x\epsilon _k}{2},-\frac{M_z\epsilon _k}{2}\leq z\leq \frac{M_z\epsilon _k}{2} \right\}  $ that is then partitioned into $M_xM_z$ sub-rectangles, each with equal area $\epsilon _k^2$. Since $\epsilon _k\ll 1$, we have $f\left( x,z \right) \approx f\left( m_x\epsilon _k,m_z\epsilon _k \right) $ for $\forall \left( x,z \right) \in \left\{ \left( x,z \right) \mid \left( m_x-\frac{1}{2} \right) \epsilon _k\leq x\leq \left( m_x+\frac{1}{2} \right) \epsilon _k,\left( m_z-\frac{1}{2} \right) \epsilon _k\leq z \right.$ $\left.\leq \left( m_z+\frac{1}{2} \right) \epsilon _k \right\}$. Based on the concept of integral, we have 
		\begin{align}
			\sum\nolimits_{m_x,m_z}{f\left( m_x\epsilon _k,m_z\epsilon _k \right) \epsilon _k^2}\approx \iint_{\mathcal{H}}{f\left( x,z \right) dxdz}.    
		\end{align}
		As a result, \eqref{d1} can be rewritten as 
		\begin{equation}\label{a6}
			\begin{split}
				\mathsf{g}_k^{\mathsf{n}}=\frac{\xi \Psi _k}{4\pi}\int_{-\frac{M_z\epsilon _k}{2}}^{\frac{M_z\epsilon _k}{2}}{\int_{-\frac{M_x\epsilon _k}{2}}^{\frac{M_x\epsilon _k}{2}}}f(x,z)dxdz.  
			\end{split}
		\end{equation}
		We can calculate the inner integral with the aid of \cite[Eq. (2.264.5)]{integral} and then the outer integral with the aid of \cite[Eq. (2.284)]{integral}, which yields the results of \eqref{channelgain_NF}. 
		
		Following similar steps to obtain \eqref{a6}, the CCF can be written as follows:
		\begin{align}
			\rho &=\prod_{k=1}^2{\frac{A\Psi _k}{4\pi r_{k}^{2}\mathsf{g}^{\mathsf{n}}_k}}
			\left| \sum_{m_x,m_z}{f_1\left( m_x\epsilon _{1},m_z\epsilon _{1} \right) g_2\left( m_x\epsilon _{1},m_z\epsilon _{1} \right)} \right|^2\nonumber\\
			&=\prod_{k=1}^2{\frac{A\Psi _k}{4\pi r_{k}^{2}\mathsf{g}^{\mathsf{n}}_k\epsilon _{1}^2}}
			\left| \int_{\frac{-M_z\epsilon _{1}}{2}}^{\frac{M_z\epsilon _{1}}{2}}{\int_{\frac{-M_x\epsilon _{1}}{2}}^{\frac{M_x\epsilon _{1}}{2}}{f_1\left( x,z \right) f_2\left( x,z \right) \mathrm{d}x}\mathrm{d}z} \right|^2.    
		\end{align}
		The above integrals can be numerically evaluated by applying the Chebyshev-Gauss quadrature rule, i.e., $\int_{-1}^1{\frac{f\left( x \right)}{\sqrt{1-x^2}}\mathrm{d}}x\approx \sum_{t=1}^T{f\left( x_j \right)}$ with $x_t=\cos \left( \frac{ 2t-1 }{2T} \pi \right) $, resulting in the expressions of \eqref{rho_NF}.
		
		\subsection{Proof of Corollary~\ref{ula_mac_cor}}\label{Appendix:C}
		With $M_x=1$ and $M_z=M$, \eqref{channelgain_NF} can be rewritten as
		\begin{equation}
			\begin{split}
				\mathsf{g}_k^{\mathsf{n}}=&\frac{\xi}{4\pi}\sum\nolimits_{x\in \left\{ \frac{\epsilon _k}{2}\pm \Phi _k \right\}}^{}\sum\nolimits_{z\in \left\{ \frac{M\epsilon _k}{2}\pm \Omega _k \right\}}^{}\\
				&\times{\arctan \bigg( \frac{xz}{\Psi _k\sqrt{\Psi _k^{2}+x^2+z^2}} \bigg)}\triangleq g(\epsilon _k).
			\end{split}
		\end{equation}
		Since $\epsilon _k\ll 1$, we can utilizing the first-order Taylor approximation $\mathsf{g}_k^{\mathsf{n}}\approx g\left( 0 \right) +g^{\prime}\left( 0 \right) \epsilon _k$, leading to the results of \eqref{channelgain_ula}. When $M\rightarrow \infty $, we can obtain $\lim_{M\rightarrow \infty} \mathsf{g}_k^{\mathsf{n}}=\frac{\xi \epsilon _k\sin \phi _k}{2\pi \sin \theta _k}$ by applying L'Hôpital's rule, which yield the results of \eqref{ula_mac}.

		\subsection{Proof of Theorem~\ref{MC_capacity_the}}\label{Appendix:F}
		The optimal solution of problem \eqref{Multicast_Capacity_Problem1} can be derived from the KKT conditions as follows:
		\setlength\abovedisplayskip{5pt}
		\setlength\belowdisplayskip{5pt}
		\begin{align}
			&\nabla(-t)+\eta\nabla(\lVert{\mathbf{w}}\rVert^2-1)+\mu_1\nabla (t-\lvert{\bar{\mathbf{h}}}_1^{\mathsf{H}}\mathbf{w}\rvert^2)\notag\\
			&\qquad \qquad \qquad \qquad \qquad \ +\mu_2\nabla (t-\lvert{\bar{\mathbf{h}}}_2^{\mathsf{H}}\mathbf{w}\rvert^2)={\mathbf{0}}, \label{KKT_1} \\ 
			&\mu_1(t-\lvert{\bar{\mathbf{h}}}_1^{\mathsf{H}}\mathbf{w}\rvert^2)=0, \ \mu_2(t-\lvert{\bar{\mathbf{h}}}_2^{\mathsf{H}}\mathbf{w}\rvert^2)=0,\\
			&\mu_1\geq0,\ \mu_2\geq0,\ \eta\in{\mathbbmss{R}},
		\end{align}
		where $\bar{\mathbf{h}}_1=\sigma_1^{-1}\mathbf{h}_1$, $\bar{\mathbf{h}}_2=\sigma_2^{-1}\mathbf{h}_2$, and $\{\eta,\mu_1,\mu_2\}$ are real-valued Lagrangian multipliers. From \eqref{KKT_1}, we can obtain
		{\setlength\abovedisplayskip{2.5pt}
			\setlength\belowdisplayskip{2.5pt}
			\begin{align}
				&\mu_1+\mu_2=1,\label{KKT_1_Dev2}\\
				&(\mu_1{\bar{\mathbf{h}}}_1{\bar{\mathbf{h}}}_1^{\mathsf{H}}+\mu_2{\bar{\mathbf{h}}}_2{\bar{\mathbf{h}}}_2^{\mathsf{H}}){\mathbf{w}}=\eta{\mathbf{w}}.\label{KKT_1_Dev1}   
		\end{align}}
		It follows from \eqref{KKT_1_Dev2} that $\mu_1$ and $\mu_2$ cannot be $0$ at the same time. Particularly, three different cases are discussed as follows.
		\subsubsection{$\mu_1=1$ and $\mu_2=0$}: In this case, we have $t=\lvert{\bar{\mathbf{h}}}_1^{\mathsf{H}}\mathbf{w}\rvert^2$, from which we can readily deduce the optimal solutions ${\mathbf{w}^\star}={{\mathbf{h}}_1}/{\lVert{\mathbf{h}}_1\rVert}$ and $t^\star=\lVert{\bar{\mathbf{h}}}_1\rVert^2$. Substituting the results into the constraint \eqref{Power_Min_Problem_Instantaneous_Cons} yields ${\lVert{\bar{\mathbf{h}}}_{1}\rVert^2\leq\rho\lVert{\bar{\mathbf{h}}}_{2}\rVert^2}$.
		\subsubsection{$\mu_1=0$ and $\mu_2=1$}: Following similar steps to the first case, we can obtain ${\mathbf{w}^\star}={{\mathbf{h}}_2}/{\lVert{\mathbf{h}}_2\rVert}$ with the condition ${\lVert{\bar{\mathbf{h}}}_{2}\rVert^2\leq\rho\lVert{\bar{\mathbf{h}}}_{1}\rVert^2}$.
		\subsubsection{$\mu_1>0$ and $\mu_2>0$}:
		In this case, we have
		\begin{equation}\label{KKT_1_Dev3}
			t=\lvert{\bar{\mathbf{h}}}_1^{\mathsf{H}}\mathbf{w}\rvert^2=\lvert{\bar{\mathbf{h}}}_2^{\mathsf{H}}\mathbf{w}\rvert^2.
		\end{equation}
		According to \eqref{KKT_1_Dev1} and \eqref{KKT_1_Dev3}, we can write $\mathbf{w}$ as a linear combination of ${\bar{\mathbf{h}}}_1$ and ${\bar{\mathbf{h}}}_2$:
		{\setlength\abovedisplayskip{2.5pt}
			\setlength\belowdisplayskip{2.5pt}
			\begin{align}\label{wab}
				{\mathbf{w}}=a{\bar{\mathbf{h}}}_1+b{\bar{\mathbf{h}}}_2    
		\end{align}}
		with $\frac{a}{b}=\frac{{\mu_1}}
		{{\mu_2}{\rm{e}}^{-{\rm{j}}\angle({\mathbf{h}}_{1}^{\mathsf{H}}{\mathbf{h}}_{2})}}$. Substituting \eqref{wab} into \eqref{KKT_1_Dev1} gives
		\begin{align}\label{mu_eta}
			{\mu_1}\Big(\lVert{\bar{\mathbf{h}}}_1\rVert^2+\frac{b}{a}{\bar{\mathbf{h}}}_{1}^{\mathsf{H}}{\bar{\mathbf{h}}}_{2}\Big)={\mu_2}\Big(\lVert{\bar{\mathbf{h}}}_2\rVert^2+\frac{a}{b}{\bar{\mathbf{h}}}_{2}^{\mathsf{H}}{\bar{\mathbf{h}}}_{1}\Big)=\eta.    
		\end{align}
		By combining \eqref{mu_eta} and \eqref{KKT_1_Dev2}, we can derive $\mu _1=\frac{\left\| \bar{\mathbf{h}}_2 \right\| ^2-\left| \bar{\mathbf{h}}_{1}^{\mathsf{H}}\bar{\mathbf{h}}_2 \right|}{\left\| \bar{\mathbf{h}}_1 \right\| ^2+\left\| \bar{\mathbf{h}}_2 \right\| ^2-2\left| \bar{\mathbf{h}}_{1}^{\mathsf{H}}\bar{\mathbf{h}}_2 \right|}$, $\mu _2=\frac{\left\| \bar{\mathbf{h}}_1 \right\| ^2-\left| \bar{\mathbf{h}}_{1}^{\mathsf{H}}\bar{\mathbf{h}}_2 \right|}{\left\| \bar{\mathbf{h}}_1 \right\| ^2+\left\| \bar{\mathbf{h}}_2 \right\| ^2-2\left| \bar{\mathbf{h}}_{1}^{\mathsf{H}}\bar{\mathbf{h}}_2 \right|}$, and $\eta =\frac{\left\| \bar{\mathbf{h}}_1 \right\| ^2\left\| \bar{\mathbf{h}}_2 \right\| ^2-\left| \bar{\mathbf{h}}_{1}^{\mathsf{H}}\bar{\mathbf{h}}_2 \right|^2}{\left\| \bar{\mathbf{h}}_1 \right\| ^2+\left\| \bar{\mathbf{h}}_2 \right\| ^2-2\left| \bar{\mathbf{h}}_{1}^{\mathsf{H}}\bar{\mathbf{h}}_2 \right|}$. Given the prerequisites $\mu_1>0$ and $\mu_2>0$, it is required that ${\lVert{\bar{\mathbf{h}}}_{1}\rVert^2\ge \rho\lVert{\bar{\mathbf{h}}}_{2}\rVert^2}$ and ${\lVert{\bar{\mathbf{h}}}_{2}\rVert^2\ge \rho\lVert{\bar{\mathbf{h}}}_{1}\rVert^2}$ in this case. Consequently, since $\lVert{\mathbf{w}}\rVert=1$, $\mathbf{w}$ can be expressed as 
		\begin{equation}
			{\mathbf{w}}=\frac{\mu_1{\bar{\mathbf{h}}}_1+\mu_2{\bar{\mathbf{h}}}_2{\rm{e}}^{-{\rm{j}}\angle({\mathbf{h}}_{1}^{\mathsf{H}}{\mathbf{h}}_{2})}}
			{\sqrt{\mu_1^2\lVert{\bar{\mathbf{h}}}_1\rVert^2+\mu_2^2\lVert{\bar{\mathbf{h}}}_2\rVert^2+2\mu_1\mu_2\lvert{\bar{\mathbf{h}}}_{1}^{\mathsf{H}}{\bar{\mathbf{h}}}_{2}\rvert}}.
		\end{equation}
		It is worth noting that $\eta=\mu_1^2\lVert{\bar{\mathbf{h}}}_1\rVert^2+\mu_2^2\lVert{\bar{\mathbf{h}}}_2\rVert^2+2\mu_1\mu_2\lvert{\bar{\mathbf{h}}}_{1}^{\mathsf{H}}{\bar{\mathbf{h}}}_{2}\rvert$. Finally, the optimal beamforming vector under the case of $\mu _1>0$ and $\mu _2>0$ is obtained as 
		\begin{align}
			\mathbf{w}^\star=\frac{\mu_1}{\sqrt{\eta}\sigma_1}\mathbf{h}_1+\frac{\mu_2}{\sqrt{\eta}\sigma_2}\mathbf{h}_2{\rm{e}}^{-{\rm{j}}\angle({\mathbf{h}}_{1}^{\mathsf{H}}{\mathbf{h}}_{2})}.
		\end{align}
		Taking the three cases together with the substitutions of $\lVert{\mathbf{h}}_{k}\rVert^2=\mathsf{g}_k^{\mathsf{n}}$ and $\rho=\rho_{\mathsf{n}}$, we can obtain the results in \eqref{Optimal_Beamformer_Solution}, which then lead to the multicast capacity in \eqref{MC_capacity}.

		\subsection{Proof of Lemma~\ref{MC_K>2}}\label{Appendix:G}
		Based on \eqref{Multicast_Capacity}, the MC capacity satisfies
		\begin{align}
			\mathsf{C}_{\mathsf{MC}}&\le \max_{\mathbf{\Sigma }\succeq 0,\mathsf{tr}(\mathbf{\Sigma })\le P} \log _2\left( 1+\frac{1}{K}\sum\nolimits_{k=1}^K{\frac{\mathbf{h}_{k}^{\mathsf{H}}\mathbf{\Sigma h}_k}{\sigma _{k}^{2}}} \right),\notag\\
			&=\max_{\mathbf{\Sigma }\succeq 0,\mathsf{tr}(\mathbf{\Sigma })\le P} \log _2\left( 1+\frac{1}{K}\mathrm{tr}\left( \mathbf{\Sigma }\sum\nolimits_{k=1}^K{\frac{\mathbf{h}_k\mathbf{h}_{k}^{\mathsf{H}}}{\sigma _{k}^{2}}} \right) \right) \notag\\
			&\le \max_{\mathbf{\Sigma }\succeq 0,\mathsf{tr}(\mathbf{\Sigma })\le P} \log _2\left( 1+\frac{1}{K}\mathrm{tr}\left( \mathbf{\Sigma } \right) \mathrm{tr}\left( \sum\nolimits_{k=1}^K{\frac{\mathbf{h}_k\mathbf{h}_{k}^{\mathsf{H}}}{\sigma _{k}^{2}}} \right) \right) \notag\\
			&= \log _2\left( 1+\frac{P}{K}\sum\nolimits_{k=1}^K{\sigma _{k}^{-2}\left\| \mathbf{h}_k \right\| ^2} \right) .
		\end{align}
		
	\end{appendix}
	
	\bibliographystyle{IEEEtran}
	\bibliography{IEEEabrv}
\end{document}